\documentclass{aa}  

\usepackage{graphicx,txfonts,lscape,natbib}
\bibpunct{(}{)}{;}{a}{}{,} 

\begin{document}

\title{New ultracool subdwarfs identified in large-scale surveys using Virtual Observatory tools \thanks{Based on observations made with ESO Telescopes at the La Silla Paranal Observatory under programmes IDs 088.C-0250(A), 090.C-0832(A); Based on observations made with the Nordic Optical Telescope, operated by the Nordic Optical Telescope Scientific Association at the Observatorio del Roque de los Muchachos, La Palma, Spain, of the Instituto de Astrof\'isica de Canarias.; Based on observations made with the Gran Telescopio Canarias (GTC), installed in the Spanish Observatorio del Roque de los Muchachos of the Instituto de Astrof\'isica de Canarias, in the island of La Palma (programs GTC44-09B, GTC53-10B, GTC31-MULTIPLE-11B, GTC36/12B, and GTC79-14A); The data presented in this paper are gathered in a VO-compliant archive at http://svo2.cab.inta-csic.es/vocats/ltsa/}}

   \subtitle{II. SDSS DR7 vs UKIDSS LAS DR6, SDSS DR7 vs UKIDSS LAS DR8, SDSS DR9 vs UKIDSS LAS DR10, and SDSS DR7 vs 2MASS}

 \author{
N.\ Lodieu\inst{1,2},
M.\ Espinoza Contreras\inst{1,2}, 
M.\ R.\ Zapatero Osorio\inst{3}, 
E.\ Solano\inst{4,5},
M.\ Aberasturi\inst{4,5},
\newline
E.\ L.\ Mart\'in\inst{3},
C.\ Rodrigo\inst{4,5}
}

\institute{Instituto de Astrof\'isica de Canarias, C/V\'ia L\'actea s/n, E-38200, La Laguna, Tenerife, Spain
\and
Departamento de Astrof\'isica, Universidad de La Laguna, E-38206, La Laguna, Tenerife, Spain
\and
Centro de Astrobiolog\'ia (CSIC-INTA), Ctra. Ajalvir km 4, E-28850, Torrej\'on de Ardoz, Madrid, Spain  
\and
Centro de Astrobiolog\'ia (INTA-CSIC), Departamento de Astrof\'isica. P.O. Box 78, E-28691, Villanueva de la Ca\~nada, Madrid, Spain  
\and
Spanish Virtual Observatory, Madrid, Spain  
}

   \date{Received \today{}; accepted \today{}}


  \abstract
  {}
   {We aim at developing an efficient method to search for late-type subdwarfs (metal-depleted dwarfs with spectral types $\geq$ M5) to improve the current statistics. Our objectives are: improve our knowledge of metal-poor low-mass dwarfs, bridge the gap between the late-M and L types, determine their surface density, and understand the impact of metallicity on the stellar and substellar mass function.}
   {We carried out a search cross-matching the Sloan Digital Sky Survey (SDSS) Data Release 7 (DR7) 
and the Two Micron All Sky Survey (2MASS), and different releases of SDSS and the United Kingdom 
InfraRed Telescope (UKIRT) Infrared Deep Sky Survey (UKIDSS) using STILTS, Aladin, and Topcat developed 
as part of the Virtual Observatory tools. We considered different photometric and proper motion criteria 
for our selection. We identified 29 and 71 late-type subdwarf candidates in each cross-correlation over 
8826 and 3679 square degrees, respectively (2312 square degrees overlap). We obtained our own low-resolution 
optical spectra for 71 of our candidates.
: 26 were observed with the Gran Telescopio de Canarias 
(GTC; R\,$\sim$\,350, $\lambda\lambda$5000--10000\,\,\AA{}), six with the Nordic Optical Telescope 
(NOT; R\,$\sim$\,450, $\lambda\lambda$5000--10700\,\,\AA{}), and 39 with the Very Large Telescope 
(VLT; R\,$\sim$\,350, $\lambda\lambda$6000--11000\,\,\AA{}). We also retrieved spectra for 30 of our candidates 
from the SDSS spectroscopic database (R\,$\sim$\,2000 and $\lambda\lambda$\,3800--9400\,\,\AA{}), nine of these 
30 candidates with an independent spectrum in our follow-up. 
We classified 92 candidates based on 101
optical spectra using two methods: spectral indices and comparison with templates of known subdwarfs.}
   {We developed an efficient photometric and proper motion search methodology to identify metal-poor M dwarfs. 
We confirmed 86\% and 94\% of the candidates as late-type subdwarfs from the SDSS vs 2MASS and SDSS vs UKIDSS 
cross-matches, respectively. These subdwarfs have spectral types ranging between M5 and L0.5 and SDSS magnitudes 
in the $r$\,=\,19.4--23.3 mag range. Our new late-type M discoveries include 49 subdwarfs, 25 extreme subdwarfs, 
six ultrasubdwarfs, one subdwarf/extreme subdwarf, and two dwarfs/subdwarfs. In addition, we discovered three 
early-L subdwarfs to add to the current compendium of L-type subdwarfs known to date. We doubled the numbers of
cool subdwarfs (11 new from SDSS vs 2MASS and 50 new from SDSS vs UKIDSS). We derived a surface 
density of late-type subdwarfs of 0.040$^{+0.012}_{-0.007}$ per square degree in the SDSS DR7 vs UKIDSS
LAS DR10 cross-match ($J$\,=\,15.9--18.8 mag) after correcting for incompleteness.
The density of M dwarfs decreases with decreasing metallicity. We also checked the Wide Field
Survey Explorer (AllWISE) photometry of known and new subdwarfs and found that mid-infrared colours of 
M subdwarfs do not appear to differ from their solar-metallicity counterparts of similar spectral types. 
However, the near-to-mid-infrared colours $J-W2$ and $J-W1$ are bluer for lower metallicity dwarfs, 
results that may be used as a criterion to look for late-type subdwarfs in future searches.
}
   {0}

\authorrunning{Lodieu et al.}
\titlerunning{New late-type subdwarfs in large-scale surveys} 

   \maketitle
%

\section{Introduction}
\label{sdM_VO:intro}
Subdwarfs have luminosity class VI in the Yerkes spectral classification system and lie below the main-sequence 
in the Hertzsprung-Russell diagram \citep{morgan43}. Subdwarfs appear less luminous than solar metallicity dwarfs 
with similar spectral types, due to the lack of metals in their atmospheres \citep{baraffe97}.  They have typical 
effective temperatures ($T_{eff}$) between $\sim$\,2500 and 4000\,K, interval dependent on metallicity \citep{woolf09}. 
Subdwarfs are Population II dwarfs located in the halo and the thick disk of the Milky Way. They are part of the 
first generations of stars and can be considered tracers of the Galactic chemical history. They are very old, with 
ages between 10 and 15 Gyr \citep{burgasser03b}. Subdwarfs have high proper motions and large heliocentric 
velocities \citep{gizis97a}.   
In the same way as ordinary main-sequence stars, stellar cool subdwarfs\footnote{We will use indistinctly the 
terms subdwarfs and cool subdwarfs when mentioning our targets.} produce their energy from hydrogen fusion 
and show strong metal-hydride absorption bands and metal lines. 
Some L dwarfs with low-metallicity features have been found over the past decade, but no specific classification exists 
for L subdwarfs yet.

\citet{gizis97a} presented the first spectral classification for M subdwarfs dividing them into two groups: 
subdwarfs and extreme subdwarfs. The classification was based on the strength of the TiO and CaH absorption bands at 
optical wavelengths. \citet{lepine07c} updated the \citet{gizis97a} classification using a parameter which 
quantifies the weakening of the strength of the TiO band in the optical as a function of metallicity; introducing a new class 
of subdwarfs: the ultrasubdwarfs. The current classification of low-mass M stars includes dwarfs and three 
low-metallicity classes: subdwarfs, extreme subdwarfs, and ultrasubdwarfs, with approximated metallicities of 
$-$0.5, $-$1.0, and $-$2.0 respectively \citep{lepine07c}. \citet{jao08} also proposed a classification for 
cool subdwarfs based on temperature, gravity, and metallicity.

The typical methods to identify subdwarfs focus on proper motion and/or photometric searches in photographic 
plates taken at different epochs \citep{luyten79,luyten80,scholz00,lepine03b,lodieu05b}.  Nowadays, the existence of
large-scale surveys mapping the sky at optical, near-infrared, and mid-infrared wavelengths
offer an efficient way to look for these metal-poor dwarfs. After the first spectral classification for 
M subdwarfs proposed by \citet{gizis97a}, other authors contributed to the increase in the numbers 
of this type of objects. New M subdwarfs with spectral types later than M7 were published in
\citet{gizis97b}, \citet{schweitzer99}, \citet{lepine03a}, \citet{scholz04c}, \citet{scholz04b}, \citet{lepine08b}, 
\citet{cushing09}, \citet{kirkpatrick10}, \citet{lodieu12b}, and \citet{zhang13}. The largest samples 
come from \citet{lepine08b}, \citet{kirkpatrick10}, \citet{lodieu12b}, and \citet{zhang13} and include 
23, 15, 20, and 30 new cool subdwarfs, respectively.

\citet{burgasser03b} published the first "substellar subdwarf", with spectral type (e?)sdL7. It was 
followed by a sdL4 subdwarf \citep{burgasser04} and years later by other seven L subdwarfs: 
a sdL3.5--4 in \citet{sivarani09}, a sdL5 in \citet{cushing09}, a sdL5 in \citet{lodieu10a} 
(re-classified in this paper as sdL3.5--sdL4), a sdL1, sdL7, and sdL8 in \citet{kirkpatrick10}, a sdL5 in \citet{schmidt10a} 
and also in \citet{bowler10a}. Our group published two new L subdwarfs \citep{lodieu12b}. In this work we 
add three more, with spectral types sdL0 and sdL0.5\@. The coolest L subdwarfs might have masses close to the 
star-brown dwarf boundary for subsolar metallicity according to models \citep{baraffe97,lodieu15a}.

The main purpose of this work is to develop an efficient method to search for late-type subdwarfs in 
large-scale surveys to increase their numbers using tools developed as part of the Virtual Observatory 
(VO)\footnote{http://www.ivoa.net} like STILTS\footnote{www.star.bris.ac.uk/$\sim$mbt/stilts} \citep{taylor06}, 
Topcat\footnote{www.star.bris.ac.uk/$\sim$mbt/topcat} 
\citep{taylor05}, and Aladin\footnote{aladin.u-strasbg.fr} \citep{bonnarel00}. We want to improve our knowledge 
of late-type subdwarfs, bridge the gap between late-M and L spectral types, determine the surface densities 
for each metallicity class, and understand the role of metallicity on the mass function from the stellar to the 
sub-stellar objects.

This is the second paper of a long-term project with several global objectives. The first paper was already 
published in \citet{lodieu12b}, where we cross-matched SDSS DR7 and UKIDSS LAS DR5, reporting 20 
new late-type subdwarfs. In this second paper, we present
the second part of our work, reporting new subdwarfs identified in SDSS DR9
\citep{york00}, UKIDSS LAS DR10 \citep{lawrence07}, and 2MASS \citep{cutri03,skrutskie06}.


\section{Sample selection of late-type subdwarfs}
\label{sdM_VO:sample_select}
We carried out two main cross-matches using different data releases of SDSS, UKIDSS, and 
2MASS: on the one hand SDSS DR9 vs UKIDSS LAS DR10, and, on the other hand SDSS DR7 vs 
2MASS\@. The area covered by these cross-matches are 3679 and 8826 square degrees, respectively. 
We emphasise that the candidates from the SDSS vs UKIDSS cross-matches in earlier releases are 
recovered in the SDSS DR9 vs UKIDSS LAS DR10 cross-correlation.
The common area between SDSS DR9 vs UKIDSS LAS DR10 and SDSSDR7 vs 2MASS
amounts for 2312 square degrees. The baseline in these cross-matches oscillate between 1 and 
7 years approximately, which corresponds to the maximum temporal separation between SDSS 
DR7 and 2MASS\@. 

All the candidates in this paper followed a search workflow that consisted in four main steps detailed 
here for the SDSS vs UKIDSS cross-correlation. We did the search in SDSS and 2MASS using the same
method with equivalent criteria:
\begin{itemize}
\renewcommand{\labelitemi}{$\bullet$}
\item Astrometric criteria:\\ 
For each SDSS source we looked for UKIDSS couterparts at radii between 1 and 5 arcsec. 
The nearest counterpart was kept. A minimum distance of 1 arcsec between the UKIDSS 
and SDSS source was required.\\
We selected point sources in SDSS ({\tt{cl}}\,=\,6).\\
We selected point sources in UKIDSS ({\tt{mergedClass}} equal to $-$1 or $-$2).
\item Quality flag criterion:\\
{\tt{ppErrBits}}\,$\leq$\,256 for $J$ and $K$ (sources with good quality flags).\\
{\tt{Xi}} and {\tt{Eta}} between $-$0.5 and 0.5 for $J$ and $K$ (these parameters refer to positional matching).
\item Photometric criteria:\\
$J$\,$>$\,10.5 mag and $K$\,$>$\,10.2 mag (to avoid bright sources).\\
$r-i$\,$\geq$\,1.0, $g-r$\,$\geq$\,1.8, $r-z$\,$\geq$\,1.6, and $J-K$\,$\leq$\,0.7 mag
\item Reduced Proper Motion criterion:\\
H$r$\,$\geq$\,20.7 mag, where H$r$\,=\,$r$\,$+$\,5$\times$\,$\log$($\mu$)\,$+$\,5, 
where $\mu$ is the proper motion (in arcsec/yr) and H$r$ the reduced proper motion.
\end{itemize}

We sought late-type subdwarfs with spectral types later than M5 in the solar vicinity. Here we consider the same 
criteria employed in \citet{lodieu12b}; these criteria are also comparable to those of previous surveys
\citep{jones72a,evans92,salim02,scholz04b,lepine05d,burgasser07b,lepine08b,lodieu09c}.

We present the final list of 100 late-type subdwarf candidates in Table \ref{Table_candidates}: 29 candidates come 
from the SDSS vs 2MASS exploration, and 71 candidates from SDSS vs UKIDSS\@. Table \ref{Table_candidates} provides the 
objects' coordinates, the optical $ugriz$ magnitudes from SDSSS DR9 for all 100 candidates, the 
near-infrared magnitudes given by the catalogues used in the corresponding searches (i.e., 2MASS for 
candidates found in the SDSS vs 2MASS, and UKIDSS for those registered from the SDSS vs UKIDSS survey),
proper motions, and reduced proper motions. We also list in the first column of Table \ref{Table_candidates}
an identification number (ID) that will be used throughout this paper to designate the candidates. Objects with 
ID between 1 and 29 were selected from the SDSS DR7 vs 2MASS search. The remaining candidates selected 
from the different SDSS vs UKIDSS cross-matches have IDs between 30 and 100 as follows:
\begin{itemize}
\renewcommand{\labelitemi}{$\bullet$}
\item SDSS DR7 vs UKIDSS LAS DR6: for ID 30 to 42
\item SDSS DR7 vs UKIDSS LAS DR8: for ID 43 to 68
\item SDSS DR9 vs UKIDSS LAS DR10: for ID 69 to 100
\end{itemize}

For the astrometric cross-match exercise we used the Aladin tool. We set a miniminum separation of 1 arcsec 
between epochs to ensure that our candidates have significant proper motion. We inspected all 100 candidates
by eye in the images from SuperCOSMOS Sky Surveys \citet{hambly01a,hambly01b,hambly01c} for additional 
epochs to exclude false positives. We calculated the proper motions by considering the direct differences in the 
right ascensions and declinations given by the catalogues and their respective observing epochs. We provide 
these values in Table \ref{Table_candidates}, which we used to determine the reduced proper motions. 
We revise some proper motion measurements in Section \ref{sdM_VO:PM_revision} mainly to identify 
solar-metallicity M dwarf contaminants in our sample. The 100 candidates exhibit total proper motions between 
$\sim$0.1 and $\sim$1.9 arcsec/yr.

\begin{figure}
   \centering
   \includegraphics[width=\hsize]{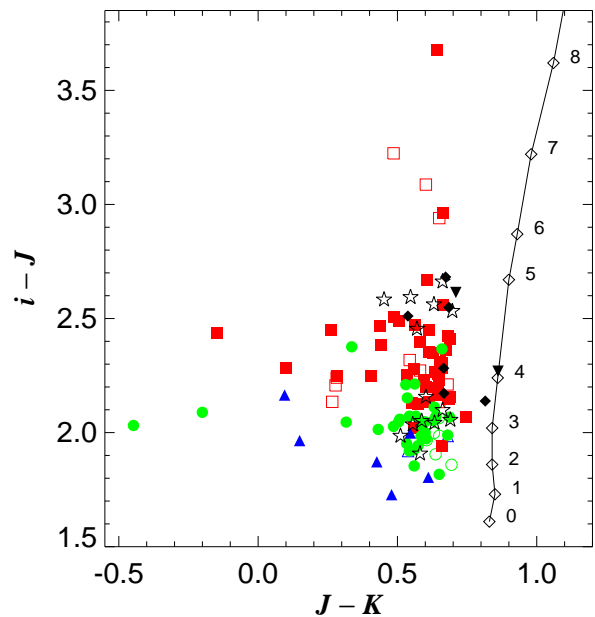}
      \caption{($i-J$ vs $J-K$) colour-colour diagram for candidates with optical spectra. Symbols
are as follows: sdM/L (filled red squares), esdM (filled green circles), usdM (filled blue triangles), 
candidates with uncertain class (filled black upside down triangles), and solar-metallicity M dwarfs 
(filled black diamonds). The empty five pointed stars represent the 14 candidates without optical spectroscopy yet. 
We also included a few known subdwarfs, extreme subdwarfs, and ultrasubdwarfs published in the literature (empty symbols). 
The empty diamonds joined with a continuum line in the right part of the diagram are solar-metallicity M dwarfs from \citet{west08}. 
The $J$ and $K$ magnitudes of candidates with ID\,=\,1--29 are in the UKIDSS Vega system \citep{hewett06}, except 
for objects with ID\,=\,1, 2, 6, 10, 11, and 19, which are plotted with their original 2MASS photometry.}
         \label{Fig_i-J_vs_J-K}
   \end{figure}

\citet{scholz04b} presented the idea of a generic photometric search for metal-poor dwarfs, 
where the ($J-K$,$i-J$) colour-colour diagram could be useful to separate subdwarfs from their 
solar-metallicity counterparts.
We placed our 100 candidates in the colour-colour diagram of Fig.\ \ref{Fig_i-J_vs_J-K}  using the UKIDSS 
Vega system photometry \citep{hewett06}. Five candidates (ID = 1, 2, 6, 10, and 19) from the SDSS vs 2MASS
survey also have UKIDSS photometry: the typical differences between the 2MASS and UKIDSS magnitudes 
are $\pm$0.05 mag, $\pm$0.06 mag, and $\pm$0.10 mag in the $J$, $H$, and $K$ filters, respectively. In 
Fig.\ \ref{Fig_i-J_vs_J-K}, we plot the UKIDSS photometry of these candidates while the objects from the 
SDSS DR7 vs 2MASS cross-match without UKIDSS photometry are plotted with their 2MASS magnitudes transformed 
to the UKIDSS system using the prescription of \citet{hewett06}. A few candidates show 
$J-K$\,$>$\,0.7 mag, which does not comply with our photometric criteria. We classified objects with 
ID\,=\,10 and ID\,=\,19 as dM/sdM, ID\,=\,2 as a confirmed solar-metallicity dM3 dwarf, and ID\,=\,1 as 
late-type sdM6 dwarf (two spectra available). We note that the candidates with the latest spectral 
classification show red $i-J$ colours as expected for late-M dwarfs \citep{hawley02,west05}.

\begin{figure}
   \centering
   \includegraphics[width=\hsize]{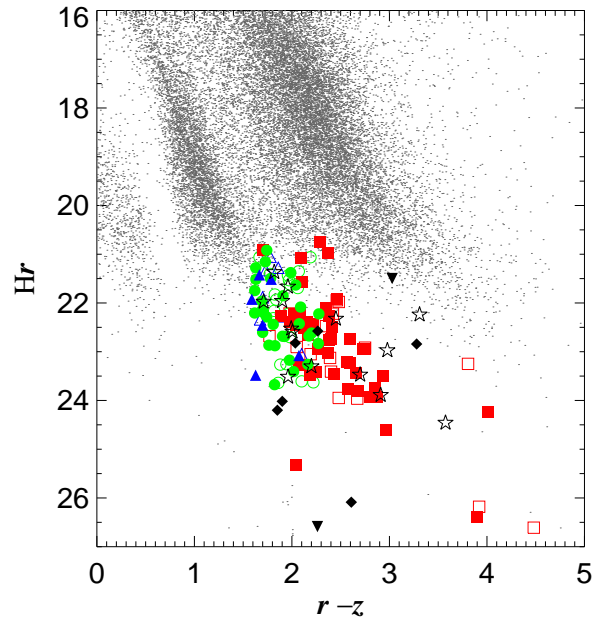}
         \caption{Reduced proper motion diagram of new subdwarfs and solar-metallicity M dwarfs. 
      H$r$ was computed using the proper motions obtained from the positions and epochs of the 
      cross-matched catalogues. Same symbology as Fig.\ \ref{Fig_i-J_vs_J-K} for our candidates. 
      Small grey dots correspond to objects from L\'epine's catalogue with SDSS photometry 
      \citet{lepine05d}.}
         \label{Fig_Hr_vs_r-z}
   \end{figure}

The reduced proper motion represents our main astrometric criterion. It is a key parameter 
to look for late-type subdwarfs. Fig.\ \ref{Fig_Hr_vs_r-z} displays the H$r$ as a function of $r-z$ 
for the objects included in the \citet{lepine05d} catalogue (using their SDSS photometry). As shown 
in \citet{lepine08b}, we can easily distinguish three sequences: white dwarfs on the left, low-metallicity 
dwarfs or subdwarfs in the middle, and solar-metallicity dwarfs on the right. In Fig.\ \ref{Fig_Hr_vs_r-z}
we overplotted our candidates and known subdwarfs from the literature with the same symbology as in 
Fig.\ \ref{Fig_i-J_vs_J-K}. The majority of the 100 candidates nicely fit the expected sequences of low 
metallicity dwarfs.

%
%
\begin{longtab}
\begin{landscape}
\scriptsize
\tiny
\begin{longtable}{rccccccccccccc}
\caption{Candidates with their SDSS coordinates (in J2000), magnitudes, proper motions, and reduced proper motions. The first 29 come from the SDSS DR7 vs 2MASS cross-match, the
remaining ones from the SDSS vs UKIDSS cross-correlation.}\\
\hline\hline
ID & RA & Dec. & $u$ & $g$ & $r$ & $i$ & $z$ & $Y$ & $J$ & $H$ & $K$ & pm & H$r$ \\
 & [hh:mm:ss.ss] & [dd:mm:ss.s] & [mag] & [mag] & [mag] & [mag] & [mag] & [mag] & [mag] & [mag] & [mag] & [$''$/yr] & [mag] \\
\hline
\endfirsthead
\caption{continued.}\\
\hline\hline
ID & RA & Dec. & $u$ & $g$ & $r$ & $i$ & $z$ & $Y$ & $J$ & $H$ & $K$ & pm & H$r$ \\
 & [hh:mm:ss.ss] & [dd:mm:ss.s] & [mag] & [mag] & [mag] & [mag] & [mag] & [mag] & [mag] & [mag] & [mag] & [$''$/yr] & [mag] \\
\hline
\endhead
\hline
\endfoot
   1 & 01:34:52.47 & $-$01:04:37.9 & 23.322$\pm$0.543 & 21.412$\pm$0.051 & 19.488$\pm$0.015 & 18.217$\pm$0.009 & 17.508$\pm$0.016 & --- & 16.024$\pm$0.076 & 15.467$\pm$0.100 & 15.478$\pm$0.188 & 0.378 & 22.397\\
   2 & 07:50:02.47 & +21:15:21.3 & 24.705$\pm$1.027 & 22.109$\pm$0.119 & 20.301$\pm$0.036 & 19.063$\pm$0.019 & 18.450$\pm$0.036 & --- & 16.976$\pm$0.152 & 16.405$\pm$0.212 & 16.345$\pm$0.305 & 0.604 & 24.200\\
   3 & 08:22:33.87 & +17:00:16.5 & 24.883$\pm$0.589 & 21.612$\pm$0.048 & 19.211$\pm$0.012 & 17.871$\pm$0.008 & 17.139$\pm$0.011 & --- & 15.717$\pm$0.059 & 15.517$\pm$0.100 & 15.619$\pm$0.218 & 0.597 & 23.078\\
   4 & 08:30:51.71 & +36:12:55.5 & 23.071$\pm$0.505 & 20.190$\pm$0.022 & 18.179$\pm$0.008 & 16.968$\pm$0.005 & 16.275$\pm$0.007 & --- & 14.907$\pm$0.036 & 14.536$\pm$0.051 & 14.338$\pm$0.059 & 0.799 & 22.682\\
   5 & 08:35:26.17 & +39:29:14.6 & 23.456$\pm$0.651 & 21.947$\pm$0.086 & 19.858$\pm$0.022 & 18.784$\pm$0.013 & 18.151$\pm$0.026 & --- & 16.763$\pm$0.148 & 16.684$\pm$0.263 & 17.242$\pm$---\space\space\space\space\space & 0.292 & 22.196\\
   6 & 08:43:58.50 & +06:00:38.6 & 22.357$\pm$0.239 & 19.710$\pm$0.014 & 17.814$\pm$0.006 & 16.721$\pm$0.005 & 16.086$\pm$0.008 & --- & 14.757$\pm$0.039 & 14.298$\pm$0.040 & 14.061$\pm$0.055 & 0.463 & 21.157\\
   7 & 08:46:48.89 & +30:28:01.8 & 23.157$\pm$0.479 & 20.504$\pm$0.025 & 18.513$\pm$0.008 & 17.467$\pm$0.006 & 16.847$\pm$0.010 & --- & 15.675$\pm$0.073 & 14.939$\pm$0.092 & 15.061$\pm$0.149 & 0.383 & 21.424\\
   8 & 08:55:00.38 & +35:41:07.5 & 25.076$\pm$0.735 & 21.671$\pm$0.052 & 19.879$\pm$0.019 & 18.541$\pm$0.010 & 17.803$\pm$0.016 & --- & 16.425$\pm$0.111 & 15.868$\pm$0.160 & 15.860$\pm$0.205 & 0.476 & 23.279\\
   9 & 08:55:48.71 & +36:36:01.4 & 24.245$\pm$1.556 & 22.083$\pm$0.141 & 19.794$\pm$0.032 & 18.527$\pm$0.017 & 17.806$\pm$0.028 & --- & 16.386$\pm$0.103 & 16.044$\pm$0.183 & 15.823$\pm$0.220 & 0.212 & 21.377\\
  10 & 08:58:38.92 & +09:19:57.7 & 22.578$\pm$0.322 & 21.695$\pm$0.068 & 19.850$\pm$0.020 & 17.841$\pm$0.007 & 16.822$\pm$0.010 & --- & 15.255$\pm$0.055 & 14.827$\pm$0.085 & 14.571$\pm$0.081 & 0.212 & 21.498\\
  11 & 09:03:07.94 & +08:42:43.1 & 22.304$\pm$0.187 & 19.017$\pm$0.009 & 17.073$\pm$0.005 & 15.961$\pm$0.005 & 15.331$\pm$0.005 & --- & 13.995$\pm$0.023 & 13.580$\pm$0.026 & 13.411$\pm$0.042 & 0.591 & 20.919\\
  12 & 09:04:23.07 & +46:38:18.6 & 24.954$\pm$1.101 & 21.492$\pm$0.067 & 19.638$\pm$0.017 & 18.350$\pm$0.010 & 17.488$\pm$0.014 & --- & 16.096$\pm$0.079 & 15.573$\pm$0.108 & 15.440$\pm$0.157 & 0.357 & 22.395\\
  13 & 09:07:41.80 & +46:20:35.1 & 24.746$\pm$0.943 & 21.882$\pm$0.086 & 20.090$\pm$0.024 & 18.571$\pm$0.011 & 17.825$\pm$0.020 & --- & 16.071$\pm$0.120 & 15.802$\pm$0.185 & 15.500$\pm$---\space\space\space\space\space & 0.318 & 22.576\\
  14 & 09:09:03.58 & +19:41:43.6 & 21.962$\pm$0.195 & 19.534$\pm$0.013 & 17.758$\pm$0.006 & 16.310$\pm$0.006 & 15.470$\pm$0.006 & --- & 14.065$\pm$0.032 & 13.575$\pm$0.034 & 13.396$\pm$0.035 & 0.388 & 20.751\\
  15 & 09:40:43.35 & +39:40:35.2 & 23.457$\pm$0.533 & 21.516$\pm$0.047 & 19.526$\pm$0.016 & 18.352$\pm$0.010 & 17.700$\pm$0.021 & --- & 16.443$\pm$0.120 & 16.129$\pm$---\space\space\space\space\space & 15.870$\pm$0.244 & 0.467 & 22.877\\
  16 & 10:12:00.28 & +20:46:11.6 & 23.630$\pm$0.547 & 21.871$\pm$0.060 & 19.668$\pm$0.015 & 18.260$\pm$0.009 & 17.466$\pm$0.013 & --- & 16.213$\pm$0.072 & 15.692$\pm$0.094 & 15.693$\pm$0.177 & 0.386 & 22.616\\
  17 & 10:27:57.77 & +34:01:46.8 & 24.693$\pm$0.982 & 22.293$\pm$0.100 & 20.352$\pm$0.027 & 18.598$\pm$0.010 & 17.677$\pm$0.016 & --- & 16.158$\pm$0.100 & 15.797$\pm$0.161 & 15.892$\pm$0.287 & 0.493 & 23.806\\
  18 & 10:44:10.01 & +30:01:42.3 & 25.174$\pm$0.762 & 22.482$\pm$0.107 & 20.535$\pm$0.033 & 19.250$\pm$0.018 & 18.713$\pm$0.038 & --- & 16.895$\pm$0.189 & 16.264$\pm$0.218 & 16.220$\pm$0.308 & 0.435 & 23.676\\
  19 & 10:46:57.93 & $-$01:37:46.4 & 25.131$\pm$0.772 & 22.025$\pm$0.080 & 20.208$\pm$0.027 & 18.812$\pm$0.014 & 17.946$\pm$0.021 & --- & 16.487$\pm$0.129 & 15.924$\pm$0.214 & 15.841$\pm$0.304 & 1.872 & 26.587\\
  20 & 11:11:47.19 & +27:25:16.7 & 24.072$\pm$0.735 & 21.798$\pm$0.058 & 20.001$\pm$0.020 & 18.905$\pm$0.014 & 18.260$\pm$0.026 & --- & 16.828$\pm$0.144 & 16.206$\pm$0.168 & 17.082$\pm$---\space\space\space\space\space & 0.288 & 22.287\\
  21 & 11:19:29.20 & +67:21:04.1 & 24.381$\pm$1.057 & 22.538$\pm$0.149 & 20.523$\pm$0.036 & 19.175$\pm$0.018 & 18.485$\pm$0.034 & --- & 16.833$\pm$0.161 & 16.196$\pm$0.212 & 16.207$\pm$0.390 & 0.923 & 25.333\\
  22 & 12:27:41.90 & +25:12:59.6 & 22.436$\pm$0.203 & 20.465$\pm$0.021 & 18.659$\pm$0.009 & 17.092$\pm$0.006 & 16.244$\pm$0.008 & --- & 14.792$\pm$0.035 & 14.259$\pm$0.054 & 14.116$\pm$0.055 & 0.587 & 22.497\\
  23 & 13:51:28.49 & +55:06:56.9 & 24.158$\pm$0.806 & 21.172$\pm$0.035 & 18.994$\pm$0.011 & 17.693$\pm$0.007 & 16.955$\pm$0.011 & --- & 15.675$\pm$0.060 & 15.135$\pm$0.085 & 15.068$\pm$0.127 & 0.337 & 21.632\\
  24 & 14:34:33.99 & +38:41:03.4 & 25.256$\pm$0.728 & 21.881$\pm$0.077 & 20.062$\pm$0.022 & 18.613$\pm$0.011 & 17.848$\pm$0.020 & --- & 16.224$\pm$0.096 & 16.227$\pm$0.198 & 15.583$\pm$0.232 & 0.297 & 22.428\\
  25 & 15:20:29.33 & +14:34:37.0 & 25.508$\pm$0.756 & 20.420$\pm$0.023 & 18.736$\pm$0.010 & 17.554$\pm$0.007 & 16.928$\pm$0.011 & --- & 15.518$\pm$0.056 & 15.006$\pm$0.083 & 14.900$\pm$0.110 & 0.551 & 22.440\\
  26 & 15:25:35.90 & +43:15:45.2 & 23.504$\pm$0.658 & 20.241$\pm$0.023 & 18.438$\pm$0.009 & 17.366$\pm$0.006 & 16.730$\pm$0.010 & --- & 15.436$\pm$0.058 & 15.055$\pm$0.090 & 14.740$\pm$0.093 & 0.309 & 20.908\\
  27 & 16:05:19.49 & +03:05:34.2 & 24.648$\pm$0.977 & 21.534$\pm$0.057 & 19.993$\pm$0.021 & 18.733$\pm$0.012 & 18.092$\pm$0.024 & --- & 16.462$\pm$0.118 & 16.003$\pm$0.161 & 15.767$\pm$0.250 & 0.566 & 24.018\\
  28 & 16:40:08.58 & +11:03:22.7 & 23.272$\pm$0.488 & 22.133$\pm$0.075 & 20.331$\pm$0.023 & 18.939$\pm$0.012 & 18.237$\pm$0.023 & --- & 16.669$\pm$0.117 & 16.204$\pm$0.166 & 16.585$\pm$---\space\space\space\space\space & 0.143 & 21.085\\
  29 & 16:57:39.57 & +39:39:48.0 & 24.436$\pm$0.750 & 21.176$\pm$0.036 & 19.386$\pm$0.013 & 17.842$\pm$0.007 & 17.017$\pm$0.011 & --- & 15.576$\pm$0.054 & 15.176$\pm$0.096 & 14.992$\pm$0.110 & 0.207 & 20.981\\
  30 & 01:04:48.47 & +15:35:01.9 & 25.499$\pm$0.781 & 24.942$\pm$0.717 & 22.245$\pm$0.167 & 20.365$\pm$0.048 & 19.284$\pm$0.064 & 18.484$\pm$0.046 & 17.929$\pm$0.052 & 18.064$\pm$0.111 & 18.077$\pm$0.167 & 0.298 & 24.610\\
  31 & 02:05:33.75 & +12:38:24.1 & 23.201$\pm$0.490 & 21.935$\pm$0.091 & 19.764$\pm$0.021 & 18.120$\pm$0.009 & 17.303$\pm$0.016 & 16.456$\pm$0.009 & 15.872$\pm$0.008 & 15.709$\pm$0.012 & 15.590$\pm$0.018 & 0.270 & 21.925\\
  32 & 02:12:58.07 & +06:41:17.6 & 23.574$\pm$0.566 & 25.232$\pm$0.558 & 23.272$\pm$0.336 & 21.104$\pm$0.082 & 19.373$\pm$0.079 & 18.204$\pm$0.029 & 17.425$\pm$0.025 & 17.058$\pm$0.033 & 16.783$\pm$0.052 & 0.422 & 26.386\\
  33 & 08:58:33.76 & +02:04:52.9 & 25.975$\pm$0.553 & 22.339$\pm$0.126 & 20.344$\pm$0.031 & 18.970$\pm$0.016 & 18.225$\pm$0.030 & 17.408$\pm$0.027 & 16.836$\pm$0.017 & 16.422$\pm$0.021 & 16.222$\pm$0.030 & 0.269 & 22.504\\
  34 & 09:32:44.46 & +01:12:59.9 & 25.926$\pm$0.662 & 24.476$\pm$0.680 & 21.714$\pm$0.106 & 20.113$\pm$0.034 & 19.281$\pm$0.068 & 18.156$\pm$0.030 & 17.645$\pm$0.029 & 17.487$\pm$0.075 & 17.208$\pm$0.077 & 0.223 & 23.462\\
  35 & 09:49:05.26 & +02:32:50.7 & 24.276$\pm$0.998 & 21.528$\pm$0.058 & 19.554$\pm$0.018 & 18.139$\pm$0.010 & 17.387$\pm$0.016 & 16.478$\pm$0.009 & 15.910$\pm$0.009 & 15.513$\pm$0.009 & 15.262$\pm$0.012 & 0.418 & 22.659\\
  36 & 10:36:58.90 & +03:36:23.2 & 26.449$\pm$0.246 & 23.318$\pm$0.234 & 21.239$\pm$0.049 & 19.362$\pm$0.015 & 18.371$\pm$0.023 & 17.410$\pm$0.022 & 16.803$\pm$0.021 & 16.426$\pm$0.028 & 16.140$\pm$0.038 & 0.346 & 23.935\\
  37 & 11:40:01.19 & +00:37:04.0 & 26.548$\pm$0.446 & 24.699$\pm$1.016 & 22.833$\pm$0.394 & 21.638$\pm$0.190 & 20.133$\pm$0.167 & 19.699$\pm$0.179 & 19.076$\pm$0.137 & 18.688$\pm$0.163 & 18.445$\pm$0.218 & 0.134 & 23.472\\
  38 & 14:30:13.20 & +01:20:19.1 & 23.601$\pm$0.808 & 23.663$\pm$0.400 & 21.571$\pm$0.084 & 20.471$\pm$0.045 & 19.873$\pm$0.104 & 19.116$\pm$0.085 & 18.506$\pm$0.077 & 18.208$\pm$0.091 & 18.357$\pm$0.221 & 0.151 & 22.455\\
  39 & 14:41:28.38 & +00:31:21.5 & 25.314$\pm$0.576 & 22.667$\pm$0.116 & 20.785$\pm$0.037 & 19.702$\pm$0.022 & 19.167$\pm$0.043 & 18.454$\pm$0.046 & 17.885$\pm$0.047 & 17.321$\pm$0.048 & 17.235$\pm$0.077 & 0.193 & 22.206\\
  40 & 14:57:43.44 & +01:27:47.4 & 24.038$\pm$0.938 & 21.963$\pm$0.075 & 20.095$\pm$0.025 & 19.021$\pm$0.015 & 18.404$\pm$0.032 & 17.594$\pm$0.024 & 16.974$\pm$0.023 & 16.652$\pm$0.027 & 16.470$\pm$0.042 & 0.235 & 21.964\\
  41 & 15:41:28.39 & +04:10:04.6 & 24.536$\pm$0.755 & 22.454$\pm$0.097 & 20.633$\pm$0.031 & 19.303$\pm$0.017 & 18.592$\pm$0.030 & 17.743$\pm$0.025 & 17.154$\pm$0.029 & 16.722$\pm$0.019 & 16.472$\pm$0.035 & 0.233 & 22.458\\
  42 & 15:48:28.37 & +00:18:10.4 & 23.365$\pm$0.449 & 23.536$\pm$0.240 & 21.597$\pm$0.076 & 20.270$\pm$0.039 & 19.522$\pm$0.084 & 18.832$\pm$0.052 & 18.318$\pm$0.065 & 17.923$\pm$0.051 & 17.785$\pm$0.121 & 0.147 & 22.425\\
  43 & 07:40:13.59 & +24:29:45.1 & 23.828$\pm$0.868 & 22.160$\pm$0.086 & 20.197$\pm$0.028 & 19.087$\pm$0.017 & 18.568$\pm$0.043 & 17.669$\pm$0.023 & 17.162$\pm$0.016 & 16.787$\pm$0.029 & 16.602$\pm$0.038 & 0.165 & 21.278\\
  44 & 08:06:05.53 & +29:28:00.9 & 26.055$\pm$0.481 & 22.868$\pm$0.153 & 20.874$\pm$0.046 & 19.598$\pm$0.023 & 18.857$\pm$0.042 & 18.097$\pm$0.023 & 17.584$\pm$0.024 & 17.326$\pm$0.047 & 17.153$\pm$0.066 & 0.321 & 23.408\\
  45 & 08:26:50.57 & +28:52:53.4 & 25.713$\pm$0.672 & 23.806$\pm$0.359 & 21.471$\pm$0.067 & 19.782$\pm$0.027 & 18.880$\pm$0.039 & 17.952$\pm$0.020 & 17.372$\pm$0.019 & 16.907$\pm$0.032 & 16.684$\pm$0.050 & 0.225 & 23.228\\
  46 & 09:45:03.44 & +10:36:00.3 & 24.205$\pm$0.616 & 23.224$\pm$0.169 & 21.411$\pm$0.054 & 19.502$\pm$0.018 & 18.473$\pm$0.028 & 17.474$\pm$0.015 & 16.832$\pm$0.010 & 16.454$\pm$0.027 & 16.225$\pm$0.028 & 0.263 & 23.504\\
  47 & 10:32:18.45 & +01:15:56.5 & 24.860$\pm$0.978 & 23.482$\pm$0.335 & 21.132$\pm$0.071 & 20.102$\pm$0.043 & 19.345$\pm$0.079 & 18.619$\pm$0.037 & 18.104$\pm$0.036 & 17.806$\pm$0.080 & 17.558$\pm$0.096 & 0.118 & 21.514\\
  48 & 10:33:18.01 & +05:30:54.9 & 25.058$\pm$0.806 & 24.461$\pm$0.599 & 22.165$\pm$0.126 & 20.493$\pm$0.044 & 19.604$\pm$0.070 & 18.828$\pm$0.051 & 18.109$\pm$0.038 & 17.795$\pm$0.077 & 17.669$\pm$0.101 & 0.167 & 23.205\\
  49 & 10:37:35.78 & +11:32:49.8 & 26.130$\pm$0.606 & 23.029$\pm$0.228 & 20.978$\pm$0.060 & 19.399$\pm$0.022 & 18.627$\pm$0.041 & 17.808$\pm$0.023 & 17.250$\pm$0.024 & 16.835$\pm$0.034 & 16.564$\pm$0.040 & 0.168 & 22.099\\
  50 & 10:42:06.18 & +09:23:19.9 & 24.696$\pm$0.960 & 23.188$\pm$0.233 & 21.691$\pm$0.100 & 19.520$\pm$0.024 & 18.412$\pm$0.036 & 17.464$\pm$0.017 & 16.838$\pm$0.015 & 16.404$\pm$0.023 & 16.165$\pm$0.028 & 0.158 & 22.842\\
  51 & 10:44:51.52 & $-$01:46:34.1 & 24.228$\pm$1.223 & 23.106$\pm$0.287 & 21.094$\pm$0.069 & 20.001$\pm$0.040 & 19.475$\pm$0.088 & 18.570$\pm$0.036 & 18.147$\pm$0.035 & 17.650$\pm$0.081 & 17.587$\pm$0.102 & 0.135 & 21.746\\
  52 & 10:50:12.58 & +08:51:22.6 & 22.983$\pm$0.370 & 23.086$\pm$0.179 & 20.862$\pm$0.043 & 19.105$\pm$0.015 & 18.064$\pm$0.021 & 17.203$\pm$0.018 & 16.597$\pm$0.019 & 16.271$\pm$0.027 & 16.109$\pm$0.032 & 0.403 & 23.932\\
  53 & 10:57:03.59 & +06:48:50.4 & 23.947$\pm$0.893 & 22.123$\pm$0.100 & 20.064$\pm$0.027 & 18.591$\pm$0.011 & 17.800$\pm$0.019 & 16.940$\pm$0.012 & 16.359$\pm$0.011 & 15.992$\pm$0.014 & 15.762$\pm$0.021 & 0.376 & 22.953\\
  54 & 11:00:17.82 & +01:12:18.7 & 23.024$\pm$0.361 & 23.797$\pm$0.283 & 21.868$\pm$0.093 & 20.254$\pm$0.038 & 19.483$\pm$0.075 & 18.578$\pm$0.043 & 18.000$\pm$0.037 & 17.628$\pm$0.057 & 17.464$\pm$0.076 & 0.119 & 22.263\\
  55 & 11:04:21.86 & +05:37:24.0 & 26.360$\pm$0.444 & 24.940$\pm$0.680 & 21.555$\pm$0.082 & 20.533$\pm$0.046 & 19.966$\pm$0.104 & 19.102$\pm$0.073 & 18.805$\pm$0.098 & 18.422$\pm$0.093 & 18.326$\pm$0.195 & 0.118 & 21.924\\
  56 & 11:13:58.73 & +03:11:37.7 & 25.773$\pm$0.462 & 23.813$\pm$0.334 & 21.691$\pm$0.081 & 20.278$\pm$0.040 & 19.415$\pm$0.069 & 18.740$\pm$0.036 & 18.206$\pm$0.035 & 17.995$\pm$0.121 & 17.640$\pm$0.122 & 0.127 & 22.224\\
  57 & 11:38:44.65 & +06:54:10.0 & 24.168$\pm$1.028 & 23.864$\pm$0.377 & 21.489$\pm$0.080 & 20.221$\pm$0.037 & 19.564$\pm$0.060 & 18.640$\pm$0.039 & 18.109$\pm$0.036 & 17.675$\pm$0.071 & 17.477$\pm$0.084 & 0.170 & 22.687\\
  58 & 11:43:28.18 & +11:22:21.9 & 22.919$\pm$0.520 & 24.496$\pm$0.749 & 21.429$\pm$0.093 & 20.041$\pm$0.039 & 19.182$\pm$0.056 & 18.413$\pm$0.046 & 17.737$\pm$0.041 & 17.335$\pm$0.079 & 17.076$\pm$0.081 & 0.249 & 23.425\\
  59 & 12:49:04.39 & +10:04:13.5 & 23.781$\pm$0.908 & 23.362$\pm$0.297 & 21.076$\pm$0.062 & 19.654$\pm$0.026 & 18.701$\pm$0.047 & 18.265$\pm$0.024 & 17.634$\pm$0.020 & 17.243$\pm$0.045 & 17.073$\pm$0.064 & 0.195 & 23.029\\
  60 & 12:51:34.45 & $-$00:55:55.5 & 23.022$\pm$0.428 & 24.097$\pm$0.526 & 21.980$\pm$0.131 & 20.328$\pm$0.044 & 19.389$\pm$0.074 & 18.823$\pm$0.056 & 18.081$\pm$0.054 & 17.805$\pm$0.084 & 17.676$\pm$0.140 & 0.140 & 22.732\\
  61 & 13:18:22.81 & $-$01:11:50.2 & 25.946$\pm$0.553 & 23.586$\pm$0.335 & 21.346$\pm$0.086 & 19.995$\pm$0.044 & 19.171$\pm$0.076 & 18.530$\pm$0.036 & 17.949$\pm$0.042 & 17.805$\pm$0.086 & 17.633$\pm$0.107 & 0.185 & 22.677\\
  62 & 13:45:55.25 & +02:22:49.4 & 24.123$\pm$1.023 & 22.288$\pm$0.109 & 20.337$\pm$0.028 & 18.992$\pm$0.015 & 18.248$\pm$0.029 & 17.342$\pm$0.017 & 16.782$\pm$0.019 & 16.437$\pm$0.024 & 16.252$\pm$0.034 & 0.223 & 22.084\\
  63 & 14:14:05.74 & $-$01:42:02.7 & 24.745$\pm$0.958 & 24.555$\pm$0.658 & 22.483$\pm$0.184 & 19.771$\pm$0.028 & 18.475$\pm$0.034 & 17.498$\pm$0.027 & 16.807$\pm$0.024 & 16.455$\pm$0.027 & 16.143$\pm$0.031 & 0.239 & 24.230\\
  64 & 15:12:17.83 & $-$01:12:35.4 & 25.155$\pm$0.607 & 21.204$\pm$0.036 & 19.368$\pm$0.013 & 17.652$\pm$0.007 & 16.704$\pm$0.009 & 15.807$\pm$0.007 & 15.202$\pm$0.006 & 14.796$\pm$0.007 & 14.522$\pm$0.009 & 0.646 & 23.445\\
  65 & 15:24:34.64 & +00:20:58.0 & 24.869$\pm$0.739 & 23.527$\pm$0.240 & 21.518$\pm$0.067 & 19.662$\pm$0.022 & 18.794$\pm$0.040 & 17.871$\pm$0.022 & 17.212$\pm$0.020 & 16.813$\pm$0.022 & 16.598$\pm$0.032 & 0.193 & 22.946\\
  66 & 15:25:39.97 & +00:24:10.0 & 24.863$\pm$0.695 & 23.401$\pm$0.214 & 21.160$\pm$0.052 & 19.737$\pm$0.024 & 19.050$\pm$0.050 & 18.074$\pm$0.027 & 17.491$\pm$0.029 & 16.989$\pm$0.024 & 16.846$\pm$0.046 & 0.125 & 21.577\\
  67 & 15:36:47.08 & +02:55:01.6 & 25.055$\pm$1.207 & 22.125$\pm$0.100 & 20.127$\pm$0.027 & 19.080$\pm$0.016 & 18.496$\pm$0.039 & 17.596$\pm$0.021 & 17.053$\pm$0.023 & 16.702$\pm$0.027 & 16.567$\pm$0.045 & 0.189 & 21.515\\
  68 & 15:41:43.81 & +01:26:31.4 & 22.991$\pm$0.420 & 23.193$\pm$0.215 & 21.112$\pm$0.055 & 19.870$\pm$0.028 & 19.203$\pm$0.065 & 18.412$\pm$0.040 & 17.799$\pm$0.045 & 17.411$\pm$0.055 & 17.161$\pm$0.077 & 0.142 & 21.852\\
  69 & 07:44:31.25 & +28:39:16.6 & 23.222$\pm$0.627 & 23.806$\pm$0.384 & 21.802$\pm$0.106 & 20.143$\pm$0.039 & 19.401$\pm$0.065 & 18.405$\pm$0.034 & 17.861$\pm$0.039 & 17.439$\pm$0.048 & 17.201$\pm$0.067 & 0.154 & 22.740\\
  70 & 09:45:52.72 & $-$00:34:32.4 & 24.566$\pm$0.931 & 23.492$\pm$0.280 & 21.803$\pm$0.092 & 20.203$\pm$0.034 & 19.268$\pm$0.054 & 18.417$\pm$0.040 & 17.654$\pm$0.047 & 17.160$\pm$0.045 & 16.969$\pm$0.063 & 0.771 & 26.087\\
  71 & 10:00:40.68 & +12:56:10.5 & 24.714$\pm$0.551 & 23.628$\pm$0.228 & 21.586$\pm$0.066 & 20.366$\pm$0.037 & 19.611$\pm$0.063 & 18.806$\pm$0.053 & 18.153$\pm$0.064 & 17.951$\pm$0.084 & 17.591$\pm$0.113 & 0.208 & 23.174\\
  72 & 10:16:26.29 & +02:41:03.9 & 24.518$\pm$0.937 & 24.274$\pm$0.487 & 22.119$\pm$0.138 & 20.722$\pm$0.064 & 19.955$\pm$0.116 & 18.950$\pm$0.065 & 18.346$\pm$0.072 & 18.039$\pm$0.120 & 18.010$\pm$0.172 & 0.169 & 23.260\\
  73 & 10:18:24.97 & +02:15:12.5 & 24.382$\pm$0.930 & 22.742$\pm$0.151 & 20.818$\pm$0.043 & 19.632$\pm$0.023 & 18.999$\pm$0.055 & 18.039$\pm$0.028 & 17.533$\pm$0.032 & 17.144$\pm$0.050 & 16.871$\pm$0.056 & 0.129 & 21.367\\
  74 & 10:29:35.81 & +11:09:01.6 & 26.188$\pm$0.368 & 23.095$\pm$0.190 & 21.174$\pm$0.059 & 19.969$\pm$0.031 & 19.274$\pm$0.071 & 18.600$\pm$0.049 & 18.061$\pm$0.043 & 17.652$\pm$0.069 & 17.480$\pm$0.086 & 0.144 & 21.959\\
  75 & 10:43:09.04 & +05:21:12.0 & 23.590$\pm$0.468 & 21.180$\pm$0.036 & 19.378$\pm$0.013 & 18.115$\pm$0.008 & 17.419$\pm$0.012 & 16.568$\pm$0.009 & 16.059$\pm$0.008 & 15.588$\pm$0.010 & 15.369$\pm$0.015 & 0.287 & 21.666\\
  76 & 10:47:53.45 & +14:59:42.0 & 24.610$\pm$0.831 & 22.481$\pm$0.103 & 20.469$\pm$0.030 & 19.220$\pm$0.018 & 18.463$\pm$0.033 & 17.718$\pm$0.019 & 17.184$\pm$0.019 & 16.812$\pm$0.025 & 16.629$\pm$0.037 & 0.266 & 22.591\\
  77 & 10:51:02.31 & +13:33:46.9 & 24.420$\pm$0.560 & 22.076$\pm$0.062 & 20.202$\pm$0.022 & 18.920$\pm$0.013 & 18.179$\pm$0.023 & 17.331$\pm$0.013 & 16.796$\pm$0.012 & 16.347$\pm$0.017 & 16.223$\pm$0.039 & 0.255 & 22.235\\
  78 & 10:53:09.90 & +13:14:30.1 & 23.602$\pm$0.506 & 23.425$\pm$0.219 & 21.433$\pm$0.055 & 20.381$\pm$0.036 & 19.669$\pm$0.071 & 18.891$\pm$0.044 & 18.438$\pm$0.050 & 18.018$\pm$0.077 & 17.864$\pm$0.146 & 0.193 & 22.866\\
  79 & 11:06:51.29 & +04:48:15.0 & 24.449$\pm$0.897 & 22.766$\pm$0.155 & 20.568$\pm$0.036 & 18.792$\pm$0.012 & 17.814$\pm$0.019 & 16.941$\pm$0.010 & 16.301$\pm$0.009 & 15.995$\pm$0.016 & 15.797$\pm$0.021 & 0.297 & 22.934\\
  80 & 11:14:47.27 & $-$01:57:23.4 & 25.243$\pm$0.715 & 21.853$\pm$0.069 & 20.038$\pm$0.022 & 18.927$\pm$0.014 & 18.328$\pm$0.027 & 17.512$\pm$0.023 & 16.941$\pm$0.024 & 16.543$\pm$0.029 & 16.430$\pm$0.048 & 0.244 & 21.975\\
  81 & 11:14:53.35 & +12:29:18.8 & 24.040$\pm$1.278 & 21.925$\pm$0.101 & 19.997$\pm$0.028 & 18.766$\pm$0.017 & 18.036$\pm$0.029 & 17.276$\pm$0.014 & 16.723$\pm$0.013 & 16.306$\pm$0.020 & 16.090$\pm$0.028 & 0.505 & 23.514\\
  82 & 11:18:14.67 & +09:41:12.1 & 24.733$\pm$0.850 & 21.131$\pm$0.038 & 19.315$\pm$0.013 & 18.032$\pm$0.008 & 17.280$\pm$0.013 & 16.417$\pm$0.008 & 15.860$\pm$0.006 & 15.453$\pm$0.011 & 15.193$\pm$0.016 & 0.503 & 22.822\\
  83 & 11:44:42.54 & +15:51:05.9 & 25.235$\pm$0.617 & 21.991$\pm$0.065 & 20.062$\pm$0.021 & 19.026$\pm$0.015 & 18.365$\pm$0.030 & 17.602$\pm$0.018 & 17.041$\pm$0.017 & 16.674$\pm$0.023 & 16.456$\pm$0.034 & 0.260 & 22.141\\
  84 & 12:13:56.96 & $-$02:55:25.5 & 25.424$\pm$0.788 & 23.982$\pm$0.377 & 22.042$\pm$0.123 & 20.593$\pm$0.052 & 19.591$\pm$0.076 & 18.795$\pm$0.058 & 18.138$\pm$0.049 & 17.653$\pm$0.044 & 17.568$\pm$0.095 & 0.114 & 22.324\\
  85 & 12:18:12.86 & +07:06:10.4 & 24.377$\pm$0.794 & 22.070$\pm$0.069 & 20.232$\pm$0.023 & 18.913$\pm$0.012 & 18.153$\pm$0.020 & 17.307$\pm$0.014 & 16.746$\pm$0.013 & 16.336$\pm$0.025 & 16.100$\pm$0.035 & 0.251 & 22.231\\
  86 & 12:44:10.11 & +27:36:25.8 & 24.474$\pm$1.128 & 24.373$\pm$0.536 & 22.506$\pm$0.197 & 20.160$\pm$0.041 & 18.931$\pm$0.060 & 18.282$\pm$0.037 & 17.577$\pm$0.027 & 17.322$\pm$0.048 & 17.125$\pm$0.063 & 0.246 & 24.458\\
  87 & 12:55:51.65 & +34:14:21.2 & 24.615$\pm$0.939 & 21.598$\pm$0.052 & 19.740$\pm$0.018 & 18.677$\pm$0.011 & 17.959$\pm$0.018 & 17.147$\pm$0.018 & 16.624$\pm$0.016 & 16.185$\pm$0.019 & 15.994$\pm$0.027 & 0.225 & 21.497\\
  88 & 12:59:30.64 & +13:16:34.2 & 24.478$\pm$1.077 & 23.693$\pm$0.328 & 21.671$\pm$0.084 & 20.186$\pm$0.035 & 19.485$\pm$0.071 & 18.531$\pm$0.043 & 17.987$\pm$0.043 & 17.520$\pm$0.067 & 17.383$\pm$0.087 & 0.231 & 23.487\\
  89 & 13:06:15.21 & +04:59:08.9 & 24.082$\pm$0.752 & 23.049$\pm$0.187 & 21.103$\pm$0.048 & 19.434$\pm$0.019 & 18.529$\pm$0.033 & 17.668$\pm$0.020 & 17.074$\pm$0.020 & 16.637$\pm$0.033 & 16.401$\pm$0.043 & 0.339 & 23.755\\
  90 & 13:09:59.60 & +05:29:38.7 & 25.213$\pm$0.635 & 24.744$\pm$0.621 & 22.853$\pm$0.211 & 20.876$\pm$0.066 & 20.003$\pm$0.105 & 18.998$\pm$0.075 & 18.406$\pm$0.063 & ---$\pm$---      & 17.844$\pm$0.137 & 0.150 & 23.737\\
  91 & 13:10:38.46 & +33:39:11.7 & 25.079$\pm$0.899 & 22.707$\pm$0.122 & 20.888$\pm$0.038 & 19.655$\pm$0.021 & 18.895$\pm$0.038 & 18.155$\pm$0.041 & 17.604$\pm$0.040 & 17.208$\pm$0.043 & 17.017$\pm$0.064 & 0.213 & 22.526\\
  92 & 13:20:12.19 & +05:39:44.0 & 24.298$\pm$0.933 & 24.279$\pm$0.512 & 22.209$\pm$0.133 & 20.851$\pm$0.061 & 19.938$\pm$0.090 & 19.327$\pm$0.087 & 18.783$\pm$0.106 & 18.208$\pm$0.100 & 18.096$\pm$0.178 & 0.133 & 22.831\\
  93 & 13:27:37.53 & +34:51:03.7 & 25.561$\pm$0.683 & 22.375$\pm$0.081 & 20.543$\pm$0.031 & 19.332$\pm$0.018 & 18.654$\pm$0.033 & 17.730$\pm$0.023 & 17.174$\pm$0.024 & 16.747$\pm$0.040 & 16.484$\pm$0.059 & 0.220 & 22.258\\
  94 & 13:27:41.76 & $-$01:29:15.8 & 23.909$\pm$0.913 & 23.946$\pm$0.445 & 21.717$\pm$0.095 & 19.771$\pm$0.029 & 18.738$\pm$0.045 & 17.838$\pm$0.023 & 17.238$\pm$0.026 & 16.789$\pm$0.040 & 16.541$\pm$0.039 & 0.178 & 22.971\\
  95 & 13:28:23.36 & +30:21:44.9 & 24.451$\pm$0.985 & 23.551$\pm$0.359 & 21.418$\pm$0.074 & 20.068$\pm$0.033 & 19.218$\pm$0.054 & 18.406$\pm$0.032 & 17.911$\pm$0.031 & 17.421$\pm$0.037 & 17.309$\pm$0.075 & 0.238 & 23.300\\
  96 & 13:50:53.40 & +23:50:24.3 & 23.747$\pm$0.620 & 21.767$\pm$0.054 & 19.965$\pm$0.019 & 18.863$\pm$0.012 & 18.265$\pm$0.021 & 17.416$\pm$0.016 & 16.890$\pm$0.015 & 16.474$\pm$0.022 & 16.287$\pm$0.033 & 0.337 & 22.603\\
  97 & 13:55:28.24 & +06:51:14.6 & 24.606$\pm$0.903 & 23.428$\pm$0.231 & 21.559$\pm$0.078 & 19.894$\pm$0.026 & 19.183$\pm$0.051 & 18.091$\pm$0.034 & 17.547$\pm$0.036 & 17.175$\pm$0.039 & 16.919$\pm$0.055 & 0.175 & 22.768\\
  98 & 14:47:29.91 & $-$01:59:50.3 & 24.899$\pm$0.984 & 24.586$\pm$0.619 & 22.487$\pm$0.183 & 20.762$\pm$0.056 & 19.577$\pm$0.086 & 18.899$\pm$0.077 & 18.168$\pm$0.081 & 17.920$\pm$0.103 & 17.621$\pm$0.147 & 0.191 & 23.888\\
  99 & 15:32:41.87 & +02:41:46.0 & 23.699$\pm$0.759 & 22.898$\pm$0.167 & 21.057$\pm$0.056 & 20.050$\pm$0.033 & 19.429$\pm$0.068 & 18.678$\pm$0.049 & 18.178$\pm$0.056 & 17.818$\pm$0.068 & 17.752$\pm$0.112 & 0.305 & 23.482\\
 100 & 20:58:19.76 & +00:01:03.9 & 24.115$\pm$0.844 & 23.482$\pm$0.240 & 21.170$\pm$0.051 & 18.962$\pm$0.014 & 17.861$\pm$0.021 & 16.870$\pm$0.013 & 16.300$\pm$0.012 & 15.916$\pm$0.019 & 15.638$\pm$0.021 & 0.164 & 22.238\\
\label{Table_candidates}
\end{longtable}
Other candidates published in in the literature with their identifier and/or spectral type are: 
\citet{lepine05d} 3, 4 (LHS-2023), 6 (LHS-2045), 7, 11 (LHS-2096), 12, 13, 14, 22, 23, 25 (LHS-3061), 
26, 29, 75, 82, 87, 97; \citet{lepine08b}, 31 (sdM7.5), 17 (sdM8.0), 79 (sdM8.5), and 23 (esdM7.0); \citet{lodieu12b}: 6 (usdM5.5) and 31 (sdM8.0). After a revision of our templates, we revised the 
the original spectral types from \citet{west08} for the following sources (Table \ref{Table_indices_SpT}): ID: 4 (M3), 5 (M4), 6 (M3), 7 (M3), 8 (M3), 9 (M4), 11 (M2), 12 (M4), 15 (M4), 17 (M4), 18 (M4), 
23 (M5), 29 (M4), 31 (M5), 33 (M5), 43 (M3), 53 (M5), 67 (M4), 77 (M4), 79 (M5), 83 (M3), 85 (M4). 
The candidate with ID\,=\,31 was also retrieved from the SDSS DR7 vs 2MASS and the 
SDSS DR7 vs UKIDSS LAS DR6 cross-matches but was only kept in the later.
The $J$ and $K$ photometry from UKIDSS exists for several candidates identified in the 
SDSS DR7 vs 2MASS cross-match: (ID\,=\,1, 2, 6, 10, 11, and 19). We used this photometry
to plot them in Fig.\ \ref{Fig_i-J_vs_J-K}: $J$\,=\,16.149$\pm$0.011, $K$\,=\,15.403$\pm$0.015; 16.925$\pm$0.014, 16.109$\pm$0.024; 14.732$\pm$0.003, 14.052$\pm$0.006; 15.227$\pm$0.004, 14.517$\pm$0.007;  13.962$\pm$0.002, 13.362$\pm$0.003; 16.543$\pm$0.010, 15.682$\pm$0.023, respectively.
\end{landscape}
\end{longtab}

\normalsize

%
%

%
%
\section{Proper motion revision: discarding solar-metallicity dwarfs}
\label{sdM_VO:PM_revision}
To reject potential false candidates in our sample, we refined the proper motions by performing 
accurate astrometric studies of the bi-dimensional images retrieved from the surveys.
We carried out this work with IRAF\footnote{IRAF is distributed by the National Optical Astronomy Observatories, which are operated by the Association of Universities for Research in Astronomy, Inc., under cooperative agreement with the National Science Foundation} \citep{tody86,tody93} with the 
tasks {\it{daofind}}, {\it{xyxymatch}}, {\it{geomap}}, and {\it{geoxytran}}. The task {\it{daofind}} selects 
the good sources in each image, {\it{xyxymatch}} matches the two lists of targets selected from each image and generated by {\it{daofind}}, {\it{geomap}} gives a transformation equation to convert one 
set of coordinates into the other set, and {\it{geoxytran}} transforms the coordinates from one epoch 
to the other. 

To obtain more accurate proper motions for our candidates, we considered their physical positions 
(X and Y, in pixels) and the observing dates of each survey. This process requires a good number of 
reference stars (at least 30) in common between the images to obtain a reasonable fit with 
{\it{geomap}}. This method assumes that reference stars are background stars with negligible
motions. Contrary to the VO, we considered the positions of the objects in the two dimensional 
images rather than the catalogue coordinates of these stars. We calculated the offsets in pixels
between the two epochs and later converted pixels into arcsec with the corresponding plate scale.
At the end of the process, we obtained new proper motion estimates (in $''$/yr) in right ascension and declination.

We revised the proper motions with this method only for the 71 candidates with ID between 
30 and 100 coming from the multiple SDSS vs UKIDSS cross-matches. The mean difference between the 
refined proper motions and the proper motions calculated with VO tools is 0.029 arcsec/yr with a rms of 
0.012 arcsec/yr. We could not refine the proper motion for the candidates in the SDSS vs 2MASS cross-match
because of a poor fit mainly due to the low spatial resolution of the 2MASS images 
and the low number of sources in common to the SDSS images.

Among the 71 candidates with refined proper motions, only the ID\,=\,70 has a refined value of H$r$ 
below 20.7 mag (column 7 in Table \ref{Table_pm_Hr}), indicating that it is a potential contaminant. We confirm
spectroscopically the solar-metallicity nature of this object and classify it as a dM5.0$\pm$0.5. 
The other object that could be considered as a contaminant is ID\,=\,37, with H$r$ between 20.26 and 21.43 mag. 
This candidate remains without optical spectrum, but we keep it as a subdwarf candidate because 
its H$r$ overlaps with our original criterion. Therefore, the object with ID\,=\,70 is the only one out of 
71 candidates that could have been rejected prior to our follow-up --- it only represents 1\% of 
our sample. We conclude that the H$r$ from the VO are very reliable for the SDSS vs UKIDSS sample.

For the 29 candidates from the SDSS vs 2MASS cross-match, we looked for proper motions in the PPMXL catalogue 
\citep{roeser10}, where we found 16 of them whose ID range from 7 to 29 (Table \ref{Table_pm_Hr}).
The mean difference between the proper motions in the PPMXL catalogue and the proper motions calculated with VO tools is 
0.128 arcsec/yr. We re-calculated their H$r$ values using the proper motions from the PPMXL catalogue and found two distinct cases:
\begin{itemize}
\renewcommand{\labelitemi}{$\bullet$}
\item Three of these 16 candidates would have H$r$\,$<$\,20.7 mag (objects with ID\,=\,14, 22, and 24), 
so they would not pass our original criterion in H$r$. Therefore, they should not be in our sample using 
the PPMXL proper motions. Nevertheless, we confirmed spectroscopically these objects as late-type subdwarfs. 
Checking the 2MASS and SDSS images, we see a clear motion for ID\,=\,22, so we trust 
the proper motion calculated by the VO. For ID\,=\,14 we see a second object close to our subdwarf 
in the SDSS image but not in the 2MASS image, which may result in a false match either in the 
PPMXL or VO catalogue. In the case of ID\,=\,24, the motion is small and the 2MASS coordinates 
are not well centred on the object, suggesting that the proper motion calculated by the VO may be overestimated.
\item Thirteen of these 16 candidates would have H$r$\,$\geq$\,20.7 mag, so they would pass our
original criterion and be in our sample. We classify spectroscopically two of these 16 candidates 
(ID\,=\,13 and 27) as solar-metallicity M dwarfs.
\end{itemize}

We collected our optical spectra before checking the PPMXL catalogue so that our sample was defined using 
the total proper motion calculated by the VO, reason why we have spectra for all 29 candidates
from the SDSS DR7 vs 2MASS cross-match.
Table \ref{Table_pm_Hr} provides the new proper motion determinations: those derived from the 
cross-correlation of 2D images for objects with ID\,=\,30 through 100, and those obtained from the PPMXL 
catalogue for candidates with ID\,=\,7 through 29\@.

%
%

\begin{longtab}
\begin{longtable}{ccccccc}
\caption{Proper motion and H$r$ values for our candidates derived from the VO (columns two and three), refined values of proper motions calculated using images of SDSS and UKIDSS (columns 
four to seven, ID between 30 and 100), and proper motions from PPMXL (columns four to seven, 
ID 7 to 29). The column H$r$ show the value of the parameter considering the refined total proper motions (or the total proper motions found in PPMXL, for ID 7 to 29).}\\
\hline\hline
ID & $\mu$(VO) & H$r$(VO) & $\mu_{\alpha}cos{\delta}$ & $\mu_{\delta}$ & $\mu_{\rm total}$ & H$r$\\
   & [$''$/yr] & [mag]    & [$''$/yr]                  & [$''$/yr]      & [$''$/yr] & [mag]\\
\hline
\endfirsthead
\caption{continued.}\\
\hline\hline
ID & $\mu$(VO) & H$r$(VO) & $\mu_{\alpha}cos{\delta}$ & $\mu_{\delta}$ & $\mu_{\rm total}$ & H$r$\\
   & [$''$/yr] & [mag]    & [$''$/yr]                  & [$''$/yr]      & [$''$/yr] & [mag]\\
\hline
\endhead
\hline
\endfoot
7  & 0.383 & 21.424 & $-$0.028$\pm$0.006 & $-$0.376$\pm$0.006 & 0.377$\pm$0.008 & 21.395$\pm$0.215 (21.180--21.610)\\  
8  & 0.476 & 23.279 & $-$0.059$\pm$0.006 & $-$0.339$\pm$0.006 & 0.345$\pm$0.008 & 22.568$\pm$0.225 (22.343--22.793)\\  
9  & 0.212 & 21.377 &  0.177$\pm$0.006 &  0.281$\pm$0.006 & 0.332$\pm$0.009 & 22.400$\pm$0.245 (22.155--22.644)\\  
10 & 0.212 & 21.498 & $-$0.077$\pm$0.005 & $-$0.178$\pm$0.005 & 0.193$\pm$0.007 & 21.278$\pm$0.281 (20.996--21.559)\\  
11 & 0.591 & 20.919 & $-$0.511$\pm$0.005 & $-$0.249$\pm$0.005 & 0.568$\pm$0.007 & 20.845$\pm$0.164 (20.681--21.008)\\  
12 & 0.357 & 22.395 &  0.128$\pm$0.006 & $-$0.279$\pm$0.006 & 0.307$\pm$0.008 & 22.074$\pm$0.238 (21.835--22.312)\\  
13 & 0.318 & 22.576 &  0.065$\pm$0.005 & $-$0.151$\pm$0.005 & 0.164$\pm$0.007 & 21.164$\pm$0.305 (20.859--21.470)\\  
14 & 0.388 & 20.751 &  0.041$\pm$0.004 & $-$0.035$\pm$0.004 & 0.053$\pm$0.005 & 16.379$\pm$0.453 (15.927--16.832)\\  
18 & 0.435 & 23.676 & $-$0.298$\pm$0.006 & $-$0.204$\pm$0.006 & 0.361$\pm$0.009 & 23.323$\pm$0.235 (23.088--23.558)\\  
20 & 0.288 & 22.287 &  0.081$\pm$0.006 & $-$0.240$\pm$0.006 & 0.254$\pm$0.008 & 22.025$\pm$0.262 (21.763--22.287)\\  
22 & 0.587 & 22.497 &  0.051$\pm$0.005 & $-$0.004$\pm$0.005 & 0.051$\pm$0.007 & 17.197$\pm$0.546 (16.651--17.743)\\  
23 & 0.337 & 21.632 & $-$0.245$\pm$0.006 & $-$0.170$\pm$0.006 & 0.298$\pm$0.008 & 21.365$\pm$0.242 (21.123--21.607)\\  
24 & 0.297 & 22.428 & $-$0.011$\pm$0.006 & $-$0.005$\pm$0.006 & 0.012$\pm$0.008 & 15.458$\pm$1.203 (14.255--16.661)\\  
26 & 0.309 & 20.908 & $-$0.057$\pm$0.004 & $-$0.329$\pm$0.004 & 0.334$\pm$0.006 & 21.057$\pm$0.198 (20.859--21.254)\\  
27 & 0.566 & 24.018 & $-$0.288$\pm$0.007 & $-$0.197$\pm$0.006 & 0.349$\pm$0.009 & 22.707$\pm$0.238 (22.470--22.945)\\  
29 & 0.207 & 20.981 & $-$0.201$\pm$0.006 &  0.003$\pm$0.006 & 0.201$\pm$0.008 & 20.902$\pm$0.294 (20.608--21.196)\\  
30 & 0.298 & 24.610 & $-$0.156$\pm$0.005 &  0.212$\pm$0.005 & 0.264$\pm$0.007 & 24.343$\pm$0.177 (24.166--24.520)\\
31 & 0.270 & 21.925 & $-$0.276$\pm$0.002 &  0.024$\pm$0.001 & 0.277$\pm$0.002 & 21.979$\pm$0.026 (21.953--22.005)\\
32 & 0.422 & 26.386 & $-$0.422$\pm$0.010 & $-$0.005$\pm$0.013 & 0.422$\pm$0.016 & 26.399$\pm$0.346 (26.053--26.745)\\
33 & 0.269 & 22.504 & $-$0.229$\pm$0.008 & $-$0.113$\pm$0.007 & 0.255$\pm$0.010 & 22.378$\pm$0.091 (22.287--22.469)\\
34 & 0.223 & 23.462 &  0.220$\pm$0.008 &  0.052$\pm$0.008 & 0.226$\pm$0.011 & 23.481$\pm$0.150 (23.331--23.631)\\
35 & 0.418 & 22.659 &  0.017$\pm$0.004 &  0.417$\pm$0.007 & 0.417$\pm$0.009 & 22.656$\pm$0.050 (22.606--22.706)\\
36 & 0.346 & 23.935 & $-$0.348$\pm$0.006 & $-$0.012$\pm$0.005 & 0.348$\pm$0.008 & 23.949$\pm$0.070 (23.879--24.019)\\
37 & 0.134 & 23.472 & $-$0.017$\pm$0.007 &  0.036$\pm$0.003 & 0.040$\pm$0.008 & 20.851$\pm$0.586 (20.265--21.437)\\
38 & 0.151 & 22.455 & $-$0.086$\pm$0.005 & $-$0.095$\pm$0.004 & 0.128$\pm$0.007 & 22.096$\pm$0.145 (21.951--22.241)\\
39 & 0.193 & 22.206 & $-$0.179$\pm$0.006 &  0.015$\pm$0.005 & 0.180$\pm$0.008 & 22.051$\pm$0.103 (21.948--22.154)\\
40 & 0.235 & 21.964 & $-$0.201$\pm$0.023 &  0.095$\pm$0.014 & 0.222$\pm$0.027 & 21.828$\pm$0.265 (21.563--22.093)\\
41 & 0.233 & 22.458 & $-$0.206$\pm$0.007 & $-$0.118$\pm$0.005 & 0.238$\pm$0.009 & 22.502$\pm$0.088 (22.414--22.590)\\
42 & 0.147 & 22.425 & $-$0.112$\pm$0.005 & $-$0.099$\pm$0.004 & 0.150$\pm$0.006 & 22.470$\pm$0.115 (22.355--22.585)\\
43 & 0.165 & 21.278 & $-$0.167$\pm$0.009 &  0.020$\pm$0.007 & 0.168$\pm$0.011 & 21.326$\pm$0.145 (21.181--21.471)\\
44 & 0.321 & 23.408 & $-$0.184$\pm$0.006 & $-$0.280$\pm$0.006 & 0.335$\pm$0.009 & 23.497$\pm$0.074 (23.423--23.571)\\
45 & 0.225 & 23.228 & $-$0.201$\pm$0.006 & $-$0.056$\pm$0.005 & 0.208$\pm$0.008 & 23.065$\pm$0.107 (22.958--23.172)\\
46 & 0.263 & 23.504 & $-$0.316$\pm$0.011 &  0.081$\pm$0.008 & 0.326$\pm$0.014 & 23.973$\pm$0.108 (23.865--24.081)\\
47 & 0.118 & 21.514 & $-$0.132$\pm$0.005 &  0.002$\pm$0.003 & 0.132$\pm$0.006 & 21.754$\pm$0.122 (21.632--21.876)\\
48 & 0.167 & 23.205 & $-$0.098$\pm$0.002 & $-$0.168$\pm$0.002 & 0.194$\pm$0.003 & 23.534$\pm$0.130 (23.404--23.664)\\
49 & 0.168 & 22.099 &  0.115$\pm$0.013 & $-$0.030$\pm$0.007 & 0.119$\pm$0.015 & 21.350$\pm$0.280 (21.070--21.630)\\
50 & 0.158 & 22.842 & $-$0.152$\pm$0.005 &  0.051$\pm$0.003 & 0.160$\pm$0.006 & 22.874$\pm$0.129 (22.745--23.003)\\
51 & 0.135 & 21.746 & $-$0.112$\pm$0.004 & $-$0.095$\pm$0.004 & 0.147$\pm$0.005 & 21.928$\pm$0.101 (21.827--22.029)\\
52 & 0.403 & 23.932 & $-$0.399$\pm$0.006 &  0.063$\pm$0.003 & 0.404$\pm$0.007 & 23.939$\pm$0.057 (23.882--23.996)\\
53 & 0.376 & 22.953 &  0.353$\pm$0.011 &  0.214$\pm$0.010 & 0.413$\pm$0.015 & 23.145$\pm$0.083 (23.062--23.228)\\
54 & 0.119 & 22.263 & $-$0.095$\pm$0.005 & $-$0.099$\pm$0.005 & 0.137$\pm$0.006 & 22.579$\pm$0.133 (22.446--22.712)\\
55 & 0.118 & 21.924 & $-$0.089$\pm$0.004 &  2E$-$4 $\pm$0.005 & 0.089$\pm$0.006 & 21.311$\pm$0.168 (21.143--21.479)\\
56 & 0.127 & 22.224 & $-$0.105$\pm$0.005 &  0.130$\pm$0.003 & 0.167$\pm$0.006 & 22.819$\pm$0.112 (22.707--22.931)\\
57 & 0.170 & 22.687 & $-$0.152$\pm$0.004 & $-$0.149$\pm$0.004 & 0.213$\pm$0.006 & 23.167$\pm$0.101 (23.066--23.268)\\
58 & 0.249 & 23.425 & $-$0.213$\pm$0.009 &  0.127$\pm$0.009 & 0.248$\pm$0.012 & 23.411$\pm$0.140 (23.271--23.551)\\
59 & 0.195 & 23.029 & $-$0.208$\pm$0.003 &  0.063$\pm$0.006 & 0.218$\pm$0.007 & 23.265$\pm$0.093 (23.172--23.358)\\
60 & 0.140 & 22.732 & $-$0.132$\pm$0.004 & $-$0.049$\pm$0.004 & 0.141$\pm$0.006 & 22.755$\pm$0.160 (22.595--22.915)\\
61 & 0.185 & 22.677 &  0.011$\pm$0.004 & $-$0.179$\pm$0.005 & 0.180$\pm$0.006 & 22.617$\pm$0.112 (22.505--22.729)\\
62 & 0.223 & 22.084 & $-$0.228$\pm$0.006 & $-$0.005$\pm$0.003 & 0.228$\pm$0.007 & 22.128$\pm$0.072 (22.056--22.200)\\
63 & 0.239 & 24.230 & $-$0.175$\pm$0.006 & $-$0.151$\pm$0.005 & 0.231$\pm$0.008 & 24.153$\pm$0.199 (23.954--24.352)\\
64 & 0.646 & 23.445 &  0.643$\pm$0.010 & $-$0.130$\pm$0.006 & 0.655$\pm$0.012 & 23.451$\pm$0.042 (23.409--23.493)\\
65 & 0.193 & 22.946 & $-$0.084$\pm$0.004 & $-$0.167$\pm$0.005 & 0.187$\pm$0.006 & 22.871$\pm$0.097 (22.774--22.968)\\
66 & 0.125 & 21.577 & $-$0.090$\pm$0.005 & $-$0.091$\pm$0.004 & 0.128$\pm$0.006 & 21.620$\pm$0.114 (21.506--21.734)\\
67 & 0.189 & 21.515 &  0.017$\pm$0.009 & $-$0.186$\pm$0.008 & 0.187$\pm$0.012 & 21.482$\pm$0.142 (21.340--21.624)\\
68 & 0.142 & 21.852 & $-$0.092$\pm$0.005 & $-$0.111$\pm$0.005 & 0.144$\pm$0.007 & 21.891$\pm$0.119 (21.772--22.010)\\
69 & 0.154 & 22.740 &  0.002$\pm$0.008 &  0.159$\pm$0.008 & 0.159$\pm$0.011 & 22.810$\pm$0.184 (22.626--22.994)\\
70 & 0.771 & 26.087 &  0.010$\pm$0.013 & $-$0.040$\pm$0.008 & 0.041$\pm$0.015 & 19.889$\pm$0.800 (19.089--20.689)\\
71 & 0.208 & 23.174 &  0.188$\pm$0.007 & $-$0.051$\pm$0.006 & 0.195$\pm$0.010 & 23.034$\pm$0.129 (22.905--23.163)\\
72 & 0.169 & 23.260 &  0.115$\pm$0.015 &  0.061$\pm$0.012 & 0.131$\pm$0.019 & 22.698$\pm$0.344 (22.354--23.042)\\
73 & 0.129 & 21.367 &  0.085$\pm$0.005 & $-$0.110$\pm$0.005 & 0.139$\pm$0.007 & 21.531$\pm$0.118 (21.413--21.649)\\
74 & 0.144 & 21.959 &  0.030$\pm$0.007 & $-$0.135$\pm$0.008 & 0.138$\pm$0.011 & 21.878$\pm$0.183 (21.695--22.061)\\
75 & 0.287 & 21.666 &  0.299$\pm$0.010 & $-$0.035$\pm$0.010 & 0.302$\pm$0.014 & 21.775$\pm$0.102 (21.673--21.876)\\
76 & 0.266 & 22.591 &  0.259$\pm$0.010 & $-$0.046$\pm$0.010 & 0.263$\pm$0.014 & 22.571$\pm$0.119 (22.452--22.690)\\
77 & 0.255 & 22.235 & $-$0.203$\pm$0.010 &  0.169$\pm$0.013 & 0.264$\pm$0.017 & 22.311$\pm$0.142 (22.169--22.453)\\
78 & 0.193 & 22.866 &  0.130$\pm$0.009 &  0.087$\pm$0.008 & 0.157$\pm$0.012 & 22.407$\pm$0.175 (22.232--22.582)\\
79 & 0.297 & 22.934 & $-$0.280$\pm$0.015 &  0.081$\pm$0.010 & 0.292$\pm$0.019 & 22.893$\pm$0.146 (22.747--23.039)\\
80 & 0.244 & 21.975 &  0.121$\pm$0.008 & $-$0.217$\pm$0.008 & 0.248$\pm$0.012 & 22.012$\pm$0.107 (21.905--22.119)\\
81 & 0.505 & 23.514 & $-$0.438$\pm$0.009 & $-$0.220$\pm$0.011 & 0.490$\pm$0.014 & 23.450$\pm$0.068 (23.382--23.518)\\
82 & 0.503 & 22.822 & $-$0.513$\pm$0.006 & $-$0.054$\pm$0.007 & 0.516$\pm$0.009 & 22.880$\pm$0.040 (22.840--22.920)\\
83 & 0.260 & 22.141 &  0.252$\pm$0.016 &  0.102$\pm$0.008 & 0.271$\pm$0.018 & 22.231$\pm$0.146 (22.085--22.377)\\
84 & 0.114 & 22.324 & $-$0.072$\pm$0.011 &  0.055$\pm$0.009 & 0.091$\pm$0.014 & 21.840$\pm$0.356 (21.484--22.196)\\
85 & 0.251 & 22.231 & $-$0.061$\pm$0.015 &  0.317$\pm$0.009 & 0.323$\pm$0.018 & 22.777$\pm$0.123 (22.654--22.900)\\
86 & 0.246 & 24.458 & $-$0.234$\pm$0.010 &  0.049$\pm$0.017 & 0.239$\pm$0.020 & 24.396$\pm$0.268 (24.128--24.664)\\
87 & 0.225 & 21.497 &  0.248$\pm$0.019 & $-$0.010$\pm$0.003 & 0.248$\pm$0.019 & 21.714$\pm$0.167 (21.547--21.881)\\
88 & 0.231 & 23.487 & $-$0.106$\pm$0.011 &  0.176$\pm$0.013 & 0.205$\pm$0.017 & 23.233$\pm$0.199 (23.034--23.432)\\
89 & 0.339 & 23.755 &  0.310$\pm$0.011 &  0.032$\pm$0.010 & 0.311$\pm$0.015 & 23.570$\pm$0.115 (23.455--23.685)\\
90 & 0.150 & 23.737 & $-$0.065$\pm$0.011 &  0.289$\pm$0.005 & 0.296$\pm$0.012 & 25.208$\pm$0.229 (24.979--25.437)\\
91 & 0.213 & 22.526 &  0.202$\pm$0.013 & $-$0.025$\pm$0.007 & 0.204$\pm$0.014 & 22.436$\pm$0.154 (22.282--22.590)\\
92 & 0.133 & 22.831 &  0.056$\pm$0.011 & $-$0.113$\pm$0.006 & 0.126$\pm$0.013 & 22.711$\pm$0.261 (22.450--22.972)\\
93 & 0.220 & 22.258 & $-$0.010$\pm$0.007 & $-$0.200$\pm$0.012 & 0.200$\pm$0.014 & 22.046$\pm$0.155 (21.891--22.201)\\
94 & 0.178 & 22.971 &  0.105$\pm$0.012 & $-$0.154$\pm$0.008 & 0.186$\pm$0.015 & 23.069$\pm$0.199 (22.870--23.268)\\
95 & 0.238 & 23.300 &  0.034$\pm$0.018 &  0.221$\pm$0.012 & 0.224$\pm$0.022 & 23.167$\pm$0.226 (22.941--23.393)\\
96 & 0.337 & 22.603 & $-$0.035$\pm$0.017 &  0.378$\pm$0.020 & 0.379$\pm$0.026 & 22.859$\pm$0.150 (22.709--23.009)\\
97 & 0.175 & 22.768 &  0.184$\pm$0.007 &  0.004$\pm$0.006 & 0.184$\pm$0.010 & 22.878$\pm$0.141 (22.737--23.019)\\
98 & 0.191 & 23.888 &  0.091$\pm$0.007 & $-$0.153$\pm$0.007 & 0.178$\pm$0.010 & 23.744$\pm$0.220 (23.524--23.964)\\
99 & 0.305 & 23.482 & $-$0.241$\pm$0.008 &  0.123$\pm$0.008 & 0.271$\pm$0.012 & 23.221$\pm$0.111 (23.110--23.332)\\
100 & 0.164 & 22.238 &  0.084$\pm$0.007 & $-$0.158$\pm$0.010 & 0.179$\pm$0.012 & 22.429 $\pm$0.154 (22.275--22.583)\\
\label{Table_pm_Hr}
\end{longtable}
\end{longtab}    

\section{Optical Spectroscopic Follow-up and Data Reduction}
\label{sdM_VO:spec_followup}

We obtained long-slit optical spectra with different telescope and instrument configurations. We observed 
in service mode under grey time, clear conditions, and at parallactic angle with 
the moon further away than 30 degrees from our targets. 
In Table \ref{Table_observations} we give exposure times as well as seeing and airmass at the time 
of the observations for our own spectroscopic follow-up of 71 candidates, excluding the spectra
downloaded from the SDSS spectroscopic database. We reduced all optical spectra under the IRAF 
environment \citep{tody86,tody93} . We removed the median-combined bias, divided by the normalised 
dome flat, extracted optimally the 
spectra, calibrated in wavelengths using arc lamps, and corrected for instrumental response with 
a spectrophotometric standard star observed on the same night as the targets. We normalised 
all the spectra displayed in Figs.\ \ref{Fig_sdM_sdL_spectra} and \ref{Fig_esdM_usdM_spectra} 
at 7500\,\,\AA{}. We note that only the SDSS spectra are corrected for telluric absorptions.

We have designed an ultracool subdwarf archive containing the new subdwarfs presented in
this paper and known subdwarfs with optical spectral types later than (or equal to) M5 from the 
literature. The archive is compliant with the VO standards as described
in Appendix \ref{sdM_VO:appendix_archive}\footnote{http://svo2.cab.inta-csic.es/vocats/ltsa/}.
We provide coordinates, photometry, proper motions, and spectra of our subdwarfs 
in Appendix \ref{sdM_VO:appendix_archive}.

%
%
\begin{table}
\centering
\scriptsize
\tiny
\caption{Logs of observations for our candidates, including telescopes, instruments, total exposure times, 
mean airmass, and seeing.}
\label{Table_observations}
\begin{tabular}{@{\hspace{0mm}}c@{\hspace{1mm}}c@{\hspace{2mm}}c@{\hspace{2mm}}c@{\hspace{2mm}}c@{\hspace{2mm}}c@{\hspace{2mm}}c@{\hspace{0mm}}}
\hline\hline
ID & Telescope & Instrument & Date-OBS & ExpT & Airmass & Seeing\\
 & & & [YYYY-MM-DD] & [seconds] &  & [$''$]\\
\hline                                  
 7 & NOT & ALFOSC & 2009-01-29 & 2100 & 1.206 & 0.67\\
11 & NOT & ALFOSC & 2009-01-29 & 1800 & 1.069 & 1.12\\
14 & NOT & ALFOSC & 2009-01-29 & 2400 & 1.056 & 0.69\\
25 & NOT & ALFOSC & 2009-08-23 & 1800 & 1.275 & 1.00\\
26 & NOT & ALFOSC & 2009-08-23 & 1800 & 1.324 & 0.74\\
29 & NOT & ALFOSC & 2009-07-27 & 4200 & 1.309 & 1.54\\
10 & VLT & FORS2 & 2012-03-22 & 799 & 1.297 & 0.92\\
19 & VLT & FORS2 & 2012-03-22 & 799 & 1.086 & 1.04\\
27 & VLT & FORS2 & 2012-03-30 & 799 & 1.21 & 0.92\\
28 & VLT & FORS2 & 2012-03-29 & 799 & 1.241 & 0.88\\
30 & VLT & FORS2 & 2012-11-07 & 1500 & 1.344 & 0.79\\   
34 & VLT & FORS2 & 2012-02-28 & 1545 & 1.111 & 0.86\\
35 & VLT & FORS2 & 2012-03-10 & 799 & 1.173 & 1.13\\
36 & VLT & FORS2 & 2013-01-08 &  660 & 1.653 & 0.80\\   
38 & VLT & FORS2 & 2012-03-30 & 1545 & 1.196 & 1.16\\
39 & VLT & FORS2 & 2012-03-30 & 799 & 1.228 & 0.82\\
40 & VLT & FORS2 & 2012-02-22 & 799 & 1.262 & 1.34\\
41 & VLT & FORS2 & 2012-03-30 & 799 & 1.323 & 1.06\\
42 & VLT & FORS2 & 2012-03-30 & 1545 & 1.207 & 0.74\\
44 & VLT & FORS2 & 2012-01-29 & 799 & 1.744 & 0.71\\
45 & VLT & FORS2 & 2012-12-15 &  660 & 1.697 & 0.52\\   
47 & VLT & FORS2 & 2013-01-17 & 1500 & 1.486 & 0.92\\   
48 & VLT & FORS2 & 2013-01-08 & 1500 & 1.480 & 0.76\\   
49 & VLT & FORS2 & 2013-01-09 &  660 & 1.645 & 0.60\\   
50 & VLT & FORS2 & 2012-03-30 & 799 & 1.374 & 1.00\\
51 & VLT & FORS2 & 2013-01-08 & 1500 & 1.111 & 1.03\\   
52 & VLT & FORS2 & 2013-01-13 &  660 & 1.660 & 0.47\\   
53 & VLT & FORS2 & 2013-01-08/17 &  840 & 1.347 & 1.13\\   
54 & VLT & FORS2 & 2013-01-17 & 1500 & 1.189 & 0.93\\   
55 & VLT & FORS2 & 2013-01-08 & 1500 & 1.172 & 1.76\\   
56 & VLT & FORS2 & 2013-01-17 & 1500 & 1.174 & 0.94\\   
57 & VLT & FORS2 & 2013-01-17 & 1500 & 1.206 & 0.97\\   
58 & VLT & FORS2 & 2013-01-17 & 1500 & 1.243 & 0.93\\   
59 & VLT & FORS2 & 2013-03-05 & 1500 & 1.333 & 0.59\\   
60 & VLT & FORS2 & 2013-02-08 & 1500 & 1.107 & 1.50\\   
61 & VLT & FORS2 & 2013-03-11 & 1500 & 1.180 & 0.62\\   
62 & VLT & FORS2 & 2012-03-30 & 799 & 1.198 & 0.88\\
63 & VLT & FORS2 & 2012-03-30 & 799 & 1.144 & 1.25\\
65 & VLT & FORS2 & 2012-03-30 & 799 & 1.141 & 0.93\\
66 & VLT & FORS2 & 2012-03-30 & 799 & 1.172 & 1.24\\
68 & VLT & FORS2 & 2012-03-30 & 799 & 1.194 & 0.98\\
85 & VLT & FORS2 & 2013-01-17 &  660 & 1.188 & 1.48\\   
89 & VLT & FORS2 & 2013-02-20 &  660 & 1.241 & 0.79\\   
97 & VLT & FORS2 & 2013-03-05 &  660 & 1.250 & 0.83\\   
99 & VLT & FORS2 & 2013-03-07 & 1500 & 1.128 & 0.49\\ 
 1 & GTC & OSIRIS & 2010-01-14 & 600 & 1.273 & 0.80\\
 2 & GTC & OSIRIS & 2011-10-12 & 660 & 1.093 & 0.95\\
 5 & GTC & OSIRIS & 2010-01-15 & 900 & 1.988 & 0.80\\
 8 & GTC & OSIRIS & 2010-01-15 & 900 & 1.412 & 0.80\\
 9 & GTC & OSIRIS & 2010-01-14 & 900 & 1.312 & 1.00\\
13 & GTC & OSIRIS & 2010-01-15 & 900 & 1.494 & 0.70\\
15 & GTC & OSIRIS & 2010-01-15 & 900 & 1.206 & 0.80\\
21 & GTC & OSIRIS & 2012-01-13 & 660 & 1.285 & 0.90\\
24 & GTC & OSIRIS & 2012-01-13 & 660 & 1.151 & 1.10\\
32 & GTC & OSIRIS & 2012-01-17 & 1980 & 1.126 & 0.90\\
46 & GTC & OSIRIS & 2013-04-27 & 900 & 1.208 & 0.76\\
64 & GTC & OSIRIS & 2014-03-03 & 900 & 1.160 & 1.0\\            
69 & GTC & OSIRIS & 2014-03-03 & 2400 & 1.041 & 2.0\\            
70 & GTC & OSIRIS & 2014-03-06 & 1200 & 1.389 & 1.2\\            
71 & GTC & OSIRIS & 2014-03-07 & 1200 & 1.223 & 1.2\\            
72 & GTC & OSIRIS & 2014-03-07 & 1800 & 1.147 & 1.3\\            
78 & GTC & OSIRIS & 2014-03-08 & 1200 & 1.116 & 1.0\\                           
86 & GTC & OSIRIS & 2014-07-20 & 1200 & 1.529 & 0.8\\            
88 & GTC & OSIRIS & 2014-03-07 & 2400 & 1.055 & 1.1\\                           
90 & GTC & OSIRIS & 2014-03-03 & 1800 & 1.113 & 1.1\\             
91 & GTC & OSIRIS & 2014-07-20 &  900 & 1.478 & 0.8\\            
92 & GTC & OSIRIS & 2014-03-03 & 1800 & 1.164 & 1.2\\            
94 & GTC & OSIRIS & 2014-07-24 &  900 & 1.813 & 1.0\\            
95 & GTC & OSIRIS & 2014-07-24 & 1200 & 1.520 & 1.0\\            
98 & GTC & OSIRIS & 2014-07-18 & 1800 & 1.345 & 0.9\\            
100 & GTC & OSIRIS & 2014-07-04 &  900 & 1.144 & 1.3\\            
\hline
\end{tabular}
\end{table}

\subsection{GTC/OSIRIS spectra}
\label{sdM_VO:spec_followup_OSIRIS}
We observed 26 candidates with the Optical System for Imaging and low Resolution Integrated Spectroscopy 
\citep[OSIRIS;][]{cepa00} instrument on the 10.4\,m Gran Telescopio de Canarias (GTC) between January 2010 
and July 2014\@. The GTC is located in the Roque de Los Muchachos Observatory, in the island of La Palma, 
Spain. OSIRIS has an unvignetted field of view of 7\,$\times$\,7 arcmin. The detector of OSIRIS consists of 
a mosaic of two Marconi CCD42-82 (2048\,$\times$\,4096 pixels) with a 74 pixel gap between them. The pixel 
physical size is 15 $\mu$m, which corresponds to a scale of 0.254 arcsec on the sky, for a detector binned 
by a factor two, used for our long-slit spectroscopic follow-up.

We obtained optical spectra with a resolution of R\,$\sim$\,350 at $\sim$720 nm using the grism R500R and 
a slit of 1 arcsec, covering the 5000--10000 wavelength range with a dispersion of 4.7\,\AA{}/pixel. 
We calibrated the spectra in wavelength with a rms of 0.5\,\AA{} using arc lamps (HgAr, Xe, Ne) acquired 
the nights when the targets were observed.  

\begin{figure}
   \centering
   \includegraphics[width=\hsize]{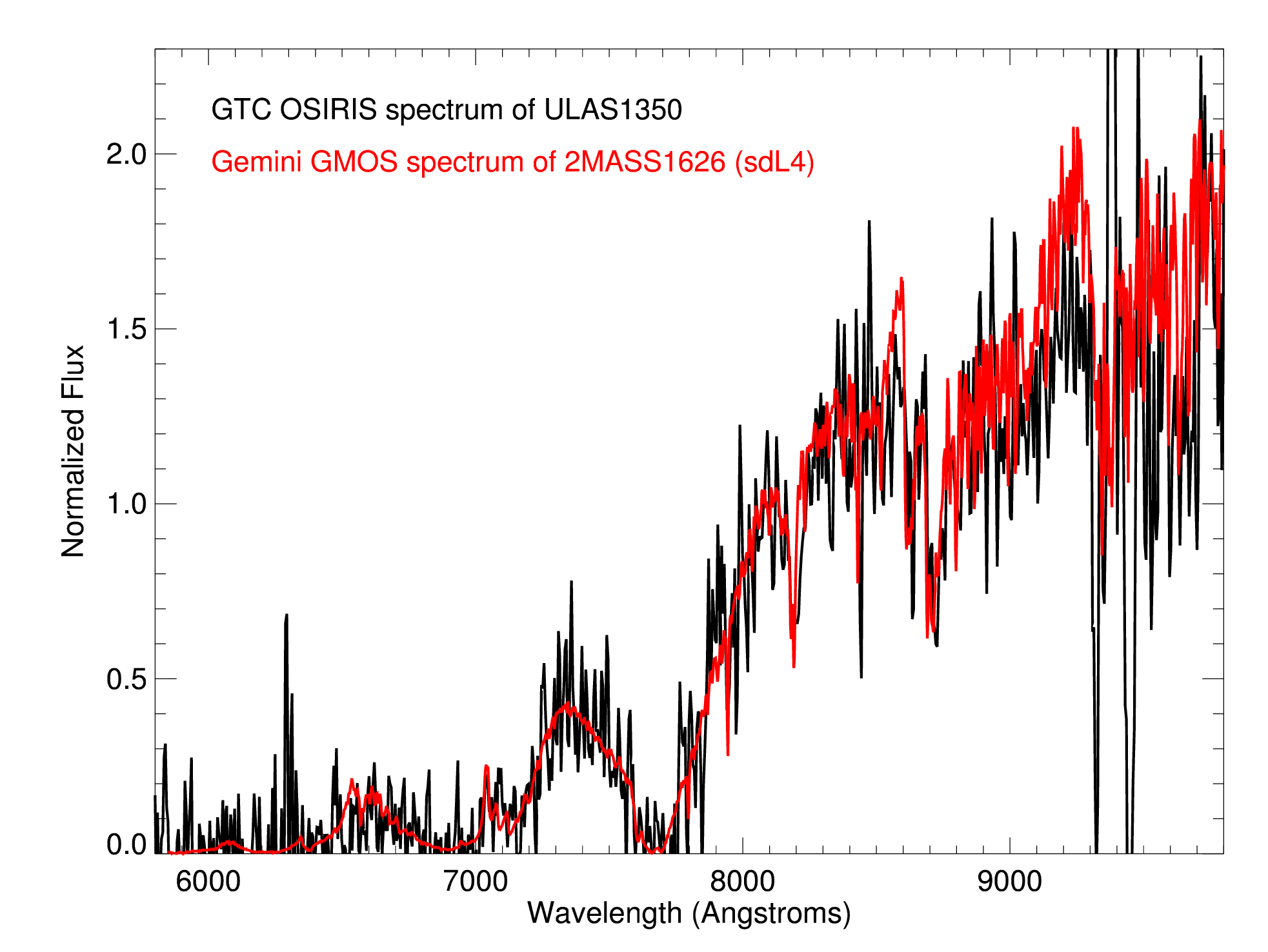}
      \caption{Improved GTC OSIRIS low-resolution optical spectrum of ULAS\,J13505886$+$0815068\@. We overplotted 
the spectrum  of the known sdL4 2MASS\,J16262035$+$3925190 from \citet{burgasser04,burgasser07b}.}
         \label{Fig_ulas1350}
   \end{figure}

In addition to the M subdwarfs, we obtained a new optical spectrum of ULAS\,J135058.86$+$081506.8 
\citep{lodieu10a} on 25 January 2014 with GTC OSIRIS and a slit of 1.5 arcsec as part of a filler 
program (GTC65-13B; PI Lodieu). The conditions were clear with dark skies but the seeing worse 
than 2 arcsec. We collected three optical spectra of 920\,s shifted along the slit to remove 
cosmic rays and detector defects. The reduced 1D spectrum is shown in Fig.\ \ref{Fig_ulas1350}, 
along with the known sdL4 subdwarf 2MASS162620.35+392519.0 \citep{burgasser04,burgasser07b}. 
Both spectra look quite similar within the spectroscopic uncertainties (Fig.\ \ref{Fig_ulas1350}), 
thus confirming the metal-depleted nature of ULAS\,J135058.86$+$081506.8 and a spectral type of 
sdL3.5--sdL4, slightly warmer than our previous classification based on a poorer spectrum \citep{lodieu10a}.

%
%
\subsection{NOT/ALFOSC spectra}
\label{sdM_VO:spec_followup_NOT}
We observed six candidates with the ALFOSC (Andalucia Faint Object Spectrograph and Camera) instrument 
on the 2.5\,m Nordic Optical Telescope (NOT) in the island of La Palma between January and August 2009\@. 
ALFOSC has a charge coupled device CCD42-40 non-inverted mode operation back illuminated of 
2048\,$\times$\,2052 pixels and has a field of view of 6.4\,$\times$\,6.4 arcmin. The pixel size and plate 
scale are 13.5\,$\mu$m and 0.19 arcsec/pixel respectively. We secured optical spectra at a resolution of 
R\,$\sim$\,450 using the grism number 5 and a slit width of 1 arcsec, except for candidates with IDs 25 and 26 
where we used a slit of 1.2 arcsec, covering a wavelength range between 5000--10700\,\,\AA{} with a dispersion
of 16.8\,\AA{}/pixel. We calibrated the spectra in wavelength with a rms better than 0.2\,\AA{} using arc 
lamps (He, Ne, Ar) obtained on the same nights as the targets.

\subsection{VLT/FORS2 spectra}
\label{sdM_VO:spec_followup_VLT}
We observed 39 candidates with the visual and near UV FOcal Reducer and low dispersion Spectrograph 
\citep[FORS2;][]{appenzeller98} instrument on the 8.2\,m (Unit Telescope 1) Very Large Telescope (VLT) 
between January 2012 and March 2013\@. The VLT is located in Cerro Paranal, in the north of Chile. 
FORS2 is equipped with a mosaic of two 2k\,$\times$\,4k MIT CCDs (with 15\,$\mu$m pixels) and has a 
field of view of 6.8\,$\times$\,6.8 arcmin with the standard resolution collimator (SR) providing an image 
scale of 0.25 arcsec/pixel in the standard readout mode (2\,$\times$\,2 binning).
We obtained optical spectra at a resolution of R\,$\sim$\,350 using the grism 150I$+$27 with a slit width 
of 1 arcsec and the order blocking filter OG590 covering a wavelength range 6000--11000\,\AA{} with a
dispersion of 3.45\,\AA{}/pixel. We used arc lamps (He, Ne, Ar) to calibrate the spectra in wavelength the 
spectra with a rms of $\sim$\,0.4--0.6\,\AA{}.

\subsection{Optical spectra from the SDSS spectroscopic database}
\label{sdM_VO:spec_followup_SDSS}
\begin{figure*}
   \resizebox{\hsize}{!}{\includegraphics[width=\hsize]{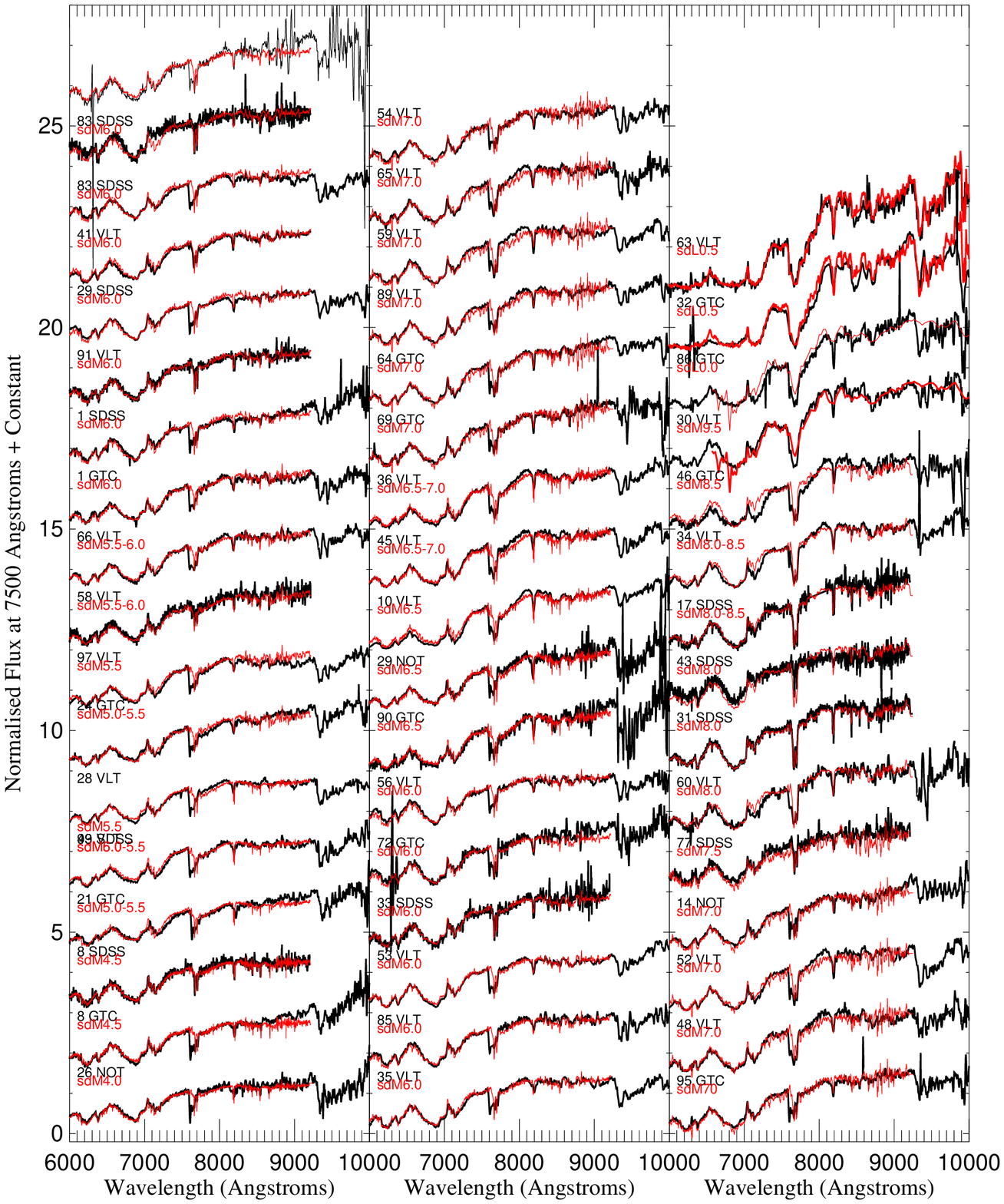}}
      \caption{Low-resolution optical spectra for 38 candidates confirmed as M and L subdwarfs (black) with 
their identifier (ID) and telescope name. Overplotted in red
are subdwarf spectral templates from SDSS used to assign the final spectral types 
(see section \ref{sdM_VO:spec_class_templates}). Candidates with ID\,=\,1, 8, and 83 appear 
twice, because they have spectra from our own follow-up and from the SDSS spectroscopic database.}
         \label{Fig_sdM_sdL_spectra}
   \end{figure*}

\begin{figure*}
   \resizebox{\hsize}{!}{\includegraphics[width=\hsize]{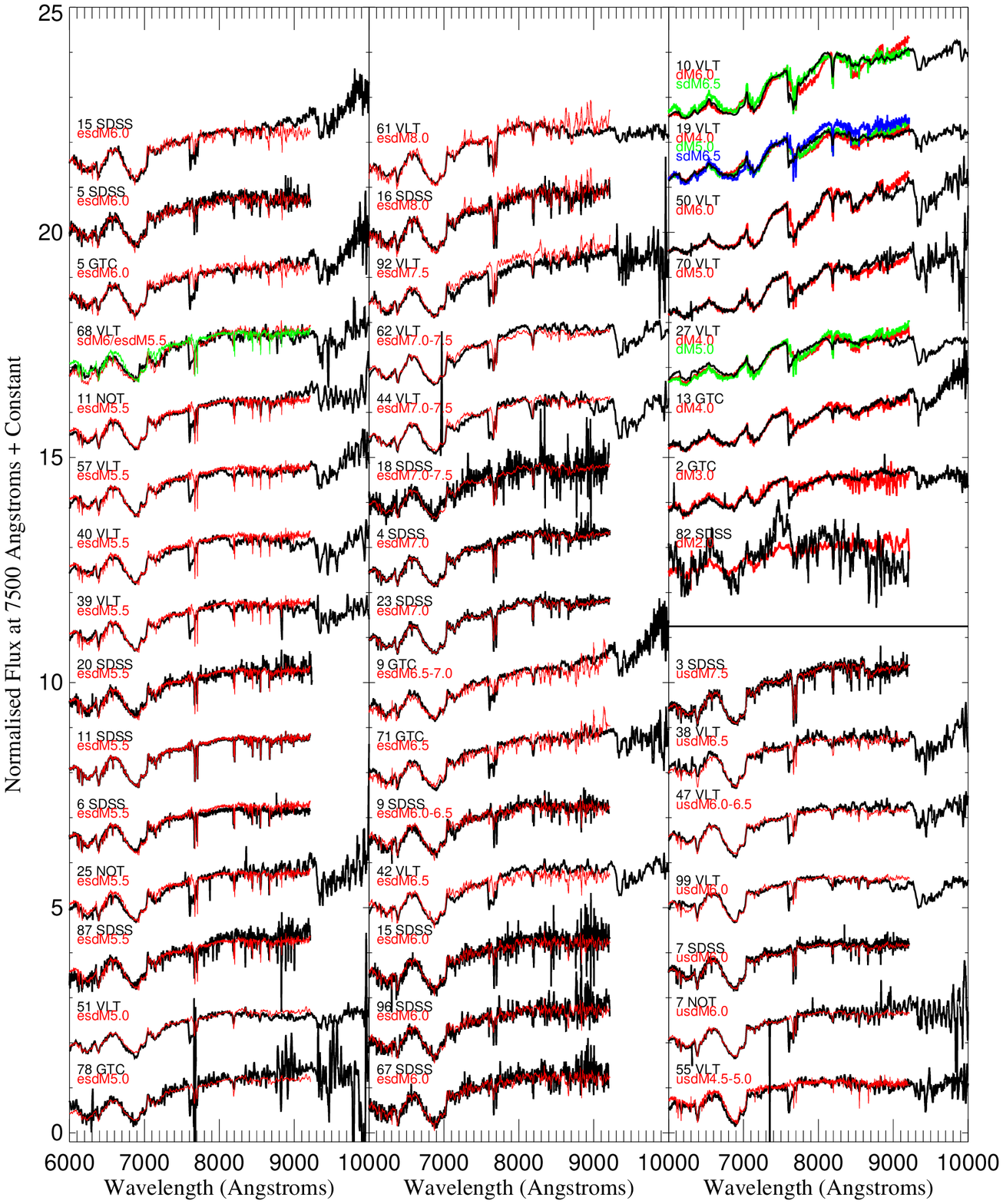}}
      \caption{Low-resolution optical spectra for our candidates (black) with identifier (ID) and 
telescope name. Overplotted in red, green, and blue are the 
spectral templates of SDSS dwarfs/subdwarfs used to assign the final spectral types (see 
section \ref{sdM_VO:spec_class_templates} and legend on the plot). {\it{First and second columns:}} 
26 candidates confirmed as extreme subdwarfs. {\it{Third column, lower part:}} 6 candidates confirmed 
as ultrasubdwarfs. {\it{Third column, upper part:}} 5 candidates classified as dM and 2 candidates 
with uncertain class, between dM and sdM. Candidates with ID\,=\,5, 9, 11, 15, and 7 appear 
twice because they were included in our spectroscopic follow-up and have a spectrum in the 
SDSS spectroscopic database.}
         \label{Fig_esdM_usdM_spectra}
   \end{figure*}

We queried the SDSS spectroscopic database to search for spectra of our candidates. These optical spectra
have R\,$\sim$\,2000 and cover the range 3800--9400\,\AA{}. We found that 30 of our candidates have a 
SDSS spectrum, nine of them in common with our own spectroscopic follow-up 
at GTC/OSIRIS, NOT/ALFOSC, or VLT/FORS2\@. These nine objects appear twice in 
Table \ref{Table_indices_SpT} and serve to double-check our spectral classification. The remaining 
21 candidates with a SDSS spectrum appear only once in Table \ref{Table_indices_SpT}.
All spectra are shown in Figs.\ \ref{Fig_sdM_sdL_spectra} and \ref{Fig_esdM_usdM_spectra}.

\section{Spectral classification}
\label{sdM_VO:spec_classification}
We used two methods to classify spectroscopically our subdwarf candidates. On the one hand, 
we classified subdwarfs with the method based on spectral indices presented by \citet{lepine07c}.
These indices measured the strength of the TiO band at 7126--7138\,\AA{} and the CaH bands
at 6380--6390\,\AA{}, 6814--6846\,\AA{}, and 6960--6990\,\AA{}.
On the other hand, we considered spectra of known M and L subdwarfs, downloaded from the SDSS 
spectroscopic database and from the literature, as templates to compare visually with the spectra 
of our candidates. However, during the process of writing this paper, \citet{savcheva14} published
templates for subdwarfs (M0--M9.5), extreme subdwarfs (esM0--esM8), and ultra-subdwarfs (usdM0--usdM7.5), 
which we will use to classify our targets.

\begin{longtab}
\tiny
\begin{longtable}{@{\hspace{0mm}}c @{\hspace{2mm}}c @{\hspace{2mm}}c @{\hspace{2mm}}c @{\hspace{2mm}}c c @{\hspace{1mm}}c @{\hspace{1mm}}c @{\hspace{1mm}}c @{\hspace{1mm}}c @{\hspace{1mm}}c @{\hspace{1mm}}c @{\hspace{1mm}}c@{\hspace{0mm}}}
\caption{Identifier (ID) of our candidates (column 1), spectral indices presented by \citet{gizis97a} 
(columns 2--5), classification based on the scheme from \citet{lepine07c} (column 6), spectral type adopted 
using SDSS spectral templates (column 7), telescope where the spectrum was obtained (column 8), spectroscopic 
distances with their errors (column 9), heliocentric velocities (column 10) compared to spectral templates 
in our sample (ID\,=\,29, 11, and 7 as sdM, esdM, and usdM templates, respectively), except for the three 
L-type subdwarf where we used the sdL3.5 subdwarf \citep{sivarani09,burgasser09a} and space motions for
sources with revised proper motions in both directions.
Some candidates appear twice because they have spectra from the SDSS database and our own spectroscopic 
follow-up.} \\
\hline\hline
ID & TiO5 & CaH1 & CaH2 & CaH3 & SpT L\'epine & SpT final & Telescope  & Distance &      Vh  & U & V & W \\
   &      &      &      &      &              &           &            &   [pc]   &    [km/s] & [km/s] & [km/s] & [km/s]  \\
\hline
\endfirsthead
\caption{continued.}\\
\hline\hline
ID & TiO5 & CaH1 & CaH2 & CaH3 & SpT L\'epine & SpT final & Telescope  & Distance &      Vh  & U & V & W  \\
   &      &      &      &      &              &           &            &   [pc]   &    [km/s]  [km/s] & [km/s] & [km/s]   \\
\hline
\endhead
\hline
\endfoot
\hline\hline
ID & TiO5 & CaH1 & CaH2 & CaH3 & SpT L\'epine & SpT final & Telescope  & Distance &      Vh  & U & V & W  \\
   &      &      &      &      &              &           &            &   [pc]   &    [km/s] & [km/s] & [km/s] & [km/s]  \\
\hline
\endhead
\hline
\endfoot
   1   &  0.63   & 0.604  & 0.366  & 0.521  & esdM4.5  & sdM6.0         & GTC        & 158.8$\pm$38.5 &       $-$106$\pm$100 & ---    &   ---    &  ---     \\
   1   &  0.615  & 0.487  & 0.296  & 0.479  & esdM5.5  & sdM6.0         & SDSS       & 158.8$\pm$38.5 &       $-$35$\pm$36   & $-$118.3$\pm$67.8 & $-$259.5$\pm$59.3 & 16.0$\pm$39.2 \\
   2   &  0.626  & 0.853  & 0.526  & 0.792  & dM1.5    & dM3.0          & GTC        & --- $\pm$ ---  &         ---$\pm$---  & ---    &   ---    &  ---     \\
   3   &  0.898  & 0.229  & 0.23   & 0.287  & usdM7.5  & usdM7.5        & SDSS       & 104.6$\pm$6.2  &       $-$5$\pm$11    & 163.5$\pm$14.2 & $-$240.7$\pm$15.7 & 58.8$\pm$13.2 \\
   4   &  0.725  & 0.433  & 0.253  & 0.402  & esdM6.5  & esdM7.0        & SDSS       & 76.3$\pm$4.5   &      $-$269$\pm$3    & 299.6$\pm$13.4 & $-$257.4$\pm$17.0 & $-$81.8$\pm$13.2 \\
   5   &  0.969  & 0.56   & 0.327  & 0.496  & usdM5.0  & esdM6.0        & GTC        & 173.7$\pm$49.5 &      $-$32$\pm$100   & ---    &   ---    &  ---     \\
   5   &  0.825  & 0.487  & 0.335  & 0.454  & esdM5.5  & esdM6.0        & SDSS       & 173.7$\pm$49.5 &      $-$138$\pm$10   & 235.5$\pm$50.7 & 12.4$\pm$17.5 & 88.3$\pm$67.3 \\
   6   &  0.769  & 0.521  & 0.366  & 0.524  & esdM4.5  & esdM5.5        & SDSS       & 79.3$\pm$8.6   &      $-$86$\pm$3     & 180.4$\pm$14.9 & $-$68.98$\pm$16.2 & $-$10.6$\pm$14.6 \\
   7   &  0.921  & 0.434  & 0.287  & 0.439  & usdM6.0  & usdM6.0        & NOT        & 122.0$\pm$7.3  &      $-$54$\pm$100   & ---    &   ---    &  ---     \\
   7   &  0.892  & 0.39   & 0.301  & 0.425  & usdM6.0  & usdM6.0        & SDSS       & 122.0$\pm$7.3  &      $-$160$\pm$8    & 124.0$\pm$12.4 & $-$182.3$\pm$15.1 & $-$163.9$\pm$10.6 \\
   8   &  0.848  & 0.706  & 0.386  & 0.606  & esdM4.0  & sdM4.5         & GTC        & 247.0$\pm$19.5 &      119$\pm$100     & ---    &   ---    &  ---     \\
   8   &  0.629  & 0.562  & 0.384  & 0.552  & sdM4.5   & sdM4.5         & SDSS       & 247.0$\pm$19.5 &      $-$19$\pm$21    & 12.0$\pm$18.0 & $-$481.0$\pm$43.2 & $-$93.0$\pm$16.9 \\
   9   &  0.849  & 0.606  & 0.346  & 0.5    & usdM5.0  & esdM6.5--7.0   & GTC        & 155.3$\pm$9.2  &      167$\pm$100     & ---    &   ---    &  ---     \\
   9   &  0.759  & 0.427  & 0.294  & 0.459  & esdM5.5  & esdM6.0--6.5   & SDSS       & $<$146.0$\pm$37.8 &   $-$13$\pm$5     & $-$33.8$\pm$20.7 & $-$106.0$\pm$39.4 & $-$74.8$\pm$22.7 \\
   10  &  0.318  & 0.616  & 0.222  & 0.451  & sdM6.5   & sdM6.5         & VLT        & 128.1$\pm$16.3 &      $-$76$\pm$100   & ---    &   ---    &  ---     \\
   11  &  1.023  & 0.524  & 0.368  & 0.54   & usdM5.0  & esdM5.0--5.5   & NOT        & 63.0$\pm$13.8  &     237$\pm$100      & ---    &   ---    &  ---     \\
   11  &  1.097  & 0.523  & 0.393  & 0.556  & usdM4.0  & esdM5.5        & SDSS       & 55.8$\pm$5.6   &     182$\pm$18       & $-$186.5$\pm$16.1 & $-$149.4$\pm$11.7 & $-$32.7$\pm$15.9 \\
   12  &  0.494  & 0.614  & 0.297  & 0.458  & sdM5.5   & esdM6.5        & SDSS       & 139.8$\pm$8.3  &       $-$100$\pm$12  & $-$32.81$\pm$11.6 & $-$173.87$\pm$12.6 & $-$163.85$\pm$11.5 \\
   13  &  0.512  & 0.846  & 0.421  & 0.721  & dM3.0    & dM4.0          & GTC        & --- $\pm$ ---  &       ---$\pm$---    & ---    &   ---    &  ---     \\
   14  &  0.593  & 0.557  & 0.29   & 0.49   & sdM5.5   & sdM7.0         & NOT        & 52.6$\pm$2.0   &     127$\pm$100      & ---    &   ---    &  ---     \\
   15  &  1.283  & 0.592  & 0.365  & 0.519  & usdM4.5  & esdM6.0        & GTC        & 149.9$\pm$40.2 &      289$\pm$100     &  ---   &   ---    &  ---     \\
   15  &  0.854  & 0.367  & 0.264  & 0.381  & usdM6.5  & esdM6.0        & SDSS       & 149.9$\pm$40.2 &      $-$63$\pm$2     &   8.64$\pm$26.0 & $-$324.3$\pm$100.0 & $-$89.5$\pm$22.7 \\
   16  &  0.795  & 0.261  & 0.244  & 0.346  & esdM7.0  & esdM8.0        & SDSS       & 160.4$\pm$19.8 &      $-$80$\pm$8     &  $-$124.7$\pm$32.0 & $-$149.5$\pm$21.5 & $-$239.9$\pm$24.9 \\
   17  &  0.527  & 0.234  & 0.172  & 0.223  & esdM8.5  & sdM8.0-8.5     & SDSS       & 156.4$\pm$21.6 &     $-$52$\pm$31     &  94.9$\pm$22.0 & $-$347.5$\pm$56.5 & $-$46.0$\pm$26.8 \\
   18  &  0.688  & 0.627  & 0.328  & 0.539  & esdM5.0  & esdM7.0--7.5   & SDSS       & 185.3$\pm$11.0 &   11$\pm$14          & $-$259.4$\pm$21.8 & $-$241.3$\pm$14.2 & $-$160.3$\pm$17.4 \\
   19  &  0.489  & 0.854  & 0.392  & 0.654  & sdM3.5   & dM4.5/sdM5.0   & VLT        & 162.4$\pm$20.3 &      $-$265$\pm$100  &  ---   &   ---    &  ---     \\
   20  &  0.778  & 0.522  & 0.382  & 0.534  & esdM4.5  & esdM5.5        & SDSS       & 205.9$\pm$33.7 &     $-$28$\pm$3      & 161.1$\pm$24.4 & $-$229.8$\pm$50.8 & $-$13.1$\pm$15.1 \\
   21  &  0.818  & 1.095  & 0.548  & 0.825  & esdM1.5  & sdM5.0--5.5    & GTC        & 190.4$\pm$27.0 &     $-$368$\pm$100   &  ---   &   ---    &  ---     \\
   22  &  0.377  & 0.536  & 0.303  & 0.473  & sdM5.5   & sdM6.0--6.5    & SDSS       & 90.1$\pm$19.7  &       $-$88$\pm$8    & $-$236.7$\pm$57.2 & $-$36.1$\pm$38.7 & $-$105.9$\pm$16.0 \\
   23  &  0.76   & 0.402  & 0.25   & 0.342  & esdM7.0  & esdM7.0        & SDSS       & 108.7$\pm$6.5  &     $-$263$\pm$4     & $-$19.6$\pm$9.7 & $-$273.4$\pm$10.9 & $-$154.7$\pm$13.6 \\
   24  &  0.464  & 0.667  & 0.394  & 0.639  & dM3.5    & sdM5.5         & GTC        & 198.3$\pm$18.0 &     $-$299$\pm$100   & ---    &   ---    &  ---     \\
   25  &  0.747  & 0.496  & 0.343  & 0.509  & esdM5.0  & esdM5.5        & NOT        & 112.6$\pm$13.2 &     84$\pm$100       & ---    &   ---    &  ---     \\
   26  &  0.681  & 0.666  & 0.45   & 0.658  & sdM3.0   & sdM4.0         & NOT        & 169.7$\pm$12.6 &     $-$398$\pm$100   & ---    &   ---    &  ---     \\
   27  &  0.561  & 0.853  & 0.492  & 0.745  & dM2.0    & dM4.0-5.0      & VLT        & --- $\pm$ ---  &       ---$\pm$---    & ---    &   ---    &  ---     \\
   28  &  0.597  & 0.641  & 0.361  & 0.585  & sdM4.0   & sdM5.0--5.5    & VLT        & 176.6$\pm$21.0 &      $-$101$\pm$100  & ---    &   ---    &  ---     \\
   29  &  0.474  & 0.763  & 0.311  & 0.593  & sdM4.5   & sdM6.5         & NOT        & 148.6$\pm$18.8 &      $-$448$\pm$100  & ---    &   ---    &  ---     \\
   29  &  0.474  & 0.763  & 0.31   & 0.593  & sdM4.5   & sdM6.0         & SDSS       & 129.2$\pm$29.7 &      $-$458$\pm$19   & $-$164.9$\pm$10.3 & $-$376.6$\pm$20.5 & $-$218.6$\pm$23.2 \\
   30  &  0.518  & 0.363  & 0.11   & 0.208  & esdM9.5  & sdM9.5         & VLT        & 228.5$\pm$28.1 &      $-$58$\pm$100   & ---    &   ---    &  ---     \\
   31  &  0.653  & 0.248  & 0.139  & 0.262  & esdM8.5  & sdM8.0         & SDSS       & 137.1$\pm$12.5 &      $-$303$\pm$36   & 303.2$\pm$25.0 & 19.6$\pm$17.1 & 178.6$\pm$26.6 \\
   32  &  0.098  & 0.041  & 0.15   & 0.39   & dM7.5    & sdL0.5         & GTC        & --- $\pm$ ---  &      $-$163$\pm$100  & ---    &   ---    &  ---     \\
   33  &  0.444  & 0.321  & 0.264  & 0.407  & sdM6.5   & sdM6.0         & SDSS       & 110.2$\pm$10.0 &      $-$66$\pm$17    & $-$11.5$\pm$13.5 &  1.5$\pm$12.1 & $-$148.4$\pm$13.0 \\
   34  &  0.451  & 0.416  & 0.166  & 0.281  & sdM8.0   & sdM8.0--8.5    & VLT        & 310.1$\pm$31.5 &      $-$84$\pm$100   & ---    &   ---    &  ---     \\
   35  &  0.706  & 0.631  & 0.351  & 0.525  & esdM4.5  & sdM6.0         & VLT        & 150.7$\pm$30.8 &        105$\pm$100   & ---    &   ---    &  ---     \\
   36  &  0.408  & 0.63   & 0.236  & 0.451  & sdM6.0   & sdM6.5--7.0    & VLT        & 261.4$\pm$28.6 &       34$\pm$100     & ---    &   ---    &  ---     \\
   38  &  1.089  & 0.685  & 0.388  & 0.5    & usdM4.5  & usdM6.5        & VLT        & 424.2$\pm$25.2 &      $-$130$\pm$100  & ---    &   ---    &  ---     \\
   39  &  0.929  & 0.514  & 0.404  & 0.558  & usdM4.0  & esdM5.5        & VLT        & 335.0$\pm$37.8 &    $-$139$\pm$100    & ---    &   ---    &  ---     \\
   40  &  0.963  & 0.592  & 0.417  & 0.55   & usdM4.0  & esdM5.5        & VLT        & 220.2$\pm$22.2 &    $-$225$\pm$100    & ---    &   ---    &  ---     \\
   41  &  0.668  & 0.617  & 0.375  & 0.57   & esdM4.0  & sdM6.0         & VLT        & 267.3$\pm$57.6 &      $-$253$\pm$5    & ---    &   ---    &  ---     \\
   42  &  0.789  & 0.499  & 0.3    & 0.413  & esdM6.0  & esdM6.0        & VLT        & 355.5$\pm$84.2 &      $-$85$\pm$100   & ---    &   ---    &  ---     \\
   43  &  0.899  & 0.337  & 0.298  & 0.47   & usdM5.5  & sdM8.0         & SDSS       & 248.3$\pm$23.6 &      44$\pm$21       & $-$125.0$\pm$21.3 & 62.9$\pm$11.3 & $-$146.8$\pm$21.7 \\
   44  &  0.88   & 0.471  & 0.24   & 0.374  & usdM7.0  & esdM7.0--7.5   & VLT        & 254.4$\pm$15.1 &      $-$285$\pm$100  & ---    &   ---    &  ---     \\
   45  &  0.399  & 0.627  & 0.249  & 0.458  & sdM6.0   & sdM6.5--7.0    & VLT        & 339.7$\pm$36.8 &    $-$354$\pm$100    & ---    &   ---    &  ---     \\
   46  &  0.51   & 0.769  & 0.207  & 0.438  & sdM6.5   & sdM8.5         & GTC        & 155.8$\pm$14.1 &       391$\pm$100    & ---    &   ---    &  ---     \\
   47  &  0.949  & 0.477  & 0.296  & 0.486  & usdM5.5  & usdM6.0--6.5   & VLT        & 362.8$\pm$21.6 &       $-$120$\pm$100 & ---    &   ---    &  ---     \\
   48  &  0.423  & 0.389  & 0.187  & 0.322  & sdM7.5   & sdM7.0         & VLT        & 338.5$\pm$13.8 &    $-$160$\pm$100    & ---    &   ---    &  ---     \\
   49  &  0.537  & 0.761  & 0.363  & 0.643  & sdM4.0   & sdM5.0--5.5    & VLT        & 230.7$\pm$16.6 &    $-$211$\pm$100    & ---    &   ---    &  ---     \\
   50  &  0.31   & 0.939  & 0.326  & 0.705  & dM3.5    & dM6.0          & VLT        & --- $\pm$ ---  &        ---$\pm$---   & ---    &   ---    &  ---     \\
   51  &  0.8    & 0.679  & 0.367  & 0.568  & esdM4.5  & esdM5.0        & VLT        & 426.5$\pm$96.1 &     $-$142$\pm$100   & ---    &   ---    &  ---     \\
   52  &  0.454  & 0.529  & 0.214  & 0.374  & sdM7.0   & sdM7.0         & VLT        & 168.7$\pm$5.4  &       50$\pm$100     & ---    &   ---    &  ---     \\
   53  &  0.596  & 0.542  & 0.294  & 0.476  & esdM5.5  & sdM6.0         & VLT        & 185.3$\pm$38.1 &     $-$197$\pm$100   & ---    &   ---    &  ---     \\
   54  &  0.587  & 0.617  & 0.283  & 0.513  & sdM5.5   & sdM7.0         & VLT        & 321.9$\pm$13.0 &     $-$24$\pm$100    & ---    &   ---    &  ---     \\
   55  &  1.197  & 0.511  & 0.366  & 0.577  & usdM4.0  & usdM4.5--5.0   & VLT        & 595.8$\pm$35.4 &       $-$68$\pm$100  & ---    &   ---    &  ---     \\
   56  &  0.724  & 0.598  & 0.326  & 0.517  & esdM5.0  & sdM6.0         & VLT        & 433.9$\pm$95.0 &      $-$215$\pm$100  & ---    &   ---    &  ---     \\
   57  &  0.992  & 0.646  & 0.433  & 0.575  & usdM3.5  & esdM5.5        & VLT        & 371.4$\pm$39.8 &     79$\pm$100       & ---    &   ---    &  ---     \\
   58  &  0.515  & 0.678  & 0.309  & 0.511  & sdM5.0   & sdM5.5--6.0    & VLT        & 349.6$\pm$77.8 &      $-$344$\pm$100  & ---    &   ---    &  ---     \\
   59  &  0.495  & 0.46   & 0.203  & 0.356  & sdM7.0   & sdM7.0         & VLT        & 272.0$\pm$8.8  &   $-$177$\pm$100     & ---    &   ---    &  ---     \\
   60  &  0.627  & 0.801  & 0.288  & 0.306  & esdM7.0  & sdM8.0         & VLT        & 379.1$\pm$43.4 &   $-$111$\pm$100     & ---    &   ---    &  ---     \\
   61  &  0.79   & 0.378  & 0.203  & 0.305  & esdM7.5  & esdM8.0        & VLT        & 276.0$\pm$16.4 &       $-$71$\pm$100  & ---    &   ---    &  ---     \\
   62  &  0.715  & 0.402  & 0.238  & 0.369  & esdM7.0  & esdM7.0--7.5   & VLT        & 175.9$\pm$10.5 &       $-$24$\pm$100  & ---    &   ---    &  ---     \\
   63  &  0.263  & 0.549  & 0.043  & 0.259  & sdM9.5   & sdL0.5         & VLT        & --- $\pm$ ---  &        97$\pm$100    & ---    &   ---    &  ---     \\
   64  &  0.376  & 0.546  & 0.218  & 0.397  & sdM7.0   & sdM7.0         & GTC        &  88.7$\pm$2.3  &      $-$83$\pm$100   & ---    &   ---    &  ---     \\
   65  &  0.411  & 0.692  & 0.248  & 0.405  & sdM6.5   & sdM7.0         & VLT        & 223.9$\pm$7.2  &     $-$346$\pm$100   & ---    &   ---    &  ---     \\
   66  &  0.572  & 0.78   & 0.386  & 0.628  & sdM3.5   & sdM5.5--6.0    & VLT        & 355.3$\pm$32.3 &       $-$439$\pm$100 & ---    &   ---    &  ---     \\
   67  &  0.942  & 0.518  & 0.351  & 0.474  & usdM5.0  & esdM6.0        & SDSS       & 198.5$\pm$42.3 &     $-$179$\pm$3     & $-$29.1$\pm$27.6 & $-$138.5$\pm$36.1 & $-$207.1$\pm$23.5 \\
   68  &  0.751  & 0.589  & 0.439  & 0.671  & esdM3.0  & sdM6.0/esdM5.5 & VLT        & 359.7$\pm$80.8 &     $-$684$\pm$100   & ---    &   ---    &  ---     \\
   69  &  0.394  & 0.512  & 0.246  & 0.458  & sdM6.0   & sdM7.0         & GTC        & 302.0$\pm$12.5 &      $-$255$\pm$100  & ---    &   ---    &  ---     \\
   70  &  0.404  & 0.850  & 0.366  & 0.677  & dM3.5    & dM5.0          & GTC        & --- $\pm$ ---  &       ---$\pm$---    & ---    &   ---    &  ---     \\
   71  &  0.838  & 0.403  & 0.265  & 0.377  & usdM6.5  & esdM6.5        & GTC        & 360.6$\pm$21.4 &      $-$50$\pm$100   & ---    &   ---    &  ---     \\
   72  &  0.578  & 0.975  & 0.305  & 0.443  & sdM5.5   & sdM6.0         & GTC        & 462.8$\pm$111.1 &     $-$237$\pm$100  & ---    &   ---    &  ---     \\
   77  &  0.64   & 0.495  & 0.326  & 0.516  & esdM5.0  & sdM7.5         & SDSS       & 201.6$\pm$13.3 &      ---$\pm$---     & ---    &   ---    &  ---     \\
   78  &  0.859  & 0.820  & 0.395  & 0.586  & esdM4.0  & esdM5.0        & GTC        & 487.7$\pm$114.0 &       126$\pm$100   & ---    &   ---    &  ---     \\
   79  &  0.493  & 0.288  & 0.152  & 0.258  & sdM8.5   & sdM6.0         & SDSS       & 180.5$\pm$36.9 &        88$\pm$17     & $-$255.8$\pm$55.2 & $-$66.9$\pm$26.9 &  6.5$\pm$29.5 \\
   82  &  0.695  & 0.15   & 0.136  & 0.54   & esdM6.5  & dM2.0          & SDSS       & --- $\pm$ ---  &       ---$\pm$---    & ---    &   ---    &  ---     \\
   83  &  1.005  & 0.338  & 0.265  & 0.504  & usdM5.5  & sdM6.0         & SDSS       & 253.7$\pm$53.0 &       40$\pm$18      & 202.9$\pm$73.8 & 218.2$\pm$44.9 & 140.4$\pm$27.9 \\
   85  &  0.699  & 0.495  & 0.299  & 0.509  & esdM5.0  & sdM6.0         & VLT        & 221.5$\pm$45.8 &      $-$128$\pm$100  & ---    &   ---    &  ---     \\
   85  &  0.753  & 0.434  & 0.317  & 0.476  & esdM5.5  & esdM5.5--6.0   & SDSS       & 198.3$\pm$19.0 &      $-$189$\pm$10   & $-$209.4$\pm$21.0 & 278.3$\pm$27.6 & $-$80.8$\pm$18.0 \\
   86  &  0.207  & 0.040  & 0.157  & 0.236  & sdM8.5   & sdL0.0         & GTC        & --- $\pm$ ---  &       ---$\pm$---    & ---    &   ---    &  ---     \\
   87  &  0.702  & 0.488  & 0.308  & 0.476  & esdM5.5  & esdM5.5        & SDSS       & 187.4$\pm$18.2 &      $-$115$\pm$5    & 194.8$\pm$24.3 & 101.4$\pm$16.3 & $-$116.8$\pm$14.9 \\
   88  &  0.546  & 0.569  & 0.210  & 0.501  & sdM6.0   & sdM6.0         & GTC        & 392.3$\pm$87.7 &      $-$8$\pm$100    & ---    &   ---    &  ---     \\
   89  &  0.485  & 0.557  & 0.224  & 0.415  & sdM6.5   & sdM7.0         & VLT        & 210.2$\pm$6.8  &      $-$142$\pm$100  & ---    &   ---    &  ---     \\
   90  &  0.318  & 0.603  & 0.251  & 0.461  & sdM6.0   & sdM6.5         & GTC        & 546.9$\pm$71.7 &      $-$90$\pm$100   & ---    &   ---    &  ---     \\
   91  &  0.604  & 0.413  & 0.305  & 0.500  & esdM5.5  & sdM6.0         & GTC        & 328.8$\pm$72.9 &      194$\pm$100     & ---    &   ---    &  ---     \\
   92  &  0.616  & 0.538  & 0.238  & 0.409  & esdM6.5  & esdM7.5        & GTC        & 429.4$\pm$25.5 &      $-$62$\pm$100   & ---    &   ---    &  ---     \\
   93  &  0.53   & 0.644  & 0.324  & 0.483  & sdM5.0   & sdM5.5         & SDSS       & 307.1$\pm$27.9 &      $-$183$\pm$13   & 161.4$\pm$22.4 & $-$269.7$\pm$27.3 & $-$140.4$\pm$15.4 \\
   94  &  0.359  & 0.140  & 0.308  & 0.652  & dM4.0    & M6.0           & GTC        & ---  $\pm$ --- &      $-$205$\pm$100  & ---    &   ---    &  ---     \\
   95  &  0.454  & 0.216  & 0.251  & 0.463  & sdM6.0   & sdM7.0         & GTC        & 309.0$\pm$11.6 &      $-$197$\pm$100  & ---    &   ---    &  ---     \\
   96  &  0.681  & 0.402  & 0.312  & 0.53   & esdM5.0  & esdM6.0        & SDSS       & 184.2$\pm$38.4 &      $-$232$\pm$6    & $-$281.3$\pm$57.2 & 215.5$\pm$69.5 & $-$195.2$\pm$16.3 \\
   97  &  0.465  & 0.718  & 0.305  & 0.535  & sdM5.0   & sdM5.5         & VLT        & 364.6$\pm$33.1 &      $-$174$\pm$100  & ---    &   ---    &  ---     \\
   98  &  0.244  & 0.092  & 0.206  & 0.488  & dM6.0    & sdM/dM6.5      & GTC        & 426.4$\pm$10.5 &      362$\pm$100     & ---    &   ---    &  ---     \\
   99  &  1.018  & 0.51   & 0.348  & 0.489  & usdM5.0  & usdM6.0        & VLT        & 364.7$\pm$21.7 &      $-$256$\pm$100  & ---    &   ---    &  ---     \\
   100 &  0.271  & 0.099  & 0.249  & 0.456  & dM6/sdM6 & sdM/dM6.5      & GTC        & 180.4$\pm$37.2 &      230$\pm$100     & ---    &   ---    &  ---     \\
\label{Table_indices_SpT}
\end{longtable}
Notes:\\
(1) Uncertainties on the distances take into account the error on the $J$-band magnitude of our
target and the error on the trigonometric distances of the subdwarf templates listed in
Table \ref{Table_known_distances}. We computed the minimum and maximum distances and
quote the largest error. \\
(2) For ID\,=19 we used the $J$-band absolute magnitude ($M_{J}$) of a M4.5 and sdM5.0, yielding distances
of 162.4$\pm$20.3 pc and 308.9$\pm$44.3 pc, respectively. \\
(3) For ID\,=\,68, we list the distances assuming a spectral type of sdM6.0\@. If we consider
the esdM5.5 classification, we find a distance of 322.0$\pm$36.0 pc.
For IDs\,=\,98 and 100, we list the distances for the metal-poor case. If we assume that
both objects are solar-metallicity M6.5 dwarfs, we find spectroscopic distances of
340.9$\pm$24.0 pc and 144.4$\pm$5.4 pc, respectively. \\
(4) For objects whose spectral types are quoted as intervals, we used the earliest spectral types
implying upper limits on the distances. extremes for the distance
estimates without including the uncertainty of half a subtype. \\
\end{longtab}

\normalsize

%
%

%
%
\subsection{Spectral classification according to indices}
\label{sdM_VO:spec_class_indices}
In columns 2--5 of Table \ref{Table_indices_SpT}, we list the values of the spectral indices calculated from 
the equations in \citet{lepine07c}. We list the associated spectral types from the 
scheme of \citet{lepine07c} in column 6 of Table \ref{Table_indices_SpT}. These spectral types 
have an uncertainty half a subtype because we approximated the values to the nearest half decimal
(e.g., a sdM6.76 was approximated to sdM7.0 and the final spectral type is sdM7.0$\pm$0.5).

%
%
   \begin{figure*}
   \resizebox{\hsize}{!}{\includegraphics[clip=true]{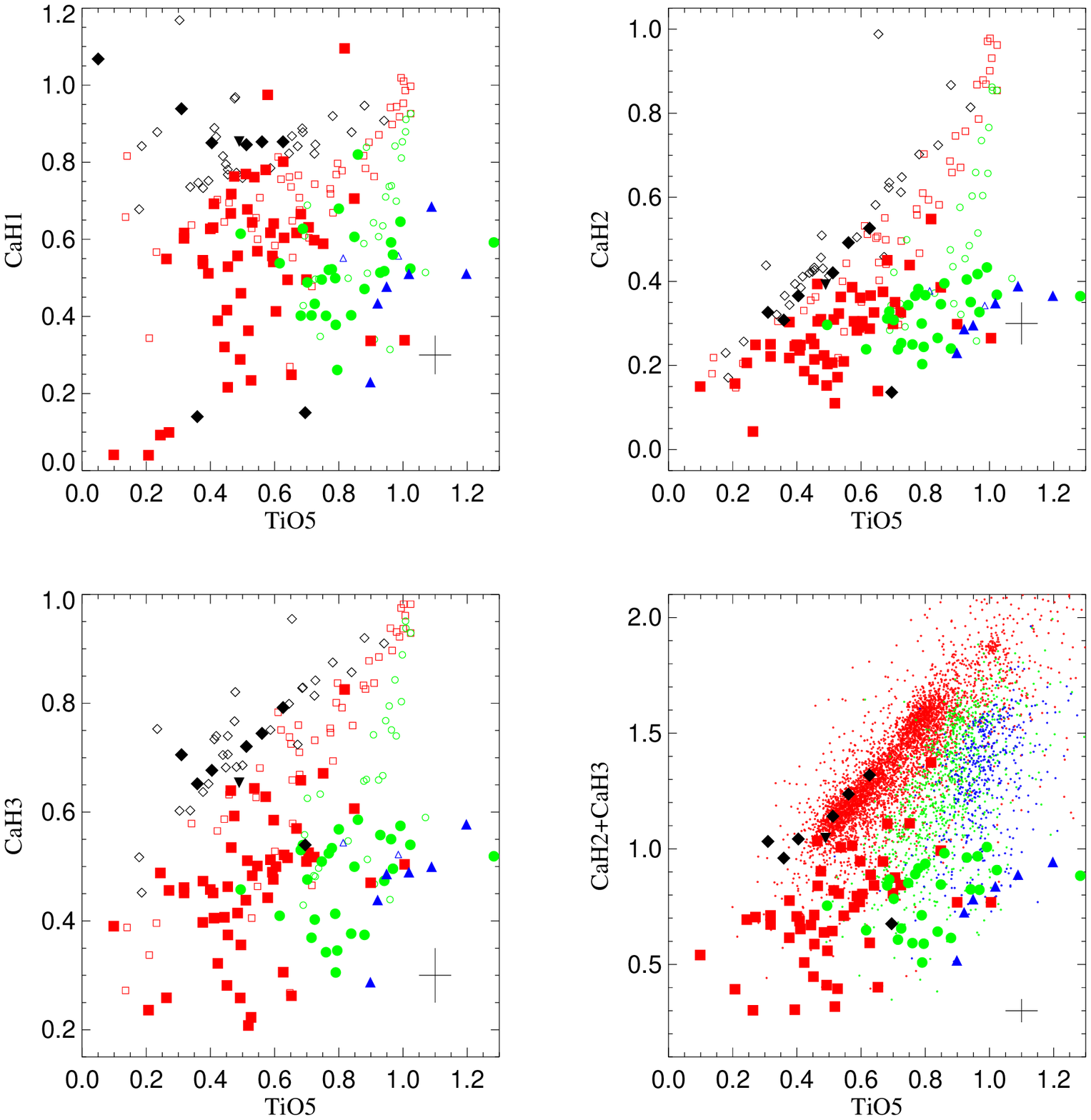}}
      \caption{Plots showing the distribution of subdwarfs as a function of their spectral indices. Symbology is the same as in Figure \ref{Fig_i-J_vs_J-K}. The bottom right plot shows the 
CaH2$+$CaH3 vs TiO5 diagram for objects classified as sdM, esdM, or usdM (small diamonds 
in red, green, and blue, respectively) in the SDSS spectroscopic database. Plots originally 
presented in \citet{gizis97a} and \citet{lepine07c}, updated with our discoveries.}
         \label{Fig_Indices_vs_TiO5}
   \end{figure*}

%
%
   \begin{figure*}
   \resizebox{\hsize}{!}{\includegraphics[width=\hsize]{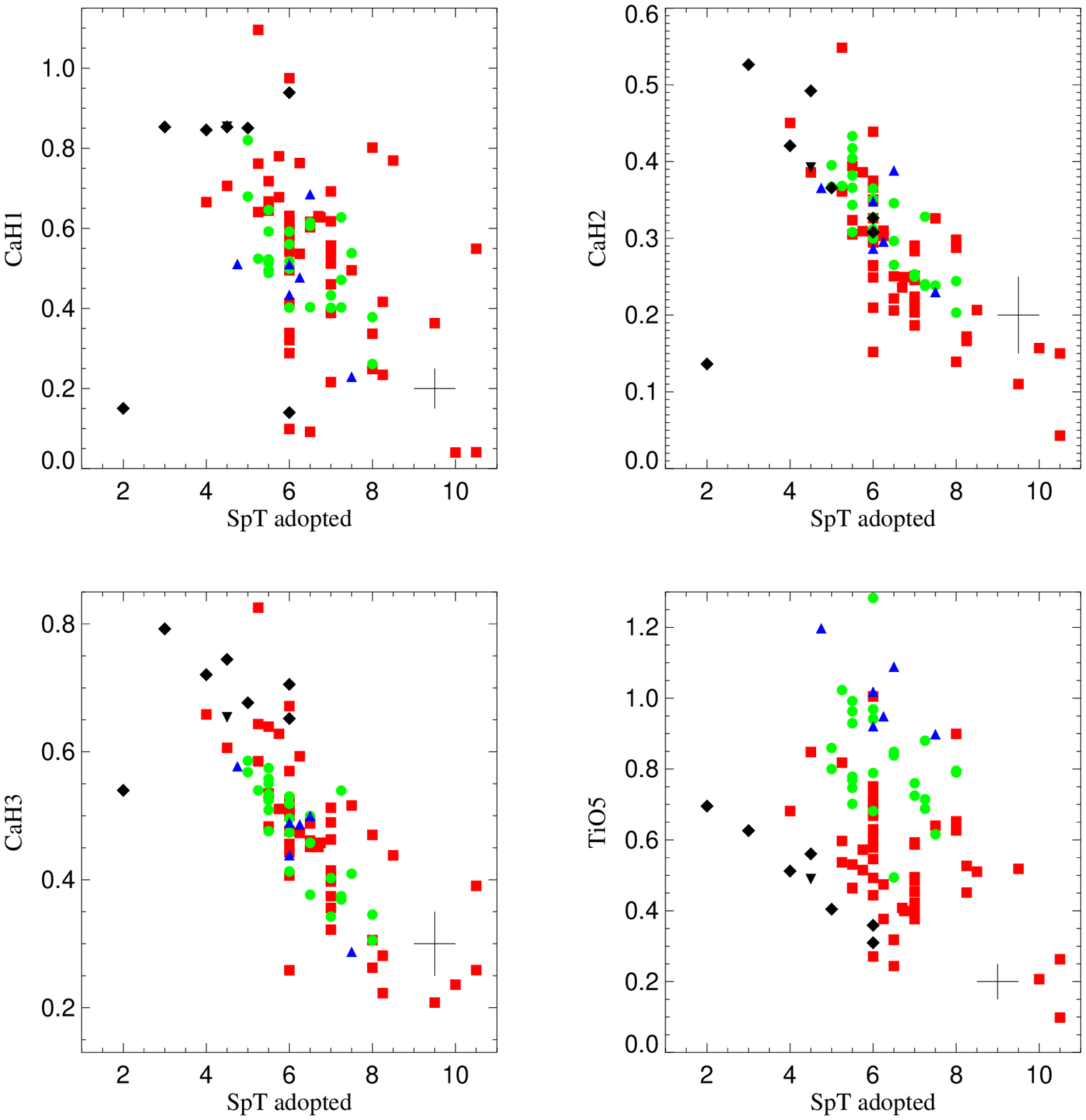}}
      \caption{Diagrams depicting the spectral indices (CaH1, CaH2, CaH3, and TiO5) vs the final spectral types for our subdwarfs, extreme subdwarfs, ultrasubdwarfs, and solar-metallicity dwarfs plotted as filled squares, circles, triangles, and diamonds respectively.}
         \label{Fig_Indices_vs_SpT}
   \end{figure*}

In Fig.\ \ref{Fig_Indices_vs_TiO5} we plot the spectral indices of \citet{gizis97a} to assign spectral types to 
our targets within the framework of the classification of \citet{lepine07c}: CaH1, CaH2, CaH3, and 
CaH2$+$CaH3 vs TiO5. In the CaH2$+$CaH3 vs TiO5 diagram, we also plot objects classified
as sdM, esdM, or usdM in the SDSS spectroscopic database (small points in red, green, and blue, respectively). 
We can distinguish three sequences in the lower right diagram in Fig.\ \ref{Fig_Indices_vs_TiO5} 
because most SDSS subdwarfs have spectral types earlier than M5\@. Our subdwarfs, on the other hand, 
are mainly late-M subdwarfs so they do not lie exactly on top of the three sequences. 
We see that some of our contaminants are located in the overlapping regions between dwarfs
and subdwarfs because the separation is not perfectly defined \citep{lepine07c}. 

We can appreciate the presence of four reasonably well-defined sequences corresponding to solar metallicity dwarfs, 
subdwarfs, extreme subdwarfs, and ultrasubdwarfs in all four panels of Fig.\ \ref{Fig_Indices_vs_TiO5}. 
According to the literature, this represents the distinct sub-solar abundances where the ultrasubdwarfs are 
the most metal-depleted stars. In our sample, the intensity of CaH seems to keep similar values for all 
subdwarf categories, which contrasts with the behaviour of TiO, which becomes less intense with decreasing metallicity. 
Actually, a TiO index of $\sim$\,1.0 implies that this oxide feature is barely seen at the resolution 
of our data. The lack of TiO absorption in high-gravity, late-type atmospheres is an excellent indicator of 
extreme sub-solar metallicity.

In Fig.\ \ref{Fig_Indices_vs_SpT} we show the spectral indices as a function of the adopted spectral type.
We argue that CaH1 is the worst indicator of spectral type and metallicity class. CaH2 and CaH3 are good indicators 
of spectral type but poor indicator of metallicity class. TiO5 is a good indicator of the metallicity class.

\subsection{Spectral classification according to visual comparison with spectral templates}
\label{sdM_VO:spec_class_templates}

To perform a classification with spectral templates, we used the templates made publicly available
by \citet{savcheva14}. Their sample include templates every subtype for sdM0--sdM9.5, esdM0--esdM8, 
and usdM0--usdM7.5 for subdwarfs, extreme subdwarfs, and ultrasubdwarfs, respectively.
In column 7 of Table \ref{Table_indices_SpT}, we list the final spectral types of our candidates with 
an uncertainty of half a subtype based on spectral templates.

We should emphasise that we performed our own spectral library before the publication of
\citet{savcheva14}. The results obtained with both libraries agree to within half a subtype or better.
We proceeded as follows: we downloaded the optical spectrum of the brightest object of each spectral type (from M0 to the latest M subtype available) for the three classes of subdwarfs and for the solar-metallicity M dwarfs from the SDSS spectroscopic database. Our spectral templates obtained from the SDSS spectroscopic database cover the following ranges: sdM0.0 to sdM8.5, esdM0.0 to esdM8.0, and usdM0.0 to usdM7.5. We smoothed some of the SDSS spectra
in particular the esdM8.0 and sdM7.5 templates. We also smoothed some of the SDSS spectra of our
candidates as well as the spectrum of our faintest candidates (ID\,=\,32). The classification of these spectra was done by SDSS under the \citet{lepine07c} scheme. 

In addition, we considered the sdL0.0 and sdL0.5 subdwarf from our previous paper 
\citep{lodieu12b} to extend the spectral library. We also downloaded spectra from the SpeX 
spectral libraries\footnote{pono.ucsd.edu/\,$\sim$\,adam/browndwarfs/spexprism/html/subdwarf.html}  
for a sdL4.0, a sdM9.5 \citep{burgasser04}, a sdM9.0 \citep{burgasser04b}, and a sdL3.5 
\citep{burgasser09a} which we used as template after smoothing  by a factor of three. These libraries 
contain roughly 1000 low-resolution, near-infrared spectra of low-temperature dwarf stars and brown dwarfs 
obtained with the SpeX spectrograph\footnote{irtfweb.ifa.hawaii.edu/\,$\sim$\,spex/} \citep{rayner03} 
mounted on the 3m NASA InfraRed Telescope Facility (IRTF) on Mauna Kea, Hawaii. They also cover part of 
the optical wavelengths, redwards of 0.8 $\mu$m, which we considered as part of our spectroscopic templates.  
We did this classification comparing visually the spectra of our candidates with all our templates of each 
class and spectral type: dM, sdM, sdL, esdM, and usdM from M0 to the latest available spectral type.

\subsection{Differences between the two classification systems}
\label{sdM_VO:spec_class_diff}
%

%
%
   \begin{figure}
   \centering
   \includegraphics[width=\hsize]{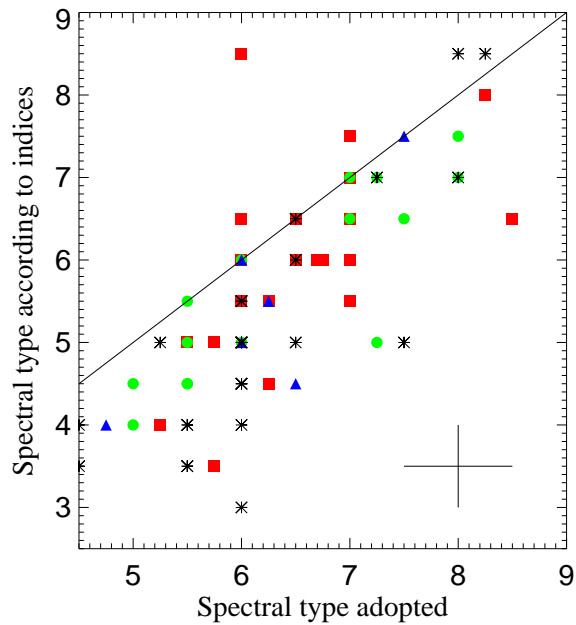}
      \caption{Differences between the adopted spectral types based on spectral templates and the spectral types derived from the indices defined by \citet{lepine07c}. We used the same symbology 
as in Figure \ref{Fig_i-J_vs_J-K}. We observe a trend where spectral templates lead to later spectral types compared to the classification based on spectral indices. We also highlighted confirmed 
metal-poor dwarfs with discrepant classes (not only spectral types) derived from both systems (asterisks).}
         \label{Fig_SpTindices_vs_SpTfinal}
   \end{figure}

As already pointed out in \citet{lodieu12b}, we find that the spectral types derived from
spectral indices tend to under-estimate the spectral type (overestimate the effective temperature) 
of the objects (Fig.\ \ref{Fig_SpTindices_vs_SpTfinal}). For this reason, we adopted the direct and 
visual comparison with SDSS templates from \citet{savcheva14} to assign spectral types to our candidates because it provides 
a more accurate and standardised classification that can be extended to cooler L-type subdwarfs. 
We note that the spectral indices are not so reliable 
to classify subdwarfs because they rely on a narrow wavelength range 
\citep[10 to 30\,\AA{} typically, see][]{gizis97a} and depend strongly on the spectral resolution, as discussed
in \citet{lepine07c}. Although we used both methods to classify our candidates, the final 
spectral types used in this work come from the direct comparison with spectral templates from \citet{savcheva14}
(column 7 in Table \ref{Table_indices_SpT}; Figs.\ \ref{Fig_sdM_sdL_spectra} and \ref{Fig_esdM_usdM_spectra}).

Comparing the spectral classification derived by the two methods (columns 6 and 7 of Table \ref{Table_indices_SpT}), we generally obtain the same metallicity class but a later spectral type using spectral templates. 
Nevertheless, some candidates turned out to have different classes in both systems. These discrepancies 
in the metallicity class occurs in \,$\sim$\,41\% of the cases (Table \ref{Table_indices_SpT}).
In Fig.\ \ref{Fig_SpTindices_vs_SpTfinal} we compare both schemes where the aforementioned trend 
can be visualised: the direct comparison with spectral templates give later spectral types for 
candidates with the same metallicity class. We also added confirmed late-type subdwarfs with different 
classes in the classification systems (asterisks in Fig.\ \ref{Fig_SpTindices_vs_SpTfinal}).

Finally, the spectral sequence shown in Figs.\ \ref{Fig_sdM_sdL_spectra} and \ref{Fig_esdM_usdM_spectra} 
appears more consistent than the sequence provided by the spectral indices. A clear example is the
classification of our new L subdwarfs, which would be classified as a M subdwarfs based on spectral indices only.

\section{Results of the search}
\label{sdM_VO:results}
\subsection{New late-type subdwarfs}
\label{sdM_VO:results_new_sdM}

We obtained our own optical spectra for 71 of our 100 candidates and 30 spectra from the SDSS 
database, including nine in common. 
Eight candidates part of the cross-matches between SDSS and UKIDSS remain without optical
spectra (ID\,=\,37, 73, 74, 75, 76, 80, 81, 84). Except for ID\,=\,37 identified in the SDSS DR7
vs UKIDSS LAS DR6 cross-match, they all come from SDSS DR9 and UKIDSS LAS DR10\@.
Twenty five of these 92 with optical spectra presented here were already reported in the literature 
with one or more spectral type estimates. Most of these have spectra in the SDSS spectroscopic 
database and we include them in this paper because they are part of our full sample
(Table \ref{Table_candidates}). Nevertheless, 22 out of 24 were published as solar-metallicity 
M dwarfs \citep{west08}, but we confirm spectroscopically their metal-poor nature in this work,
some of them already reported as metal-poor by other groups too.
We summarize below the publications and the associated spectral types (given in parenthesis):
\begin{itemize}
\renewcommand{\labelitemi}{$\bullet$}
\item One candidate from \citet{gizis97a}: ID\,=\,25 (esdM5.0). The spectral type agrees with ours: we 
classify this object as an esdM5.5$\pm$0.5
\item Five candidates from \citet{lepine08b}: ID\,=\,16 (esdM7.5), 31 (sdM7.5), 17 (sdM8.0), 79 (sdM8.5), 
and 23 (esdM7.0). We classify these objects with SDSS templates (Table \ref{Table_indices_SpT}).
Our spectral types agree with the aforementioned classification (within the uncertainty of 0.5),
except for ID\,=\,79 which we classify as a sdM6.0
\item Two candidates from our previous work \citep{lodieu12b}: ID\,=\,6 (usdM5.5) and 31 (sdM8.0). 
After a revision of our templates, the spectral type of ID\,=\,6 was modified, having now a spectral 
type adopted of esdM5.5 (Table \ref{Table_indices_SpT}). We did a revision of the spectra used as 
templates and we noticed differences in some cases (different metallicity but equal spectral type).
The differences in the spectra were not so clear due to the signal-to-noise, so, when necessary, we 
defined new known subdwarfs with better quality spectra as templates. The spectra that make up 
the new sample of templates (used to do the visual classification) show clearer differences between 
the same spectral type for the different metallicity classes, and also clearer differences every 
0.5 spectral type for the same metallicity. The new SDSS spectra made public by \citet{savcheva14}
improve our classification too
\item 22 candidates in \citet{west08}: ID\,=\,4 (M3), 5 (M4), 6 (M3), 7 (M3), 8 (M3), 9 (M4), 11 (M2), 12 (M4), 15 (M4), 17 (M4), 18 (M4), 23 (M5), 29 (M4), 31 (M5), 33 (M5), 43 (M3), 53 (M5), 67 (M4), 77 (M4), 79 (M5), 83 (M3), 85 (M4). The spectral classification was done with the HAMMER stellar spectral-typing facility \citep{covey07}, which does not include low-metallicity M dwarfs. In this work 
we confirm their low-metallicity (Table \ref{Table_indices_SpT})
\item Five candidates in \citet{zhang13}: ID\,=\,7 (usdM6), 33 (sdM6), 43 (esdM6), 67 (esdM6), and 85 (sdM6). 
These spectral types agree with our final classification for these objects, except for ID\,=\,43
and ID\,=\,85 which we classify as a sdM8.0 and esdM5.5--6.0, respectively
\end{itemize}

Regarding the SDSS DR7 vs 2MASS cross-correlation: we confirmed 26 out of 29 candidates as subdwarfs, 
extreme subdwarfs, ultra subdwarfs, or dwarfs/subdwarfs, yielding a success rate of 
90\%. We consider in this percentage all confirmed M and L subdwarfs of any spectral type as well 
as one object (ID\,=\,19) with intermediate metallicity class (dM/sdM). We also include the two objects
which we classify as early-type subdwarf: ID\,=\,8 (sdM4.5$\pm$0.5) and ID\,=\,26 (sdM4.0$\pm$0.5). 
The 24 candidates confirmed as late-type subdwarfs ($\geq$M5) in this cross-match are divided up into 
9 sdM, 1 dM/sdM, 12 esdM, and 2 usdM.
Three candidates of this cross-match turn out to be solar-metallicity M dwarfs, hence contaminants. 
Since we are looking for subdwarfs with spectral types equal or later than M5, we could consider the 
object with ID\,=\,26 (sdM4.0$\pm$0.5) as an outlier in our sample although not really a contaminant 
because it turns out to be metal-poor). If we consider all the subdwarfs of the three classes with 
spectral type $\geq$M5 (including the subdwarf with ID\,=\,8 classified as sdM4.5$\pm$0.5), and the 
one dM/sdM object (since they show features of a sdM6.5), then we have 25 confirmed late-type subdwarfs, 
yielding a success rate of \,$\sim$\,86\%. If we want to be more strict, and not consider the dM/sdM object,
class dM/sdM, then we have 24 objects and a success rate of \,$\sim$\,83\%.

From the SDSS DR9 vs UKIDSS LAS DR10 cross-correlation: we confirmed 59 out of 63 candidates 
as late-type subdwarfs, yielding a success rate of \,$\sim$\,94\%. The 59 candidates confirmed as 
late-type subdwarfs include 34 sdM, 3 sdL, 13 esdM, 4 usdM, one sdM/esdM (ID\,=\,68), and one 
(ID\,=\,85) classified as a subdwarf and extreme subdwarf from its VLT and SDSS spectra, respectively. 
The four remaining candidates have solar metallicity: ID\,=\,50 (dM6), ID\,=\,70 (dM5), ID\,=\,82 (dM2),
and ID\,=\,94 (dM6).

The success rates in the previous paragraphs come from the initial sample before refining the 
VO proper motions. If we consider the revised proper motions for the candidates with optical spectra
from the SDSS vs UKIDSS LAS cross-match, three M dwarfs (ID\,=\,50, 82, 94) would not be discarded 
but one (ID\,=\,70) would be according to the H$r$ values calculated with the refined proper motions.   

To summarize, we identified a total of 49 sdM, 3 sdL, 25 esdM, 6 usdM, 1 sdM/esdM, 2 dM/sdM, and 7 dM 
from the 92 candidates with optical spectra. We have confirmed spectroscopically 84 late-type subdwarfs 
(24 from SDSS and 2MASS, 60 from SDSS and UKIDSS LAS), two objects with uncertain class 
(dM/sdM), and one subdwarf with spectral type sdM4.0$\pm$0.5\@. Considering 84 confirmed subdwarfs
out of 92 candidates with optical spectra, we infer a success rate of \,$\sim$\,91\%, which would increase 
to \,$\sim$\,93.5\% if we consider the two dM/sdM objects. 
Of the total of 84 spectroscopically-confirmed metal-depleted dwarfs of all spectral types in this paper 
(not considering dM/sdM), the three subdwarf categories are populated as follows: 62--63$\pm$9\% (50--53 sdM/sdL), 
30--31$\pm$6\% (25--26 esdM), and 7$\pm$3\% (6 usdM), indicating that ultrasubdwarfs are about ten
times less frequent than subdwarfs in the solar neighbourhood.

Among our confirmed late-type subdwarfs, we report three L subdwarfs: one sdL0.0 and two sdL0.5\@.
Their spectral types were defined considering as templates the two previously L subdwarfs
presented in \citet{lodieu12b}. These five subdwarfs should be added to the growing number of 
L subdwarfs reported in the recent years 
\citep{burgasser03b,burgasser04,sivarani09,cushing09,lodieu10a,schmidt10a,bowler10a}.

We developed a very efficient photometric and proper motion method to look for late-type 
subdwarfs. We doubled the number of known late-type subdwarfs and we added five new L subdwarfs.

\subsection{Contamination}
\label{sdM_VO:results_contamination}
We have carefully measured the proper motions of the SDSS-UKIDSS cross match and we do 
not find significant differences with those derived using the positions in catalogues (VO). Therefore, 
we conclude that the VO proper motions (and reduced proper motions) are reliable and can be confidently 
use for the SDSS-2MASS cross-match. There are some exceptions though, detailed below for objects 
that turned out to be contaminants:
\begin{itemize}
\renewcommand{\labelitemi}{$\bullet$}
\item ID\,=\,70: this is a subdwarf candidate from the SDSS DR9 vs UKIDSS LAS DR10 cross-match 
that was predicted as a solar-metallicity M dwarf (hence, a contaminant) after revision of its 
proper motion. The origin of the incorrect proper motion calculated by the VO was due to a shift 
in the SDSS DR9 position, overestimating its proper motion. We confirm this fact after checking 
the SDSS image and loading the coordinates. We confirm its solar-metallicity nature
with a spectral type of dM5.0$\pm$0.5 from a GTC spectrum
\item ID\,=\,82: another candidate from SDSS DR9 and UKIDSS LAS DR10\@. This source passed the 
reduced proper motion criterion even after the astrometric revision of the proper motion. Yet, we classified its 
SDSS spectrum as dM2.0$\pm$0.5 although the fit is not as good as other dM\@. We cannot
explain why this candidate is a contaminant. However, we can discard the origin as being
an erroneous proper motion because we see a clear displacement blinking the SDSS and UKIDSS
images. The proper motion from the VO is similar to the refined proper motion, as
are the reduced proper motion values 
\item ID\,=\,50: a candidate from the SDSS DR7 vs UKIDSS LAS DR9 cross-match. This source passed the 
reduced proper motion criterion even after the astrometric revision of the proper motion. Its total proper motion 
is low though. Nonetheless, its spectrum indicates that it is a solar-metallicity M6 dwarf, hence, a contaminant
\item ID\,=\,2: a subdwarf candidate from the 2MASS vs SDSS DR7 cross-match without refined 
proper motion. We cannot confirm its proper motion as for the candidates from SDSS and UKIDSS 
because the error bars in the 2MASS coordinates are quite large given its faint magnitude.
This is compatible with this object being a clear contaminant, which we classify as a dM3.0$\pm$0.5
\item ID\,=\,13: another subdwarf candidate from the 2MASS and SDSS without refined proper motion.
We can explain why this object turned out to be a contaminant though: we can see another object 
very close to our candidate in the SDSS image, not present neither in the 2MASS catalogue nor in
the image. Therefore, we think that the VO used the coordinates of the neighbour object,
overestimating of proper motion of this candidate. We confirmed spectroscopically its 
solar-metallicity and classify it as a dM4.0$\pm$0.5
\item ID\,=\,27: another candidate from 2MASS and SDSS without refined proper motion. We see
that the SDSS catalogue position differs from the centroid on the SDSS image, yielding an erroneous
proper motion calculated by the VO\@. We confirmed spectroscopically its solar-metallicity nature
and classify it as a dM4.5$\pm$0.5
\end{itemize}
%

%
%
\subsection{Spectroscopic distances to our new subdwarfs}
\label{sdM_VO:results_new_dist}
We estimated spectrophotometric distances for our confirmed subdwarfs (column 9 of 
Table \ref{Table_indices_SpT}). We find a large range in distances, ranging from 50--60 pc for the closest 
subdwarfs and 500--600 pc for the furthest ones. We did not consider the effect of binarity on the 
spectrophotometric distances.

%
%
\begin{table}[!h]
\caption{Subdwarfs in the literature with known distances.}
\label{Table_known_distances}
\begin{tabular}{@{\hspace{0mm}}c@{\hspace{1mm}}c@{\hspace{2mm}}c@{\hspace{2mm}}c@{\hspace{2mm}}c@{\hspace{0mm}}}
\hline\hline
SpType & Dist &  $J$      & Name & Refs \\
              & pc           &  mag  &      & \\
\hline
sdM4.0  & 56.6$\pm$2.6   & 13.052  &   LP\,869-24         &  12 \\
sdM4.5  & 20.0$\pm$0.5   & 10.967  &   LP\,141-1          &  2, 3 \\
sdM5.0  & 30.0$\pm$1.8   & 12.820  &   LP\,803-27         &  12 \\
sdM6.0  & 85.7$\pm$17.1  & 14.684  &   LHS\,1074          &  8,7 \\
sdM6.5  & 81.0$\pm$8.0   & 14.259  &   LHS\,1166          &  12 \\
sdM7.0  & 35.2$\pm$0.8   & 13.194  &   LP\,440-52         &  12 \\
sdM7.5  & 46.5$\pm$2.8   & 13.611  &   LSR\,J2036$+$5059    &  4, 9 \\
sdM8.0  & 82.7$\pm$7.2   & 14.775  &   LSR\,J1425$+$7102    &  6, 1, 9 \\
sdM9.5  & 49.8$\pm$4.8   & 14.621  &   SSSPM\,J1013$-$1356  &  10, 9 \\
esdM5.0 & 53.5$\pm$11.0  & 13.639  &   LHS\,515           &  12 \\
esdM5.5 & 75.2$\pm$6.7   & 14.641  &   LP\,417-42         &  2, 12   \\
esdM6.0 & 73.9$\pm$14.8  & 14.907  &   LHS\,2023          &  7 \\
esdM6.5 & 106$\pm$---    & 15.717  &   LSR\,J0822$+$1700    &  5 \\
esdM7.0 &  70$\pm$---    & 14.887  &   APMPM\,J0559$-$2903  &  11 \\
\hline
\end{tabular} \\
The numbers in the references correspond to the following papers:
1\,=\,\citet{burgasser08a}; 2\,=\,\citet{gizis97a}; 3\,=\,\citet{gliese95}; 4\,=\,\citet{lepine02}; 
5\,=\,\citet{lepine03c}; 6\,=\,\citet{lepine03d}; 7\,=\,\citet{riaz08a}; 8\,=\,\citet{salim03a}; 
9\,=\,\citet{schilbach09}; 10\,=\,\citet{scholz04c}; 11\,=\,\citet{schweitzer99}; 12\,=\,\citet{vanAltena95}.
\end{table}

We estimated spectrophotometric distances for part of our sample using the $J$-band absolute 
magnitudes of subdwarfs with known trigonometric distances. In Table \ref{Table_known_distances}
we list the spectral types, distances, $J$-band magnitudes, and names of the subdwarf
templates employed to determine spectrophotometric distances of our sample.
Trigonometric distances are missing for ultrasubdwarfs as well as for the following
spectral types: sdM5.5, esdM7.5, esdM8.0, sdM8.5, sdL0.5\@.
We used the 2MASS and UKIDSS photometry for ID\,=\,1--29 and ID\,=\,30--100, respectively. We did not 
convert the UKIDSS photometry in the 2MASS system because the correction for late-M dwarfs is of the 
order of 0.02--0.06 mag, much lower than the uncertainty due to spectral typing (half a subtype).
To calculate the errors in distances, we propagated the errors on the distance of the subdwarf templates 
and the error on the $J$-band photometry but not on spectral type determination (typically half a subtype). 

For our confirmed subdwarfs whose spectral types are not covered above, we inferred their
spectrophotometric distances using the polynomial fits of the $J$-band in Table 2 of \citet{zhang13}.
These polynomial are only valid up to spectral types as late as M9.5 so we do not quote distances 
for our L subdwarfs although we can guess that the two brightest are most likely within 100 pc.
We distinguished between subdwarfs and extreme subdwarfs. For the ultrasubdwarfs, we used the
fits of the extreme subdwarfs, keeping in mind that their spectrophotometric distances will be
upper limits if there is no inversion in the magnitude vs spectral type relation in $J$.
The errors quoted in Table \ref{Table_indices_SpT} are lower limit because they 
only take into account the uncertainty of half a subclass on our optical spectral classification.

%
%
\begin{table*}
\caption{Potential companions to our subdwarfs. The columns are: ID of the subdwarf with potential companion, proper motion in RA for the subdwarf $\pm$30\%, proper motion in Dec for the subdwarfs $\pm$30\%, catalogue where the potential companion was found, name of the potential companion, separation between the subdwarf and the potential companion, proper motion in R.A. for the potential companion, and proper motion in Dec. for the potential companion.}
\centering
\begin{tabular}{cccccccc}
\hline\hline
ID & $\mu_{\alpha}cos{\delta}\pm$30\% & $\mu_{\delta}\pm$30\% & Catalogue & Name & Separation & $\mu_{\alpha}cos{\delta}$ & $\mu_{\delta}$\\ 
   & [mas/yr]         & [mas/yr]        &                   &                & [arcmin]   & [mas/yr] & [mas/yr]\\
\hline
24 &  -11.0$\pm$3.3   &   -5.0$\pm$1.5  & Hipparcos-Tycho   & TYC 3036 301 1 &  8.0       &    -8.4  &     6.5 \\    
24 &  -11.0$\pm$3.3   &   -5.0$\pm$1.5  & Hipparcos-Tycho   & TYC 3036 276 1 & 16.8       &   -17.2  &    13.1 \\   
55 &  -89.0$\pm$26.7  &    0.2$\pm$0.1  & Hipparcos-Tycho   & TYC 268 395 1  & 27.8       &   -95.9  &   -17.8 \\  
\hline
\end{tabular}
\label{Table_potential_companions}
\end{table*}

%
%
\subsection{Search for wide companions to our subdwarfs}
\label{sdM_VO:results_new_companions}

It is likely that there could be some true binaries in our sample of subdwarfs because the frequency 
of wide binaries is about 15\% amongst metal poor dwarfs \citep*{zapatero04a}.
We searched for wide companions to our confirmed subdwarfs in the Hipparcos-Tycho and Gliese databases,
as well as in the catalogue of \citet{laird88b}. The original search was done within a
radius of 10 arcmin, which we later increased to 30 arcmin. We looked for bright stars of earlier
spectral types with similar proper motions (using a maximum difference of 30\%), distances,
and metallicities to identify potential benchmark subdwarfs in our sample.

We did not find any bright star satisfying all criteria neither in the Gliese catalogue nor in
\citet{laird88b}. In the Hipparcos catalogue, we identified a potential companion
within 10 arcmin of our sdM5.5 subdwarf with ID\,=\,24 most likely due to its low revised proper
motion. The potential companion, TYC 3036-301-1, has $V$\,=\,9.35 and is classified as
a F5\@. No parallax distance exist for that source but we derive a spectroscopic distance of
140--150 pc, assuming Hipparcos distances for nearby F5V stars \citep{vanLeeuwen07}.
This distance is consistent with the spectroscopic distance of ID\,=\,24 estimated to 143.7$\pm$8.5 pc
(Table \ref{Table_indices_SpT}). However, no metallicity is available in the literature for
TYC 3036-301-1 so its remains as a potential companion to our subdwarf.

We identified two other potential companions to ID\,=\,25 and 55 within 30 arcmin based on
proper motion only: TYC 3036-276-1 and TYC 268-395-1 (Table \ref{Table_potential_companions}).
However, no information is available in Simbad on their spectral types, spectroscopic distances,
and metallicities. Thus, we cannot draw any conclusion on companionship at this stage.

%
%
%
\subsection{Radial velocities}
\label{sdM_VO:results_new_RV}

In this section we estimate heliocentric radial velocities for our sample of subdwarfs with SDSS
optical spectra using some one target as reference. Due to the lack of cool subdwarfs with well-known 
radial velocities, we proceeded as follows. 

We downloaded from the SDSS spectroscopic database a M6 dwarf (SDSS\,J08373760$+$3809585) with high 
signal-to-noise around the H$\alpha$ line ($>$50) with a known radial velocity (1.1 km/s) from \citet{west08}. 
We picked up a few other M6 dwarfs in this sample to check that our measurements give similar values to 
the ones reported by \citet{west08} within the uncertainties. We compared one of our subdwarf with the 
highest quality SDSS spectrum (ID\,=\,29) to that M6 dwarf and considered it as our RV reference for all 
other subdwarfs in our sample with SDSS spectra. We inferred a radial velocity of $-$458$\pm$19 km/s for 
ID\,=\,29 from its SDSS spectrum (Table \ref{Table_indices_SpT}).

Because the optical spectra of dwarfs and extreme/ultra subdwarfs are very different, we
considered ID\,=\,29 as our template to derive (relative) radial velocity for all sources with
SDSs spectra.
We compute radial velocities via the Fourier cross-correlation using the IRAF task {\tt{fxcor}}.
We used wide regions of the optical spectra (typically 6000--8800\,\AA{}) and
the best gaussian fits to infer the final Doppler shifts (tenth column in Table \ref{Table_indices_SpT}).
We list the error bars from the sole gaussian fits (Table \ref{Table_indices_SpT}). 
We should add in quadrature to these uncertainties the error on the radial velocity of the template
and the SDSS wavelength accuracy of 5 km/s. Nontheless, due to the spectral dispersion of the SDSS
spectra, we estimate a lower limit of 15 km/s on the RV errors.
We do not derive RVs for our subdwarfs with GTC, NOT, and VLT spectra whose very low spectral
resolutions (4--6 times worse than SDSS spectra) would most likely translate into error bars of 
the order of 100 km/s.

%
%
\subsection{Space motions}
\label{sdM_VO:results_new_Space_Motion}

We estimated the Galactic space velocities of a sub-sample of our subdwarfs considering their coordinates, 
proper motions, estimated radial velocities, and spectroscopic distances (Table \ref{Table_indices_SpT})
using the equations of \citet{johnson87}.
The U, V, and W components are defined as positive toward the Galactic anti-center, positive in the 
direction of Galactic rotation, and positive toward the North Galactic Pole, respectively.
We corrected the values for the Local Standard of Rest, where the solar motion is assumed to
be (8.50, 13.38, 6.49) km/s \citep{coskunoglu11}. 

We only considered targets with SDSS spectra because of the higher spectral resolution and lower
error bars. We fixed the errors on the SDSS vs 2MASS proper motions to 10 mas/yr and the errors on
the RVs to 15 km/s for sources with error bars lower than these values. We considered the largest
error bars on the distances to be conservative. 

Most of the targets exhibit large space motions in nearly all three $UVW$ components, confirming their 
membership to the thick disk and/or halo. Using the orientative flags defined by \citet{eggen90a}
and and \citet{leggett92}, two of our SDSS subdwarfs are classified as young-old-disk 
(YOD; Table \ref{Table_indices_SpT}) although the error bars may locate them within old-disk (OD) 
category. From the $UVW$ determinations shown in Table \ref{Table_indices_SpT}, we derive the following 
average space velocities of $<U>$\,=\,2, $<V>$\,=\,$-$124, $<W>$\,=\,$-$78 km\,s$^{-1}$ with 
velocity dispersions of $\sigma_U$\,=180, $\sigma_V$\,=\,188, and $\sigma_W$\,=\,109 km\,s$^{-1}$.
The mean space velocities of our sample are typical of 
halo stars, where the $U$ component is around null velocity while the velocity in the direction of the 
Galactic rotation is quite negative, consistent with the numbers reported in Table 7 of \citet{savcheva14}.
We note that our average value of the $W$ component is lower than the one in \citet{savcheva14} but
the dispersions are quite large and our sample is too small to draw any statistically-significant conclusion.
The Galactic velocity dispersions are significantly greater (by factors of 5--8) 
than those corresponding to solar metallicity stars up to a kiloparsec distance from the Sun 
\citep[e.g.][and references therein]{fuchs09a}. These large dispersions, particularly that of the $W$
component, give credit to the low metallicity nature of our sample as a whole.

Finally, two subdwarfs in our sample with SDSS spectra (ID\,=\,8 and 29) appear to have very high 
Galactic motions with velocities slightly above $\sim$450 km\,$s^{-1}$, which is the accepted escape 
velocity threshold in the halo \citep{kenyon08a,favia15a}. Whether these stars are runaway candidates 
is to be confirmed. We caution that a reliable trigonometric parallax and accurate radial velocities 
need to be measured before concluding that these stars are potentially unbound from the Galaxy gravity.
The Gaia mission should provide accurate proper motions and distances for both of them.

%
%
\subsection{Surface density of subdwarfs}
\label{sdM_VO:results_new_density}

Here we estimate the surface density (i.e.\ numbers of objects per square degree) of low-metallicity dwarfs 
with spectral types equal or later than M5 found in the common area of each cross-correlation.

For the SDSS DR7 vs 2MASS cross-match, we found a total common area of 8826 square degrees. 
We confirmed 24 out of 29 objects as late-type subdwarfs: 10 subdwarfs (sdM5$-$sdM8.5; including
the dM/sdM source ID\,=\,19), 12 extreme subdwarfs (esdM5$-$esdM8), and two usdM (usdM6$-$usdM7.5). 
Each metallicity subclass represents 42$\pm$13\%, 50$\pm$14\%, and 8$\pm$6\% of our total 
sample, respectively. These late-type subdwarfs have SDSS$r$\,=\,17.073$-$20.535 mag, 
$J$\,=\,13.995$-$16.895 mag, spectral types between M5 and M8.5, and proper motions in the
range 0.143$-$1.872 arcsec/yr. We derive surface densities of 0.0011$\pm$0.0004, 0.0014$\pm$0.0004, 
and 0.0002$\pm$0.0002 per square degree for subdwarfs, extreme subdwarfs, and ultrasubdwarfs respectively.

For the SDSS DR9 and UKIDSS DR10 cross-match, we confirmed 59 out of 63 candidates as 
late-type subdwarfs including three L-type dwarfs in 3679 square degrees.
Our sample is divided into 43 subdwarfs (M5$-$L0.5), 13 extreme subdwarfs (M5$-$M8), and 
4 ultrasubdwarfs (M5$-$M6.5). Each metallicity class represents 72$\pm$11\%, 22$\pm$6\%, 
and 7$\pm$3\% of our sample. These confirmed subdwarfs have SDSS$r$\,=\,19.6$-$23.3 mag,
$J$\,=\,15.9$-$18.8 mag, spectral types between M5 and L0.5, and proper motions of 0.09$-$0.66 arcsec/yr. 
We derive a surface density of late-type subdwarfs of 0.016$\pm$0.002 per square degrees,
divided up into densities of 0.012$\pm$0.002, 0.004$\pm$0.001, and 0.001$\pm$0.001 per square
degree for subdwarfs, extreme subdwarfs, and ultrasubdwarfs, respectively.

Following our previous analysis \citep{lodieu12b}, we consider the photometric sample from \citet{bochanski10} 
for field M5 dwarfs: $i$\,$<$\,22 mag, $r-z$\,$\geq$\,2.5 mag, $i-z$\,$\geq$\,0.2 mag, and $r-i$\,$\geq$\,0.3 mag.
These criteria give us a total of 653,625 photometric M5 dwarfs and later in 8000 square degrees surveyed 
by SDSS DR5, yielding a density of \,$\sim$\,82 late-M dwarfs per square degree. 
This density is \,$\sim$\,5100 times higher than the value obtained for late-type subdwarfs.
We observe that the density of late-type dwarfs drops with decreasing metallicity.
Due to the incompleteness of our survey particularly at the faintest magnitudes and coolest types, 
the aforementioned surface density are lower limits to the true density of subdwarfs. 

In addition, we are aware of potential losses of candidates when the selection involves proper
motion and photometric criteria. It is extremely hard to provide an estimate
on the losses but we know that percentages can be of the order of 30\% in the case of clusters
\citep{barrado02a,moraux03}. To give a more reliable estimate of completeness, we compiled a list 
of 114 known $\geq$\,M5 subdwarfs and attempted to recover them in the latest SDSS and
UKIDSS data releases using our criteria detailed in Section \ref{sdM_VO:sample_select}.
We found that 45 of these 114 late-type subdwarfs lie in the common area between SDSS 
and UKIDSS (45 are in UKIDSS and 95 in SDSS). We recovered 18 of these 45 
known subdwarfs in the SDSS DR9 vs UKIDSS LAS DR10 area using the exact same criteria (see 
Section \ref{sdM_VO:sample_select}), suggesting that our sample is only 40$\pm$9\% complete.
We repeated the same exercise for the 29 candidates in the SDSS DR7 vs 2MASS cross-correlation, 
and found 6 lying in the common area. We recovered one of them, suggesting that our sample 
is complete at the $\sim$\,17\% level keeping in mind the low number statistics (range of 0--33\%).

We looked at the main reasons for this low recovery rate and found the following conclusions:
\begin{itemize}
\item [$\bullet$] 16 out of 45 sources have separations outside our 1--5 arcsec range, including
15 with lower separations \citep{burgasser04c,lepine08b,lodieu10a,schmidt10a,kirkpatrick10,lodieu12b,zhang13}
and one with higher \citep[SDSS125637$-$022452;][]{sivarani09}. 
This loss represents about one third of the full sample and more than half of the losses.
It also represents the majority of losses in the 2MASS vs SDSS with three sources outside
our interval. We tested the selection of objects with separations in the 0.5--1.0 arcsec but
the numbers of candidates and contaminants increases quickly, making the spectroscopic
follow-up more complicated
\item [$\bullet$] 2 out of 45 sources are not classified as stars or stellar by SDSS (one object
with {\tt{cl}}\,=\,3) and UKIDSS (one source with {\tt{mergedClass}}\,=\,1) surveys 
\citep{lodieu12b,kirkpatrick10}
\item [$\bullet$] 4 out of 45 sources have differences outside our limits for the positional matching
in UKIDSS: {\tt{Xi}} and {\tt{Eta}} between $-$0.5 and 0.5 for $J$ and $K$
\citep{lepine08b,burgasser04c,zhang13,kirkpatrick10}
\item [$\bullet$] 5 out of 45 sources are lost due to their colours: one is lost because it has
$r-i$ of 0.96 mag, two have $g-r$ colours bluer than 1.8 mag, and two have $J-K$ colours above 0.7 mag
\citep{lodieu12b,zhang13}. In the case of the SDSS vs 2MASS cross-match, we lose one source because 
of its $J-K$ colour (sdM5.5)
\item [$\bullet$] 1 out of 6 sources in the SDSS vs 2MASS cross-match has a H$r$ value below
our limit of 20.7 mag (20.64 mag; esdM5.5)
\end{itemize}

As a consequence, we should revise the aforementioned density of cool subdwarfs per square degree
in SDSS and UKIDSS from 0.016$\pm$0.002 to 0.040$^{+0.012}_{-0.007}$ for the intervals quoted earlier. 
We do not correct the observed surface density of the 2MASS/SDSS cross-match because of the small
number statistics in the recovery rate.
The comparison between densities of M dwarfs and late-type subdwarfs is consistent with the 
upper limit of 0.68\% derived from the SDSS M dwarf sample \citep{covey08a}.

%
%
%
\section{Mid-infrared photometry of M subdwarfs}
\label{sdM_VO:results_new_WISE}

The Wide-field Infrared Survey Explorer mapped the sky at 3.4 ($W1$), 4.6 ($W2$), 12 ($W3$), 
and 22 ($W4$) $\mu$m \citep[WISE\footnote{wise.ssl.berkeley.edu};][]{wright10}.
We present mid-infrared photometry for our subdwarfs from AllWISE, which 
combines data from the WISE cryogenic and NEOWISE post-cryogenic survey phases
\citep{mainzer11}. We cross-matched our sample of subdwarfs with the AllWISE database using 
a matching radius of 12 arcsec to take into account the large motion of some subdwarfs.

\begin{figure*}
   \resizebox{\hsize}{!}{\includegraphics[clip=true]{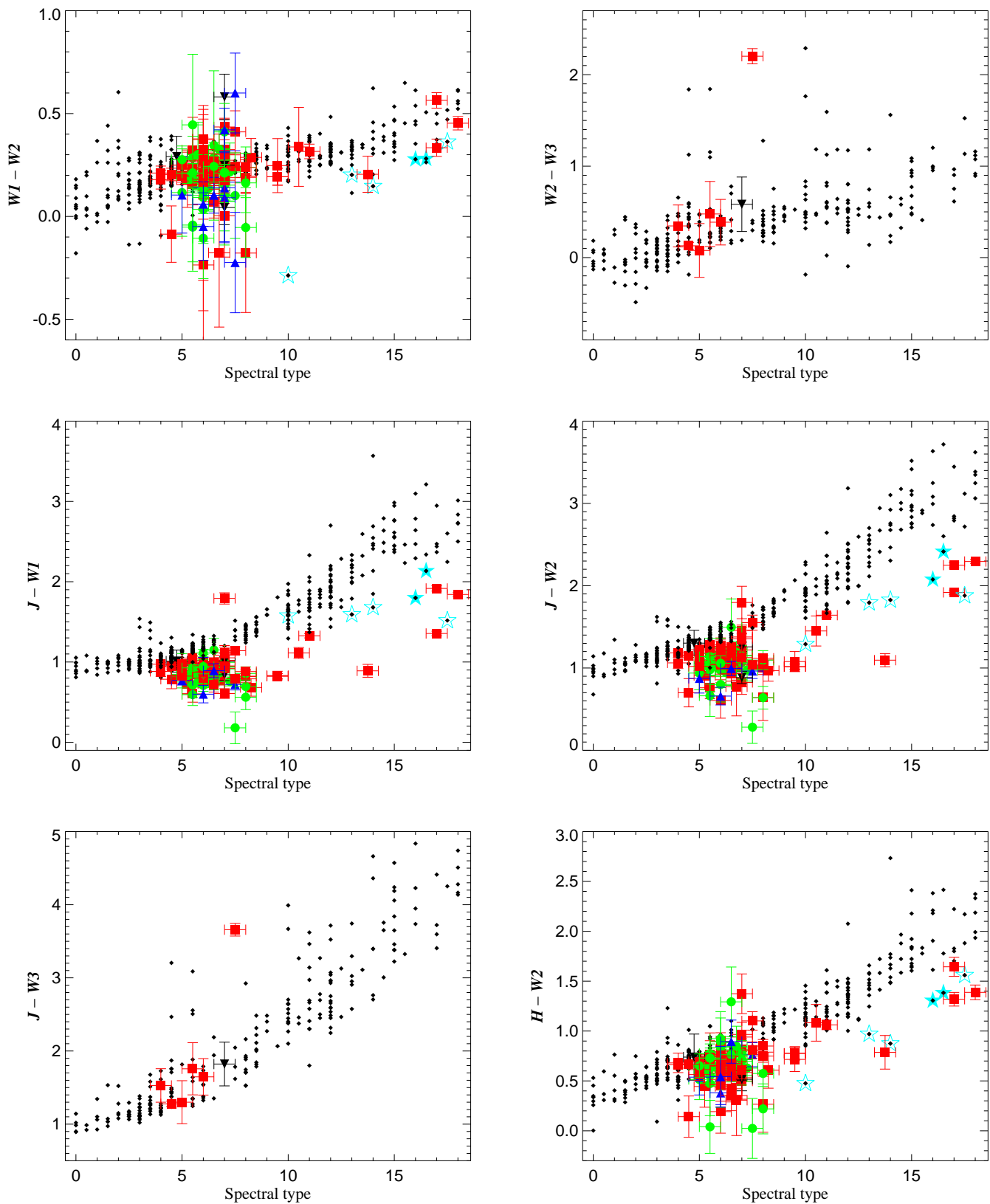}}
      \caption{Infrared colours of subdwarfs as a function of spectral types (increasing spectral
types to the right, where 0$\equiv$M0 and 10$\equiv$L0). We plot new and known subdwarfs, extreme subdwarfs, and ultrasubdwarfs as red squares, green circles, and blue triangles respectively. Small black diamonds are solar-metallicity M dwarfs from the DwarfArchives.org website with their AllWISE photometry, as well as M and L dwarfs from \citet{kirkpatrick11} with their WISE photometry. We highlighted as cyan stars field L dwarfs with features of L subdwarfs 
(Section \ref{sdM_VO:results_new_WISE}).
}
         \label{Fig_WISE}
   \end{figure*}

In Fig.\ \ref{Fig_WISE}, we plot several infrared colours as a function of spectral types 
showing our sample (coloured symbols) on top of the sequence of field M and L dwarfs 
\citep[black symbols;][]{kirkpatrick10}. This field dwarf sample contains 229 M dwarfs out 
of 536 listed in the DwarfArchives.org website as well as the M0--L8 dwarfs in \citet{kirkpatrick11}. 
We included only sources with detections above 3$\sigma$ and 
quoted error bars in each WISE passband. We overplotted our confirmed subdwarfs, extreme subdwarfs,
and ultrasubdwarfs as red squares, green circles, and blue triangles, respectively.

In Fig.\ \ref{Fig_WISE}, we see that subdwarfs do not appear to differ from those of solar-metallicity 
dwarfs of similar spectral types in $J-W3$ and $W2-W3$ colours. They appear slightly bluer in 
$W1-W2$ and $H-W2$ towards later spectral types, the latter being the most noticeable. The 
same trend becomes more obvious in the diagrams depicting the $J-W1$ and $J-W2$ colours, 
especially for metal-poor dwarfs with spectral types later than M7 that appear bluer than their 
solar-abundance counterparts.

We observe that one source, ID\,=\,14 (sdM7.0), lies above the sequence in most 
diagrams (Fig.\ \ref{Fig_WISE}). In particular, its $W2-W3$ and $J-W3$ colour appear
significantly redder than the sequence of M and L dwarfs. However, we caution this appearance
because its mid-infrared photometry might be contaminated by the presence of a close object
at the spatial resolution of WISE\@.

We checked the WISE images of all our subdwarfs. We confirmed that only one of them (ID\,=\,14)
is detected in the $W3$ and $W4$ bands and has a $W3-W4$ colour of 1.9$\pm$0.4 mag.
Among all M dwarfs in the DwarfsArchive.org website, no M7.5 dwarf has AllWISE photometry
with $W3$ and $W4$ detections with signal-to-noise higher than 3 (snr\,$>$\,3). However, the
mean $W3-W4$ colour solar-metallicity M dwarfs independent of their spectral types is
0.10 $\pm$ 0.05 mag (considering M dwarfs with snr$>$3 in both filters). Our subdwarf is
about two magnitudes redder in $W3-W4$ and deviate from the mean by more than 5$\sigma$.
Although this fact is based only on one late-M dwarf whose mid-infrared photometry might
be contaminated by the presence of a nearby source at the resolution of WISE, we might further 
investigate the role of metallicity between 10 and 20 $\mu$m either photometrically or spectroscopically.

We would like to add a special note here: in \citet{espinoza_contreras13}, we showed these
diagrams without imposing a constraint on the signal-to-noise ratio in the WISE bands considering
only subdwarfs from this paper, from \citet{lodieu12b}, and \citet{lepine08b}. However, if we include
only objects with snr\,$>$\,3, two objects pop up: ID\,=\,14 with $W2-W3$\,=\,2.2 mag and ID\,=\,16
with $W2-W3$\,=\,4.0 mag \citep{lodieu12b}. However, ID\,=\,16 does not have an error associated
to its $W3$ magnitude so we removed it from the revised version of the $W2-W3$ vs spectral type 
diagram displayed in the top right panel of Fig.\ \ref{Fig_WISE}.

\begin{figure}
   \resizebox{\hsize}{!}{\includegraphics[clip=true]{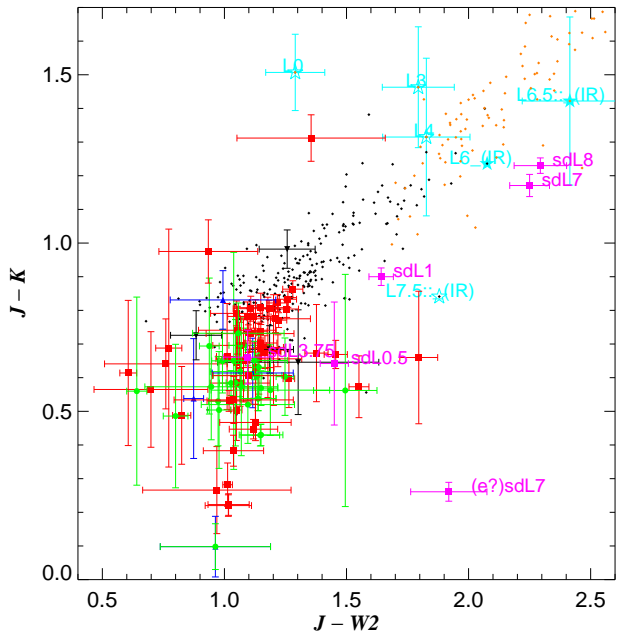}}
      \caption{We plot new and known M dwarfs and subdwarfs with different metallicity classes 
with the same symbology as in Fig.\ \ref{Fig_WISE}. We added field L dwarfs as small orange 
diamonds and known L subdwarfs with AllWISE photometry as filled magenta squares with their spectral type.}
         \label{Fig_colour-colour}
   \end{figure}

In Figures \ref{Fig_WISE} and \ref{Fig_colour-colour} we highlighted six objects that deviate 
from the main sequence of L dwarfs (cyan star symbols). All come from the sample
of \citet{kirkpatrick11} and appear bluer than field L dwarfs. We found that these six objects
have public spectra in the SpeX archive. We compare them to low-resolution
spectra of known subdwarfs with similar spectral types to clarify their nature. We see that 
WISEP\,J103322.01$+$400547.8, classified as a L6 in the near-infrared by \citet{kirkpatrick11} 
looks like a normal mid-L dwarf. The other two objects, WISEP\,J142227.23$+$221558.3 and
WISEP\,J102552.58$+$321231.5, classified in the near-infrared as L6.5:: and L7.5::, respectively
\citep{kirkpatrick11} show low-metallicity features, in particular stronger FeH lines and absence 
of the CO band around 2.3 $\mu$m. Thus, we argue that both are bona-fide metal-poor L dwarfs. 
We also checked the SDSS spectroscopic database and found optical spectra for 
WISEP\,J144938.12$+$235536.3 (L0) and WISEP\,J141011.08$+$132900.8 (L4). Both appear to 
exhibit some features typical of L subdwarfs in the optical, but higher signal-to-noise spectra 
over a wider wavelength range are needed to fully assess their nature. We are not able to
check the metal-poor nature of the remaining object (WISEP\,J143535.75-004347.4) classified as a L3\@. 
The blue nature of the $J-W1$ and $J-W2$ colours can be explained by the onset of 
collision-induced H$_{2}$ opacity operating at near-infrared wavelengths, typically beyond 2 $\mu$m. 
The collision-induced absorption is dependent on temperature and surface gravity \citep{saumon94} 
and dominates the opacities at high density and low temperatures \citep{lenzuni91}.
We propose that the $J-W1$ and $J-W2$ colours could be used to define new criteria to 
find cooler late-type subdwarfs in near- and mid-infrared surveys.

%
%
%
   
%
%
%

%
%
\section{Conclusions}
\label{sdM_VO:conclusions}

We have demonstrated that cross-correlations between optical and near-infrared large-scale surveys represent 
a very powerful tool to identify cool subdwarfs. Combining this study with \citet{lodieu12b}, we have increased 
the number of late-type subdwarfs by a factor of two and confirmed spectroscopically four new L subdwarfs. 
In this work, we report 68 new metal-poor M dwarfs, divided up into 36 subdwarfs, 26 extreme subdwarfs, and six 
ultrasubdwarfs, to which we should add two L subdwarfs and two dM/sdM, Our photometric and astrometric search 
shows success rates beyond the 80\% mark. The spectrophotometric distances of our new late-type subdwarfs range 
from 50 to \,$\sim$\,500 pc for subdwarfs and extreme subdwarfs. We inferred a surface density for
M-type subdwarfs of 0.033--0.052 per square degrees in the SDSS/UKIDSS cross-match in the 
$J$\,=\,15.9--18.8 mag range after correcting for incompleteness.

We also note that the proper motions calculated by the catalogue positions and epochs can sometimes be erroneous.
It is therefore necessary to check these proper motions and improve them to optimise further photometric and 
proper motion searches.

We searched for wide companions of early spectral types to our subdwarfs in different catalogues 
using our refined proper motions for those in SDSS vs UKIDSS and the PPMXL proper motions
for those in 2MASS and SDSS\@. We found one potential bright companion in the
Hipparcos-Tycho catalogue based on proper motion and spectroscopic distance. However the lack
of metallicity estimate does not lead to a strong conclusion. 

We also cross-matched our sample of new late-type subdwarfs as well as known subdwarfs with 
the AllWISE database to investigate the role of metallicity in the mid-infrared. We conclude that 
subdwarfs with spectral types later than M7 appear bluer than their solar-abundance counterparts
in the $J-W1$ and $J-W2$ colours, most likely due to the onset of the collision-induced H$_{2}$ opacity
beyond 2 $\mu$m. We suggest this as new colour criteria to look for ultracool subdwarfs in the future.

The main objective of this large project is to increase the number of metal-poor dwarfs, determine the 
space density of metal-poor dwarfs, improve the current classification of M subdwarfs, and expand it to 
the L subdwarfs (and later T) regime. We are now able to optimise our photometric and proper motion 
criteria and apply them to future searches in new data releases of optical, near-infrared, and mid-infrared 
large-scale surveys. This will allow us to increase the census of metal-poor dwarfs, 
especially at the coolest temperatures and lowest metallicities.
 
\begin{acknowledgements}
NL and MEC were funded by the Ram\'on y Cajal fellowship number 08-303-01-02, and supported by 
the grants numbers AYA2010-19136 and AYA2015-69350-C3-2-P from Spanish Ministry of 
Economy and Competitiveness (MINECO). ELM is supported by the MINECO grant number AYA2015-69350-C3-1-P\@.
The data presented here were obtained [in part] with ALFOSC, which is provided by the Instituto de Astrofisica de Andalucia (IAA) under a joint agreement with the University of Copenhagen and NOTSA.\\
Funding for the SDSS and SDSS-II has been provided by the Alfred P. Sloan Foundation, the Participating Institutions, the National Science Foundation, the U.S. Department of Energy, the National Aeronautics and Space Administration, the Japanese Monbukagakusho, the Max Planck Society, and the Higher Education Funding Council for England. The SDSS Web Site is http://www.sdss.org/. The SDSS is managed by the Astrophysical Research Consortium for the Participating Institutions. The Participating Institutions are the American Museum of Natural History, Astrophysical Institute Potsdam, University of Basel, University of Cambridge, Case Western Reserve University, University of Chicago, Drexel University, Fermilab, the Institute for Advanced Study, the Japan Participation Group, Johns Hopkins University, the Joint Institute for Nuclear Astrophysics, the Kavli Institute for Particle Astrophysics and Cosmology, the Korean Scientist Group, the Chinese Academy of Sciences (LAMOST), Los Alamos National Laboratory, the 
Max-Planck-Institute for Astronomy (MPIA), the Max-Planck-Institute for Astrophysics (MPA), New Mexico State University, Ohio State University, University of Pittsburgh, University of Portsmouth, Princeton University, the United States Naval Observatory, and the University of Washington.\\
This publication makes use of data products from the Two Micron All Sky Survey, which is a joint project of the University of Massachusetts and the Infrared Processing and Analysis Center/California Institute of Technology, funded by the National Aeronautics and Space Administration and the National Science Foundation.\\
The UKIDSS project is defined in \citet{lawrence07}. UKIDSS uses the UKIRT Wide Field Camera \citet[WFCAM;][]{casali07}. The photometric system is described in \citet{hewett06}, and the calibration is described in \citet{hodgkin09}. The pipeline processing and science archive are described in Irwin et al (2009, in prep) and \citet{hambly08}.\\
This research has made use of the Spanish Virtual Observatory (svo.cab.inta-csic.es) supported from the Spanish MICINN/MINECO through grants AyA2008-02156, AyA2011-24052.\\
This research has benefitted from the M, L, T, and Y dwarf compendium housed at DwarfArchives.org. This research has benefitted from the SpeX Prism Spectral Libraries, maintained by Adam Burgasser at http://pono.ucsd.edu/$\sim$adam/browndwarfs/spexprism.\\
This research has made use of data obtained from the SuperCOSMOS Science Archive, prepared and hosted by the Wide Field Astronomy Unit, Institute for Astronomy, University of Edinburgh, which is funded by the UK Science and Technology Facilities Council.\\	
\end{acknowledgements}

%
%
\bibliographystyle{aa}
\bibliography{../../AA/mnemonic,../../AA/biblio_old} 

%
%
\begin{appendix}

\section{The SVO subdwarf archive}
\label{sdM_VO:appendix_archive}

In order to help the astronomical community on using the catalogue of subdwarfs used in this paper,
we have developed an archive system that can be accessed from a 
webpage \footnote{http://svo2.cab.inta-csic.es/vocats/ltsa/} or through a Virtual Observatory 
ConeSearch\footnote{e.g. http://svo2.cab.inta-csic.es/vocats/ltsa/cs.php?RA=0\&DEC=0\&SR=100\&VERB=2}. 
We decided to include in the archive not only the subdwarfs reported in the paper (100) but 
all known ultracool subdwarfs with spectral types later than M5 from the literature (202). 

\subsection{Web access}
\label{sdM_VO:appendix_archive_web}

\begin{figure*}
 \centering
 \includegraphics[width=\linewidth]{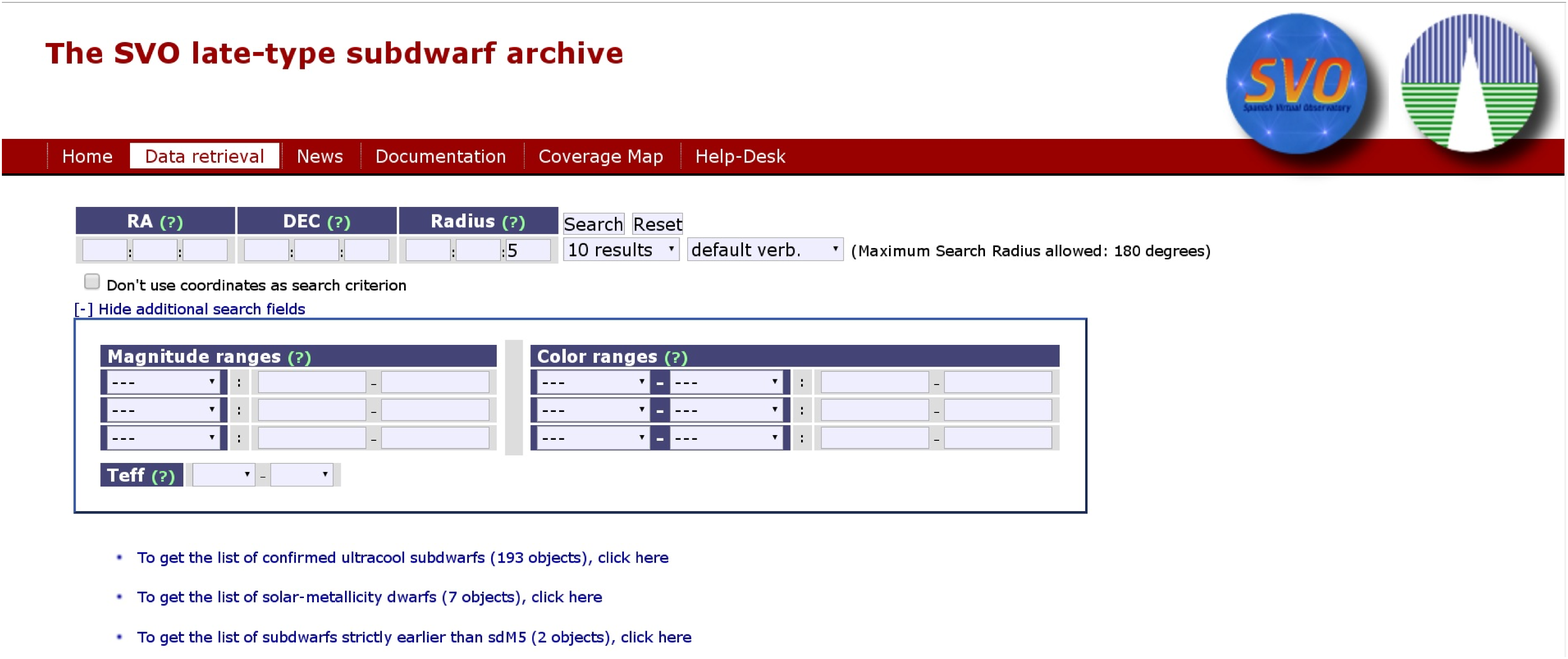}
      \caption{Screenshot of the archive search interface that permits simple queries.}
      \label{sdM_VO:appendix_figure_search}
\end{figure*}

The archive system implements a very simple search interface that permits queries by coordinates/radius 
and/or range of magnitudes, colours and effective temperatures. The default search radius is set to 
5 arcsec. The user can also select the maximum number of sources to return (with values ranging from 
10 to unlimited) (Fig.\ \ref{sdM_VO:appendix_figure_search}).

The result of the query is a HTML table with all the sources found in the archive fulfilling the search 
criteria. Detailed information on the output fields can be obtained placing the mouse over the question 
mark ("?") located close to the name of the column. The system returns the coordinates, both in decimal 
and sexagesimal degrees, the object identifier, the spectral type, direct access to the spectra
(in fits and ascii formats), the temperature obtained from
VOSA\footnote{http://svo2.cab.inta-csic.es/theory/vosa/} as well a visualization of the SED fitting by 
just cliking on the T${\rm eff}$ value, and the SDSS, 2MASS, UKIDSS (LAS and GCS surveys), VISTA 
Hemisphere Survey (VHS) and WISE magnitudes. The system also includes a link to the finderchart 
capability developed at IRSA\footnote{http://irsa.ipac.caltech.edu/applications/finderchart/}.

\begin{figure*}
 \centering
 \includegraphics[width=\linewidth]{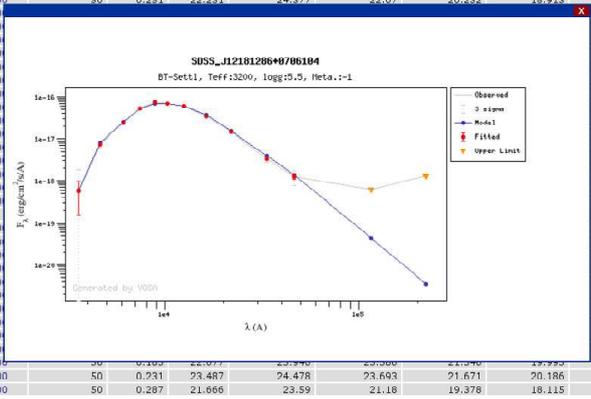}
      \caption{Screenshot of the typical window returned after a query.}
      \label{sdM_VO:appendix_figure_result}
\end{figure*}

The archive implements the SAMP\footnote{http://www.ivoa.net/documents/SAMP/} (Simple Application 
Messaging Protocol). SAMP allows applications to communicate with each other in a seamless and 
transparent way for the user. This way, the results of a query can be easily transferred to other 
VO applications, such as, for instance, Topcat (Fig. \ref{sdM_VO:appendix_figure_result}). 

%
%
%

%
%
\subsection{Virtual Observatory access}
\label{sdM_VO:appendix_archive_VO}

The Virtual Observatory (VO)\footnote{http://www.ivoa.net} is an international initiative whose 
primary goal is to provide an efficient access and analysis of the information hosted in astronomical 
archives and services. Having a VO-compliant archive is an important added value for an astronomical 
project to guarantee the optimum scientific exploitation of their datasets.

Our archive system has been designed following the IVOA standards and requirements. In particular, 
it implements the Cone Search protocol, a standard defined for retrieving records from a catalogue 
of astronomical sources. The query made through the Cone Search service describes a sky position and 
an angular distance, defining a cone on the sky. The response returns a list of astronomical sources 
from the catalogue whose positions lie within the cone, formatted as a VOTable.
\end{appendix}

\end{document}